 \documentclass[twocolumn]{aastex62}

\usepackage{graphicx}
\usepackage[caption=false]{subfig}
\usepackage{gensymb}
\usepackage{mathtools}
\usepackage{amssymb}
\usepackage{amsmath}
\usepackage{amsfonts}
\usepackage{xcolor}
\usepackage{afterpage}

\newcommand{\eg}{e.g.\@~}

\newcommand{\ie}{i.e.\@~}
\newcommand{\tbl}{Table~}

\newcommand{\feii}{[Fe \textsc{ii}]}
\newcommand{\sivi}{[Si \textsc{vi}]}
\newcommand{\nii}{[N \textsc{ii}]}
\newcommand{\sii}{[S \textsc{ii}]}

\newcommand{\oi}{[O \textsc{i}]}
\newcommand{\oiii}{[O \textsc{iii}]}

\newcommand{\hei}{He \textsc{i}}
\newcommand{\brg}{Br$\gamma$}
\newcommand{\brd}{Br$\delta$}
\newcommand{\molhy}{H$_2$}
\newcommand{\paa}{Pa$\alpha$}
\newcommand{\pab}{Pa$\beta$}
\newcommand{\hb}{H$\beta$}
\newcommand{\ha}{H$\alpha$}
\newcommand{\nev}{[Ne \textsc{v}]}

\shorttitle{KOALA: Feedback in (U)LIRGs}
\shortauthors{U et al.}

\begin{document}

  \title{Keck OSIRIS AO LIRG Analysis (KOALA):\\
    Feedback in the Nuclei of Luminous Infrared Galaxies} 
  
\correspondingauthor{Vivian U}
\email{vivianu@uci.edu}

\author[0000-0002-1912-0024]{Vivian U}
\affiliation{Department of Physics and Astronomy, 4129 Frederick Reines Hall, University of California, Irvine, CA 92697, USA}

\author[0000-0001-7421-2944]{Anne M. Medling}
\altaffiliation{Hubble Fellow}
\affiliation{Ritter Astrophysical Research Center, University of Toledo, Toledo, OH 43606, USA}
\affiliation{Cahill Center for Astronomy \& Astrophysics, California Institute of Technology, 1200 E. California Blvd., Pasadena, CA 91125, USA}
\affiliation{Research School of Astronomy and Astrophysics, The Australian National University, Canberra, ACT 2611, Australia}

\author[0000-0003-4268-0393]{Hanae Inami}
\affiliation{Centre de Recherche Astrophysique de Lyon, Observatoire de Lyon, 9 avenue Charles Andr\`{e}, 69230 Saint-Genis-Laval, France}

\author[0000-0003-3498-2973]{Lee Armus}
\affiliation{Spitzer Science Center, California Institute of Technology, Pasadena, CA 91125, USA}

\author[0000-0003-0699-6083]{Tanio D\'{i}az-Santos}
\affiliation{N\'{u}cleo de Astronom\'{i}a de la Facultad de Ingenier\'{i}a y Ciencia, Universidad Diego Portales, Av. Ej\'{e}rcito Libertador 441, Santiago, Chile}

\author[0000-0002-2688-1956]{Vassilis Charmandaris}
\affiliation{Department of Physics, University of Crete, GR-71003, Heraklion, Greece}
\affiliation{Institute for Astronomy, Astrophysics, Space Applications \& Remote Sensing, National Observatory of Athens, GR-15236, Athens, Greece}

\author[0000-0002-5924-0629]{Justin Howell}
\affiliation{Spitzer Science Center, California Institute of Technology, Pasadena, CA 91125, USA}

\author[0000-0002-2596-8531]{Sabrina Stierwalt}
\affiliation{IPAC, MC 100-22, California Institute of Technology, Pasadena, CA 91125, USA}

\author[0000-0003-3474-1125]{George C. Privon}
\affiliation{Department of Astronomy, University of Florida, 211  Bryant Space Sciences Center, Gainesville, FL 32611, USA}

\author[0000-0002-1000-6081]{Sean T. linden}
\affiliation{Department of Astronomy, University of Virginia, 530 McCormick Road, Charlottesville, VA 22904, USA}

\author[0000-0002-1233-9998]{David B. Sanders}
\affiliation{Institute for Astronomy, University of Hawaii at Manoa, 2680 Woodlawn Drive, Honolulu, HI 96822, USA}

\author[0000-0003-0682-5436]{Claire E. Max}
\affiliation{Department of Astronomy and Astrophysics, University of California, Santa Cruz, CA 95064, USA}

\author[0000-0003-2638-1334]{Aaron S. Evans}
\affiliation{Department of Astronomy, University of Virginia, 530 McCormick Road, Charlottesville, VA 22904, USA}
\affiliation{National Radio Astronomy Observatory, 520 Edgemont Road, Charlottesville, VA 22903, USA}

\author[0000-0003-0057-8892]{Loreto Barcos-Mu\~{n}oz}
\affiliation{National Radio Astronomy Observatory, 520 Edgemont Road, Charlottesville, VA 22903, USA}
\affiliation{Department of Astronomy, University of Virginia, 530 McCormick Road, Charlottesville, VA 22904, USA}
\affiliation{Joint ALMA Observatory, Alonso de C\'{o}rdova 3107, Vitacura, Santiago, Chile}

\author[0000-0002-0668-7865]{Charleston W. K. Chiang}
\affiliation{Department of Preventive Medicine, Keck School of Medicine, University of Southern California, Los Angeles, CA 90033}

\author{Phil Appleton}
\affiliation{IPAC, MC 100-22, California Institute of Technology, Pasadena, CA 91125, USA}

\author[0000-0003-4693-6157]{Gabriela Canalizo}
\affiliation{Department of Physics and Astronomy, University of California, Riverside, 900 University Avenue, Riverside, CA 92521, USA}

\author[0000-0002-0670-0708]{Giovanni Fazio}
\affiliation{Harvard-Smithsonian Center for Astrophysics, 60 Garden St., Cambridge, MA 02138, USA}

\author[0000-0002-4923-3281]{Kazushi Iwasawa}
\affiliation{ICREA and Institut del Ci\`{e}ncies del Cosmos, Universitat de Barcelona, Barcelona, Spain}

\author{Kirsten Larson}
\affiliation{Spitzer Science Center, California Institute of Technology, Pasadena, CA 91125, USA}

\author[0000-0002-8204-8619]{Joseph Mazzarella}
\affiliation{IPAC, MC 100-22, California Institute of Technology, Pasadena, CA 91125, USA}

\author[0000-0001-7089-7325]{Eric Murphy}
\affiliation{National Radio Astronomy Observatory, 520 Edgemont Road, Charlottesville, VA 22903, USA}

\author[0000-0002-5807-5078]{Jeffrey Rich}
\affiliation{Observatories of the Carnegie Institution for Science, 813 Santa Barbara St., Pasadena, CA 91101}

\author[0000-0001-7291-0087]{Jason Surace}
\affiliation{IPAC, MC 100-22, California Institute of Technology, Pasadena, CA 91125, USA}

 \begin{abstract}
The role of feedback in triggering or quenching star formation and hence
driving galaxy evolution \replaced{is being better established with the
advent of emission line modeling and}{can be directly studied with}
high resolution integral field 
observations. The manifestation of feedback in shocks is particularly
important to examine in galaxy mergers, where violent
interactions of gas takes place in the interstellar medium during the
course of the galactic collision. 
As part of \replaced{the Great Observatories All-Sky LIRG Survey's effort to
understand the local population of luminous infrared galaxies on the
galaxy transformation path}{our effort to systematically study
the local population of luminous infrared galaxies within the
Great Observatories All-Sky LIRG Survey}, we undertook \replaced{an observing
campaign, the Keck OSIRIS AO LIRG Analysis Survey,}{the Keck OSIRIS AO
LIRG Analysis observing campaign} to study the gas
dynamics \replaced{at the obscured but active}{in the} inner kiloparsec regions of these
systems at \deleted{small, resolved} spatial scales \added{of a few
  10s of parsecs}. 
With high-resolution near-infrared adaptive optics-assisted
integral-field observations taken with OSIRIS on the Keck Telescopes,
we employ near-infrared diagnostics such as \deleted{the} \brg~and the
ro-vibrationally excited \molhy~lines to quantify the nuclear star
formation rate and identify feedback
associated with shocked molecular gas seen in 21 nearby luminous
infrared galaxies. Shocked molecular gas is preferentially found in the
ultraluminous infrared systems, but may also be triggered at a
lower-luminosity, earlier merging stage. \added{On circumnuclear
  scales, AGN have a strong effect on heating the surrounding molecular
gas, though their coupling is not simply driven by AGN strength but
rather is complicated by orientation, dust shielding, density, and other
factors. 
%High extinction levels (up to $A_V \sim 40$ mag) were
%    measured, leading to a factor of $10-30$ enhancement in the
%    corrected star formation rate in these nuclear regions.
} We find that the nuclear 
star formation correlates with \deleted{global infrared luminosity,}
merger class and diminishing 
projected nuclear separations. These trends are largely consistent
with the picture of merger-induced starbursts within the center of
galaxy mergers. 
 \end{abstract}


   \section{Introduction}
A principal process that regulates the stellar content of a galaxy and
drives its evolution is \replaced{feedback, usually from star formation and/or
the accreting black hole at its center}{energetic feedback from stars,
supernovae, and accreting supermassive black holes}. Feedback injects energy and
momentum into a galaxy's interstellar medium (ISM) and can drive
powerful winds from the nucleus to galactic scales. These winds
play a key role in the chemical evolution of galaxies through
metal enrichment, metal redistribution, and the suppression or
enhancement of star formation throughout the galaxy. 

Galactic outflows are prevalent in (ultra-)luminous infrared galaxies
((U)LIRGs), which in the local universe ($z < 0.08$) are mostly galaxy
mergers~\cite[][]{Armus87,Sanders88,Larson16}. Mergers funnel gas to
the central black hole or nuclear starburst, which in turn transfers
momentum to the surrounding dense gas and potentially drives an outflow. This
phenomenon has been seen both in
simulations~\cite[][]{Mihos94,Springel03,Hopkins05,Narayanan06,Narayanan08,Torrey12,Muratov15,Nims15} and in
observations~\cite[][]{Heckman90,vanderWerf93,Veilleux95,Veilleux05,Veilleux13,Moran99,Rupke02,Rupke11,Rupke13,Martin06,Dasyra11a,Dasyra11b,Rich11,Rich15,Soto12,Spoon13}. 

Because feedback \replaced{is manifested in shocks, turbulent, or high
  velocity gas, emission line ratios and gas dynamics are pivotal
  detection tools for these mechanisms}{usually manifests itself in
  terms of high fractions of shocked and/or outflowing atomic gas,
  unusual optical and infrared emission line ratios, blueshifted and
  redshifted emission lines, and broad emission lines
  signalling turbulent gas 
  are pivotal detection tools}. For instance, optical emission
line ratios provide a critical diagnostic that illuminates the
excitation and ionization conditions, \eg through the use of
\replaced{[S II], H$\alpha$, [N II], and [O I]}{\oiii, \hb, \sii, \ha,
  \nii, and \oi} for distinguishing AGN from starbursts 
with the so-called ``BPT/VO87
diagrams''~\cite[][]{Baldwin81,Veilleux87,Kewley01,Yuan10}. The
interpretation of these optical line ratios in the context of
photoioniziation and radiative shocks is further facilitated by 
\replaced{fast and shock models predicted
  from}{self-consistent codes such as} MAPPINGS
III~\cite[\eg][]{Allen08} and others \deleted{shock-modelling codes}
\cite[][]{Farage10,Rich10,Rich11}. However, such
optical-based tools encounter challenges at the physical and technical
levels: \ie that shocks and photoionization from the AGN are difficult
to differentiate using optical ratios alone, and these emission
lines may suffer high levels of dust attenuation~\cite[$A_V >
10$;][]{Stierwalt13,Piqueras13}, particularly in the nuclei of (U)LIRGs. It is
nontrivial to investigate the \replaced{shock
  contribution}{contribution of shocks} to the total energy
budget of the feedback when the optical line ratios are muddled by
mixtures of shocks, starbursts, AGN, and clumpy dust. 

The mid-infrared counterparts of these AGN and star formation tracers
include high excitation lines like \added{\molhy,} [Ne V] or [O IV], and Polycyclic
Aromatic Hydrocarbons or Unidentified Infrared Bands~\cite[PAHs,
UIBs;][]{Genzel98,Laurent00,Armus07,Armus09,Spoon07,Groves08,Dale09,Petric11,Alonso12,Stierwalt14}.  
However, most mid-infrared and ground-based
seeing-limited observations are hampered by \added{a lack of high}
angular resolution, which 
is needed to \replaced{provide direct evidence for the effect of
  positive or negative feedback at local star-forming sites, \ie if
  star formation is enhanced or suppressed at the scales of giant
  molecular clouds ($\leq 0\farcs1$).}{resolve the regions from which
  winds originate because they are often complex and involve multiple
  ionization sources.}

In the near-infrared, using line diagnostics for studying the ISM excitation
and ionization has been challenging. Theoretical models \replaced{for
understanding molecular physics at these wavelengths}{that integrate
the complexities of molecular physics to radiative transfer} have been
limited. Nonetheless, empirical studies have produced 
BPT-like diagnostic diagrams 
in order to assess the excitation and ionization conditions of the
gas. Since iron in the ISM is highly depleted onto grains, strong
\feii~emission in the near-infrared is often associated with
shock-excited
gas~\cite[][]{Larkin98,Rodriguez04,Rodriguez05,Riffel06,Riffel13}
where the grains have been processed \replaced{from}{by} winds, supernovae, or other
sources.  Line ratios such as \feii \added{(1.26$\mu$m)}/\pab~and \molhy~\added{1$-$0 S(1)}/\brg~are thus helpful
in differentiating between starbursts, AGN, 
low-ionization nuclear emission-line regions (LINERs), and Seyferts.

In many systems, multiple mechanisms may contribute to the ISM
conditions to different degrees, \replaced{high angular-resolution
integral-field spectroscopy (IFS)}{and integral-field spectroscopy
(IFS) with high angular resolution} is crucial to distinguish them. 
The integrated measurements from large apertures or coarse resolution
observations tend to blend signals from interesting starburst, AGN,
and shocked regions, presenting a median measurement representative of
the general excitation and ionization conditions of the diffuse ISM in these LIRGs.
Differences between line ratios measured from integrated versus
spatially-resolved regions may be as large 
as a factor of two in the case of LIRGs~\cite[][]{Colina15}. 
In order to discern the true underlying contribution from the
ionizing sources, we need to look into the near-infrared with
adaptive-optics (AO)-assisted IFS
for a \replaced{better, high}{higher} resolution view of the central gas properties. Near-infrared
studies at high spatial resolution form the basis for our
understanding of the mechanisms powering 
emission lines in dusty ULIRGs~\cite[][]{U13,Medling15_ir17207} and
will pave the way for the upcoming \emph{James Webb Space Telescope
  \added{(JWST)}} era.  

Here as part of the Keck OSIRIS AO LIRG Analysis (KOALA)
Survey\footnote{\url{https://koala-goals.github.io}}, we
present the \added{warm} molecular and atomic gas dynamics in the nuclear
regions of 21 nearby (U)LIRG systems (22 nuclei), approximately
40\% of \deleted{all} the local (U)LIRGs suitable to \replaced{be observed}{observe} with the
\added{current} Keck AO system 
\added{due to limitations in the availability of tip-tilt stars}. 
The larger campaign and early results have  
been introduced in our previous work: in~\cite{Medling14} (hereafter
Paper I) we presented the gas and stellar morphology and kinematics
\replaced{, which, when fitted with GALFIT,}{and} characterized the properties of the
nuclear disks that were found to be nearly ubiquitous in these
(U)LIRGs. ~\cite{Medling15_bh} (hereafter Paper II) built upon these
nuclear disks, \deleted{and} computed the dynamical masses for the central
supermassive black holes, and \replaced{finds}{found} that \deleted{the} late-stage mergers have
central masses that are overmassive relative to the $M_{\rm BH}-\sigma_\star$
relation of normal galaxies. Here, \replaced{our campaign continues with an in-depth
exploration of the identification of molecular outflows and the effect
of feedback on the host galaxy}{we present our search for outflows and
the effects of feedback in the nuclei of the galaxies in the KOALA survey}.

This paper is organized as follows: \S \ref{sample} describes the
subsample from our larger Keck campaign analyzed for gas 
properties, as well as an overview of the acquisition and processing
of Keck OSIRIS data presented here. \S \ref{analysis} describes the
line-fitting analysis while \S \ref{results} presents the resulting emission line
maps and line ratio analysis. Various
diagnostics using these near-infrared emission lines to discern ionization
sources and excitation mechanisms are subsequently discussed in \S
\ref{discussion}. Our summary is presented in \S \ref{summary}. The kinematics of the
gas will be presented in a forthcoming paper. Throughout the paper, we have adopted
$H_0 = 70$\,km\,s$^{-1}$\,Mpc$^{-1}$, $\Omega_{\rm m}$ = 0.28, and 
 $\Omega_\Lambda$ = 0.72~\cite[][]{Hinshaw09}.

  \section{Data}
  \label{sample}

  \subsection{The KOALA-GOALS Survey}
  The Great Observatories All-sky LIRGs
  Survey~\cite[GOALS;][]{Armus09} consists of 201 of the brightest and
  closest (U)LIRGs in the local universe, a complete subset of the
  flux-limited IRAS Revised Bright Galaxy Sample~\cite[$f_{60\mu m} >
  5.24$ Jy and galactic latitude $|b| > 5\degree$;][]{Sanders03}. 
  \replaced{It includes a rich ancillary data reservoir}{The survey
    incorporates a wealth of ancillary data} spanning the entire
  electromagnetic spectrum from \emph{Chandra} \emph{X-ray} to
  the \emph{Very Large Array} radio regime.  Our near-infrared 
  Keck sample is limited to selecting from among the most luminous
  ($L_{\rm IR} > 10^{11.4}$) 88 objects
  \replaced{within GOALS's \emph{HST}-ACS coverage that host}{with
    coverage by \emph{HST}-ACS observations that have} suitable guide stars
  for the AO system.  The \emph{HST}-ACS criterion ensures
  that we have high spatial resolution (0\farcs05) images with
  precision astrometry for planning the AO observations, which is
  critical given the small field-of-view (FOV) of OSIRIS. At the
  distance of our sample ($z < 0.08$, though mostly $z < 0.05$), we
  resolve the inner kiloparsec region of each source with $\sim
  20-80$ parsec per resolution element.

  Incorporating observing feasibility~\cite[\eg observable from
  Mauna Kea and with available guide stars that fulfill the tip-tilt
  requirements of the Keck LGS/NGS AO
  systems;][]{Wizinowich06,vanDam06,Wizinowich00}, 
  approximately half the \emph{HST}-GOALS sample may be followed up with
  the Keck AO system.  \added{Since one of our program's goals is to
    trace feedback properties along the merging sequence, our
    observing priorities took into account the merger classification
    scheme for the GOALS sample as adopted
    from~\cite{Haan11} and~\cite{Kim13}: (0) single galaxy with no obvious major
    merging companion; (1) separate galaxies 
with symmetric disks and no tidal tails; (2) distinguishable
progenitor galaxies with asymmetric disks and/or tidal tails; (3) two
distinct nuclei engulfed in a common envelop within the
merger body; (4) double nuclei with visible tidal tails; (5) single
or obscured nucleus with prominent tails; and (6) single or obscured
nucleus but with disturbed morphology and short faint tails signifying
post-merger remnant.}

\begin{figure}[hbt]
  \centering
  \includegraphics[width=.48\textwidth]{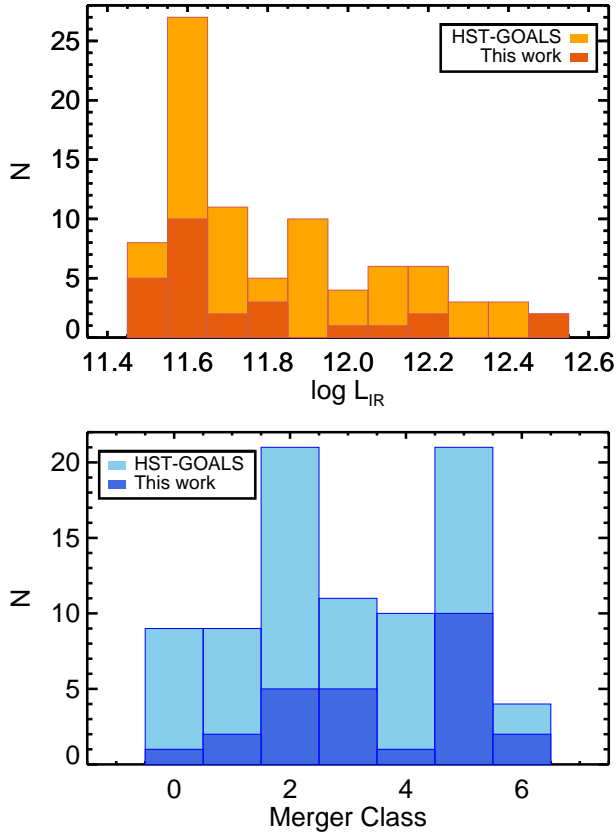}
  \caption{Distributions in infrared luminosity (top) and in merger
    class (bottom) for the \emph{HST}-GOALS sample (light orange,
    light blue) and for the galaxy systems in this work (dark orange,
    dark blue), respectively. See explanation for the merger
    classification scheme in text. Only approximately half of the
    \emph{HST}-GOALS sources may be observed with Keck AO. The
    existing KOALA sample is not yet complete with regards to the
    observing limitations within the \emph{HST}-GOALS sample, but
    spans a representative range.}
  \label{fig:sampcomp}
\end{figure}

 We initially prioritized late-stage
  mergers with one nucleus or two \replaced{closeby}{close} nuclei on the verge of
  coalesence (Papers I \added{and II}). Our campaign has since been extended to
  include systems at earlier stages of merging \added{that tend to be
    at lower luminosities and thus more closely resemble normal
    star-forming galaxies in other nearby
    studies~\cite[e.g.][]{Kennicutt03}}. Our current study   
  comprises 21 interacting systems (22 nuclei) (see \tbl
  \ref{tbl:obs}) for which $K$-band data targeting atomic and
  molecular hydrogen transitions have been
  gathered. \added{Comparisons of our current KOALA sample with the
    entire \emph{HST}-GOALS sample in infrared luminosity and in
    merger stage are shown in 
  Figure \ref{fig:sampcomp}. The KOALA sample as presented here is not
  yet complete with respect to the limitations imposed by the
Keck AO system requirements. However, it has achieved a representative
range in $L_{\rm IR}$ and merger stage relative to the
\emph{HST}-GOALS sample, which is complete in infrared 
luminosity within GOALS down to $L_{\rm IR} \geq 10^{11.4}$. }

  \subsection{Observations and Data Processing}
  The observations were taken with \added{the OH-Suppressing Infra-Red
    Imaging Spectrograph}~\cite[OSIRIS;][]{Larkin06} on the
  Keck II \replaced{Telescope during 2010$-$2012 and on the Keck I
    Telescope after 2013}{and Keck I Telescopes before and after
  January 3, 2012, respectively}. 
  The weather conditions varied from run to run but were sufficiently good (with
  typical FWHM $\sim$ 0$\farcs$06 estimated from imaging the tip-tilt star). 
  We employed adaptive optics corrections with both
  natural and laser guide stars, the tip-tilt requirements for which
  are R magnitude $<$ 15.5 and separation $r < 35\arcsec$, and R
  magnitude $<$ 18.5 and $r < 65\arcsec$, respectively.

  For this study, we focused on the observations
  obtained with the broad $K$-band filter `Kbb' (and `Kcb', which has
  identical \added{spectral} specifications as `Kbb' but \replaced{the
    difference is only in its designation when}{with a pupil} paired \replaced{with}{to}
  the 100 mas plate scale) covering 
  wavelengths 1965 nm $< \lambda <$ 2381 nm with spectral resolution
  $R \sim 3800$. This wavelength range
  enables us to target primarily the molecular hydrogen ro-vibrational
  transitions and atomic hydrogen recombination lines
 as tracers for shock excitation and \replaced{star-forming gas}{star formation},
 respectively. Our choice of observing mode, balancing the size of the
 FOV and the angular resolution, depended upon each target's redshift and
 spatial features. The FOV is  
  $0\farcs56 \times 2\farcs24$ for the 35 milliarcsec (mas) plate
  scale, $0\farcs8\times3\farcs2$ for the 50 mas plate scale, and
  $1\farcs6 \times 6\farcs4$ for the 100 mas plate scale. We aimed to center
  the observations at our best guess of the kinematic nucleus or
  nuclei based on high-resolution \emph{HST}-ACS $F435W$ and $F814W$
  observations~\cite[][Evans et al., in prep.]{Kim13}, aligned at
  position angles that would best capture multiple nuclear star clusters
  and/or the kinematics of the nuclear region (see Paper I for
  \deleted{detailed GALFIT results for} both the gaseous and stellar
  distributions).  Standard A0V stars were imaged 
  throughout the observing runs for telluric corrections.  A summary
  of the observation details for the individual galaxy systems can be
  found in \tbl \ref{tbl:obs}.

  Most of the data sets were processed using the standard OSIRIS
  pipeline~\cite[][]{Krabbe04} that incorporates dark-frame
  subtraction, channel level adjustment, crosstalk removal, glitch
  identification, cosmic ray cleaning, data cube assembly, dispersion
  correction, scaled sky subtraction for enhanced OH-line suppression,
  and telluric correction. The extracted spectra were subsequently
  processed \deleted{so} to remove bad pixels.  More details regarding the
  reduction process for specific galaxy systems may be found
  in~\cite{U13,Medling15_ir17207,Davies16}, Paper I, and Paper II.

  \section{Emission Line Fitting and Continuum Extraction}
  \label{analysis}
 
  We have detected five ro-vibrational transitions of \molhy~observable within the
  OSIRIS $K$ broadband filter \deleted{ranges} with varying signal-to-noise
  ratios \added{(S/N)} in all of our galaxies: $1-0$
  S(3) ($\lambda_{\rm rest}$ = 1.9576$\mu$m), $1-0$ S(2)
  ($\lambda_{\rm rest}$ = 2.0338$\mu$m), $1-0$ S(1) ($\lambda_{\rm
    rest}$ = 2.1218$\mu$m), $1-0$ S(0) ($\lambda_{\rm  rest}$ =
  2.2235$\mu$m), and $2-1$ S(1) ($\lambda_{\rm rest}$ = 2.2477$\mu$m).
  In addition, \replaced{a}{our} typical $K$ broadband spectrum
  \deleted{(\eg Figure~\ref{fig:1dspec})} also 
  features \brg~($\lambda_{\rm rest}$ = 2.166 $\mu$m) and
  \brd~($\lambda_{\rm rest}$ = 1.945 $\mu$m) lines, the ratio of
  which provides a measure of the dust extinction in the
  region. The median \deleted{representative integrated} line fluxes and ratios
  for these \molhy~and Brackett line transitions within the OSIRIS FOV
  are compiled in \tbl~\ref{tbl:fluxes}.  
  \hei~($\lambda_{\rm rest}$ = 2.059 $\mu$m) and
  \sivi~($\lambda_{\rm rest}$ = 1.965 $\mu$m) are also 
  within our spectral coverage, and will be presented in a forthcoming
  paper. \added{The entire suite of integrated 1D spectra for our sample is
    presented in Figure~\ref{fig:1dspec}, featuring a variety of continuum
    levels and line ratios of the aforementioned emission lines.} From
  our IFS data cubes we map the 
  morphology and kinematics of these emission lines with the 
  line-fitting method presented in~\cite{U13} and Paper I. Our method
  can be summarized as follows. 

\begin{figure*}[htbp]
  \centering
%  \includegraphics[width=.9\textwidth]{paper_1dspec_all.eps}
%\subfloat[]{
%\begin{subfigure}[h]{\textwidth}
\includegraphics[width=.9\textwidth]{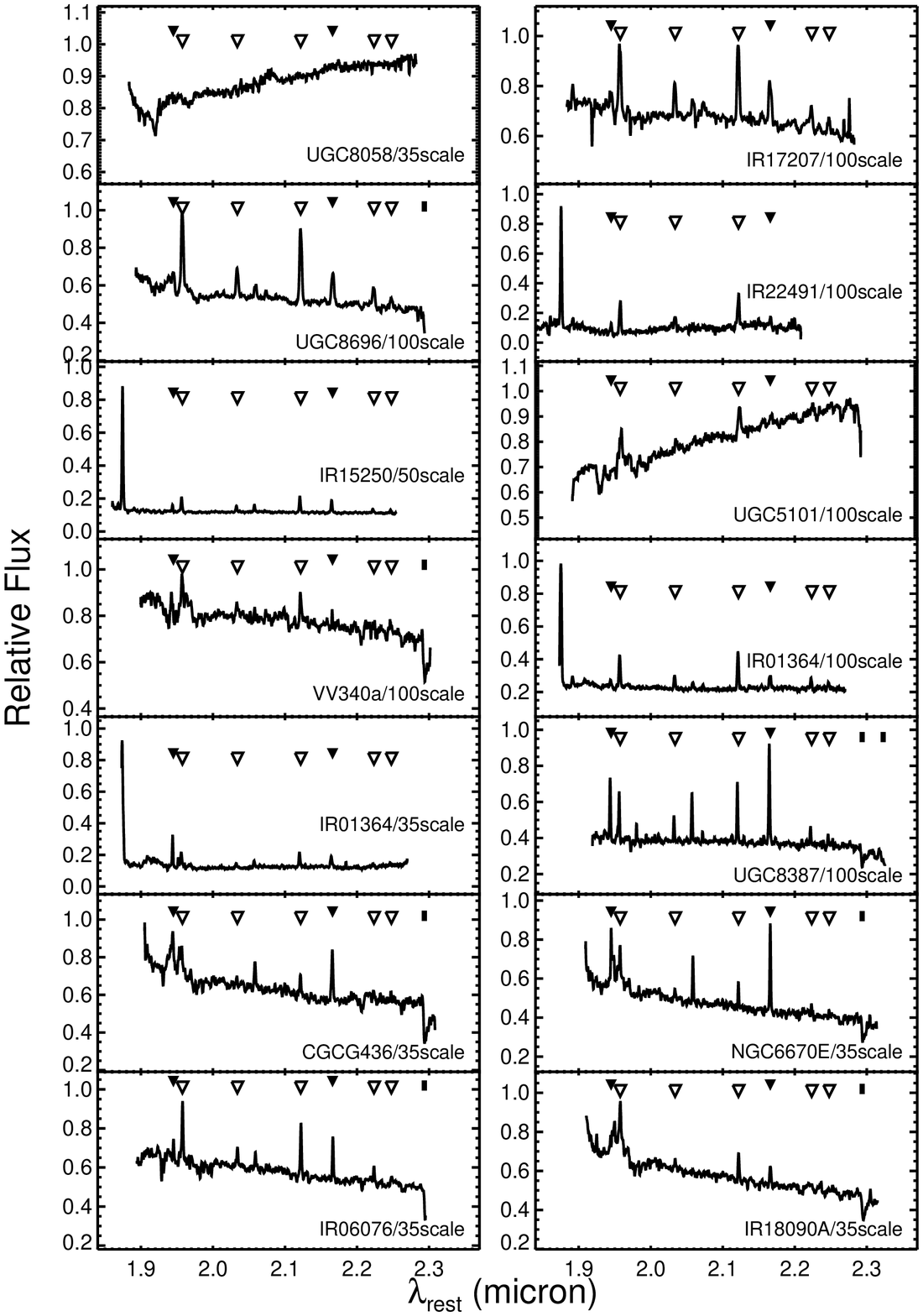}
%\end{subfigure}
%}
%  \includegraphics[width=.9\textwidth]{paper_1dspec_multipg2.eps}
%  \label{fig:1dspec}
\end{figure*}

\setcounter{figure}{2}
\begin{figure*}[htbp]
  \ContinuedFloat
 \centering
%  \subfloat[]{
%\begin{subfigure}[h]{\textwidth}
\includegraphics[width=.9\textwidth,trim={0cm 3cm 0cm 0cm}]{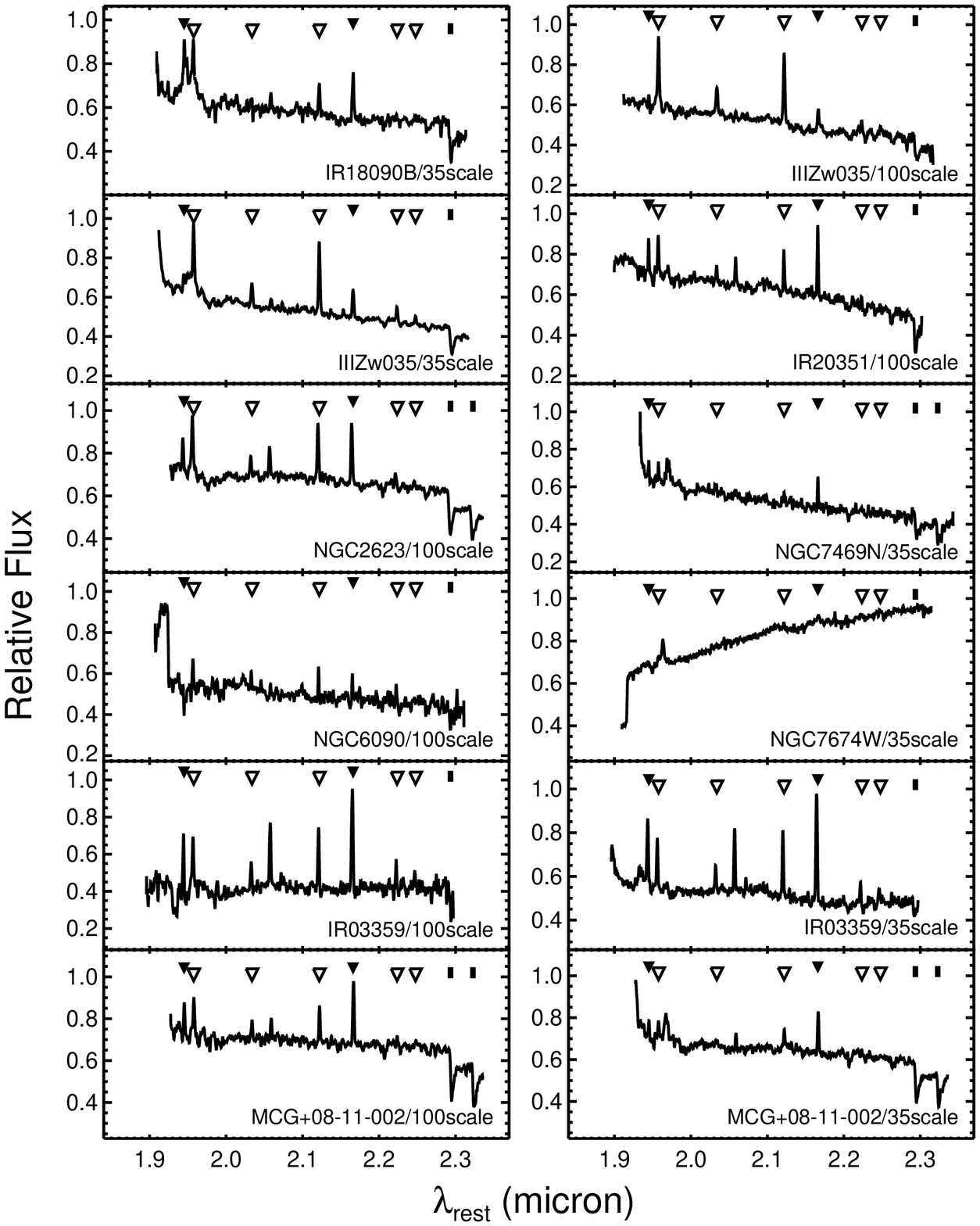}
%\end{subfigure}
%}
%  \phantomcaption
  \caption{1D $K-$broadband spectrum averaged over the nuclear region
    for all the galaxies in this sample. \added{The spectra are
      normalized to a common scale for better presentation of the
      various emission and absorption features.} The open downward triangles
    mark the five \molhy~transitions, the filled downward triangles mark
    the two Br transitions, while the filled bars mark the CO \added{absorption}
    bandheads. Not marked on the plots are \paa~($\lambda_{\rm rest}$
    = 1.875 $\mu$m), \hei~($\lambda_{\rm rest}$= 2.059 $\mu$m), and
    \sivi~($\lambda_{\rm rest}$ = 1.964 $\mu$m). The sample exhibits a
    variety of continuum slopes, including the positive slopes
      typical of strong Seyfert cases. \added{Object names and observed plate
        scales are labeled accordingly.}
\explain{We decided to enlarge the spectra for clarity purposes at the
  expense of expanding this figure into two pages.}
}
  \label{fig:1dspec}
\end{figure*}
%\end{subfigures}
  
  For each of the reduced data cubes, we extracted the underlying power-law
  continuum and subtracted it from the spectra in each spaxel. We then
  fit the lines with a Gaussian profile to determine the flux, velocity, and
  velocity dispersion. \added{The choice of using a single
    component Gaussian fit was motivated by low S/N within individual
    bins. In some cases, multi-component fits might be justified over
    integrated areas. For example, in Mrk 
  273, two Gaussian components were fit to the \sivi~line from
  different integrated regions within the OSIRIS FOV that exhibited
  different kinematics arising from two distinct parts of the
  outflow.~\cite[See][for more details.]{U13} In IRAS F17207$-$0014,
  the integrated \molhy~flux from the shocked region was decomposed
  into two Gaussian fits that were consistent with the velocities of
  the western and eastern nuclei,
  respectively~\cite[][]{Medling15_ir17207}. Multi-component spectral
  fitting will be incorporated for integrated fluxes over specific
  regions in the outflow candidates in our follow-up work on the
  detailed kinematic analysis of winds.}

  In the case of the ro-vibrational molecular hydrogen transitions and
  the hydrogen recombination lines, we fit the lines simultaneously by
  species. \replaced{For the benefit of increasing the S/N for regions with
  inadequate signal}{To increase the signal-to-noise ratio in the
  individual line maps}, we adopted optimal Voronoi binning using
  the~\cite{Cappellari03} code so that the emission line in each
  spatial bin achieves a S/N of at least 3. In order to make proper
  comparisons of the \molhy~and \brg~lines within each galaxy, we
  further imposed the spatial bins computed from the the strongest
  \molhy~transition within this regime, \molhy~1$-$0 S(1), onto the
  \brg~line while generating the flux, velocity, and velocity
  dispersion maps of the latter. This binning constraint allows us to have 
  bin-matched line ratio maps for all the galaxies. The errors for the
  line and line ratio maps were determined from refitting the line
  parameters to a synthetic data cube with added random noise 500
  times and propagated accordingly.  We present our extracted
  continuum and line emission, line ratio, \deleted{kinematic} and other derived
  property maps for \deleted{III Zw 035 ($L_{\rm IR} = 11.62$, $z = 0.0278$)}
  \added{our sample} in Figure~\ref{fig:6panel}.
\deleted{and for the rest of the sample in the Appendix.}

\figsetstart
\figsetnum{3}
\figsettitle{OSIRIS $K$-band Line Maps}

\figsetgrpstart
\figsetgrpnum{2.1}
\figsetgrptitle{III Zw 035 (35mas)}
\figsetplot{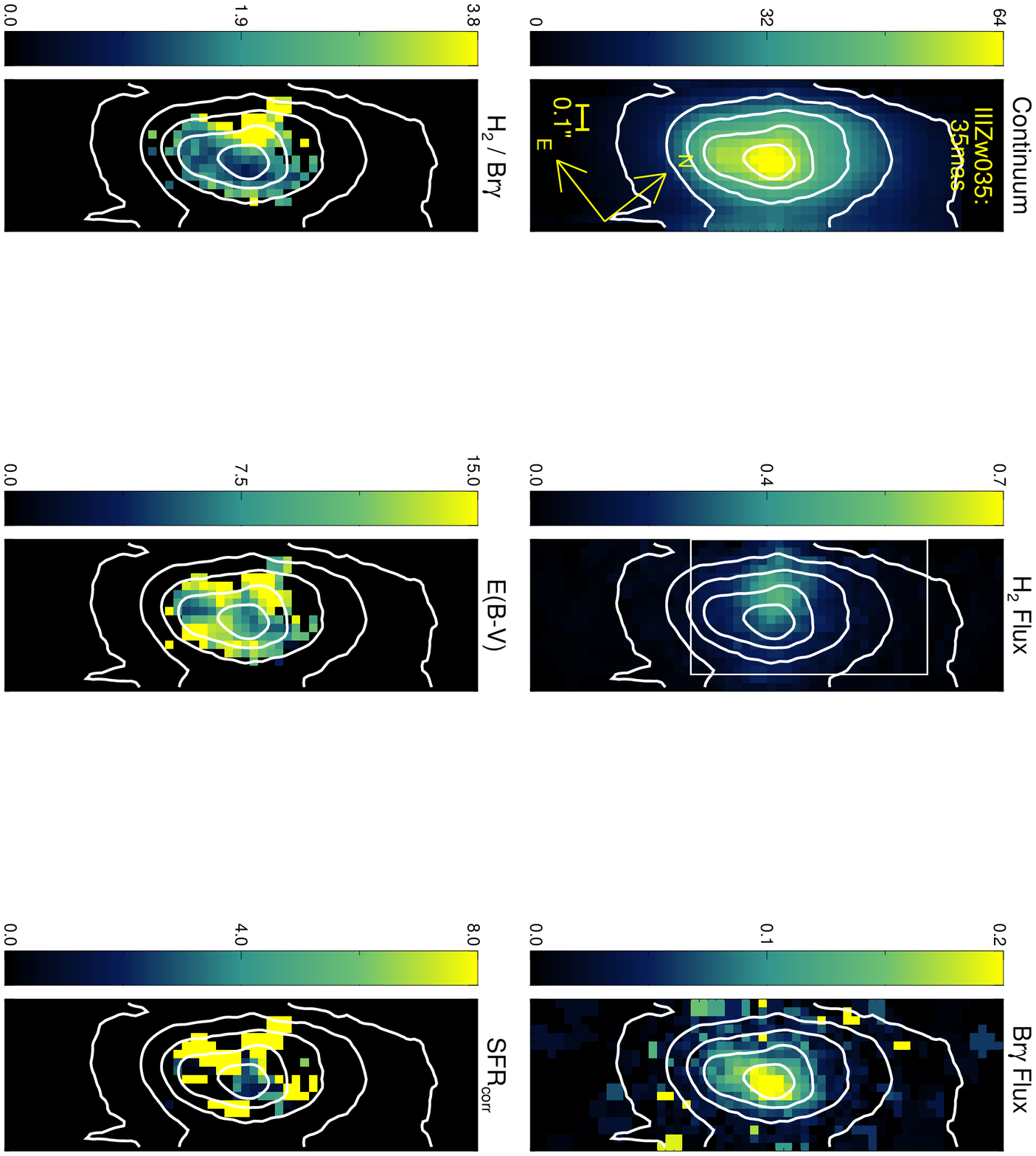}
\figsetgrpnote{Six-panel figure showing, from left to right, top to
  bottom, 1. $K$-band continuum map in relative flux units and shown
  with 0.1\arcsec scale bar and compass rose; the continuum contours
  are shown in all subsequent panels. 2. \molhy~1$-$0 S(1) flux map in
  10$^{-16}$ erg s$^{-1}$ cm$^{-2}$, with box highlighting the
  ``\molhy-dominated region"; 3. \brg~flux map in 10$^{-16}$ 
  erg s$^{-1}$ cm$^{-2}$; 4. \molhy~1$-$0
  S(1)/\brg~(\molhy/\brg~hereafter) map showing the reliable spaxels
  (with S/N $>$ 3 in both \molhy~and \brg); 5. E(B-V) map; and
  6. dust-corrected star formation rate map (M$_\odot$ yr$^{-1}$). The
dust extinction and star formation rate maps have been smoothed for
presentation purposes.}
\figsetgrpend

\figsetgrpstart
\figsetgrpnum{2.2}
\figsetgrptitle{UGC 08058 }
\figsetplot{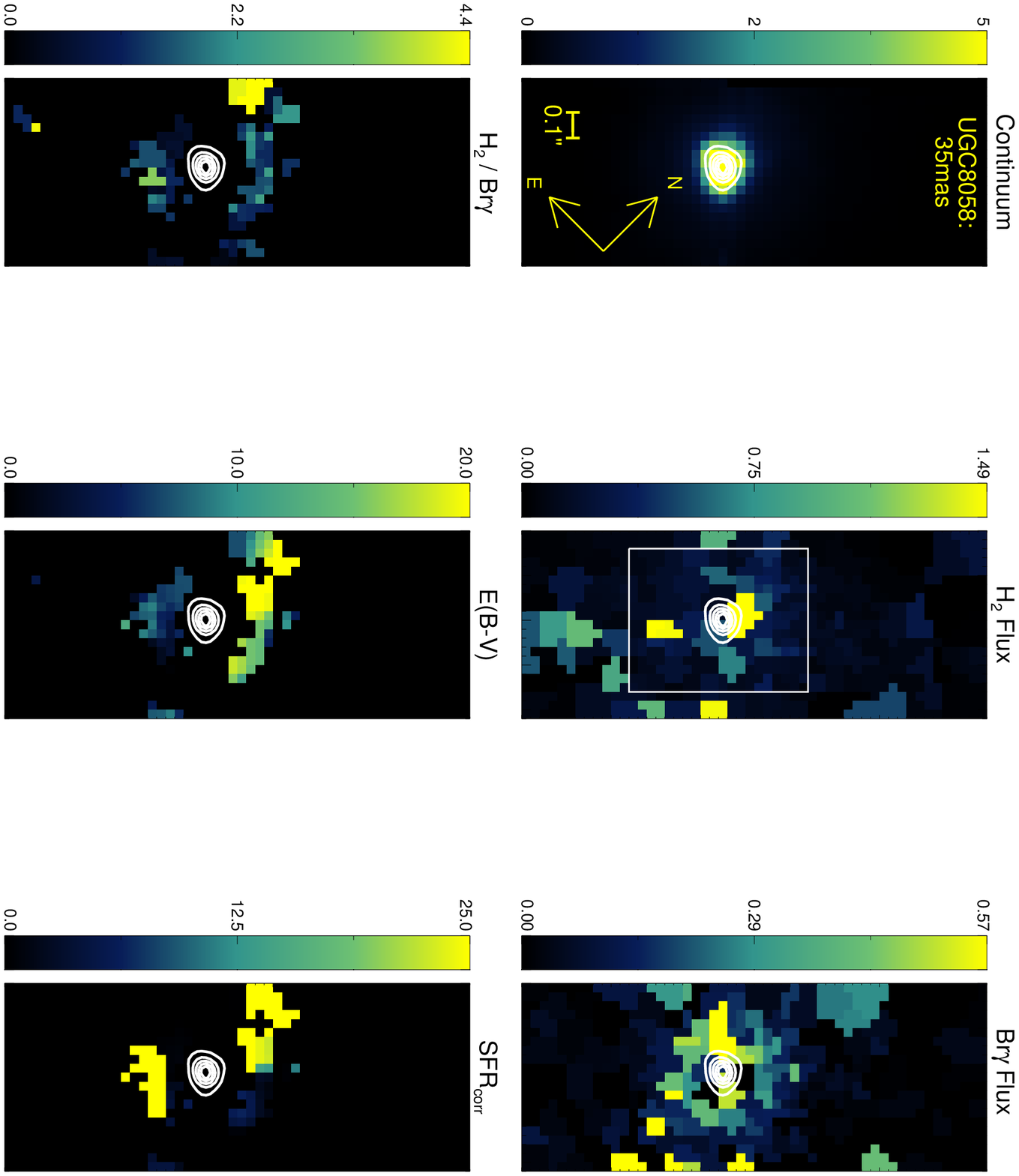}
\figsetgrpnote{Six-panel figure showing, from left to right, top to
  bottom, 1. $K$-band continuum map in relative flux units and shown
  with 0.1\arcsec scale bar and compass rose; the continuum contours
  are shown in all subsequent panels. 2. \molhy~1$-$0 S(1) flux map in
  10$^{-16}$ erg s$^{-1}$ cm$^{-2}$, with box highlighting the ``\molhy-dominated region"; 3. \brg~flux map in 10$^{-16}$
  erg s$^{-1}$ cm$^{-2}$; 4. \molhy~1$-$0
  S(1)/\brg~(\molhy/\brg~hereafter) map showing the reliable spaxels
  (with S/N $>$ 2 in both \molhy~and \brg); 5. E(B-V) map; and
  6. dust-corrected star formation rate map (M$_\odot$ yr$^{-1}$). The
dust extinction and star formation rate maps have been smoothed for
presentation purposes.}
\figsetgrpend

\figsetgrpstart
\figsetgrpnum{2.3}
\figsetgrptitle{IRAS F17207$-$0014 }
\figsetplot{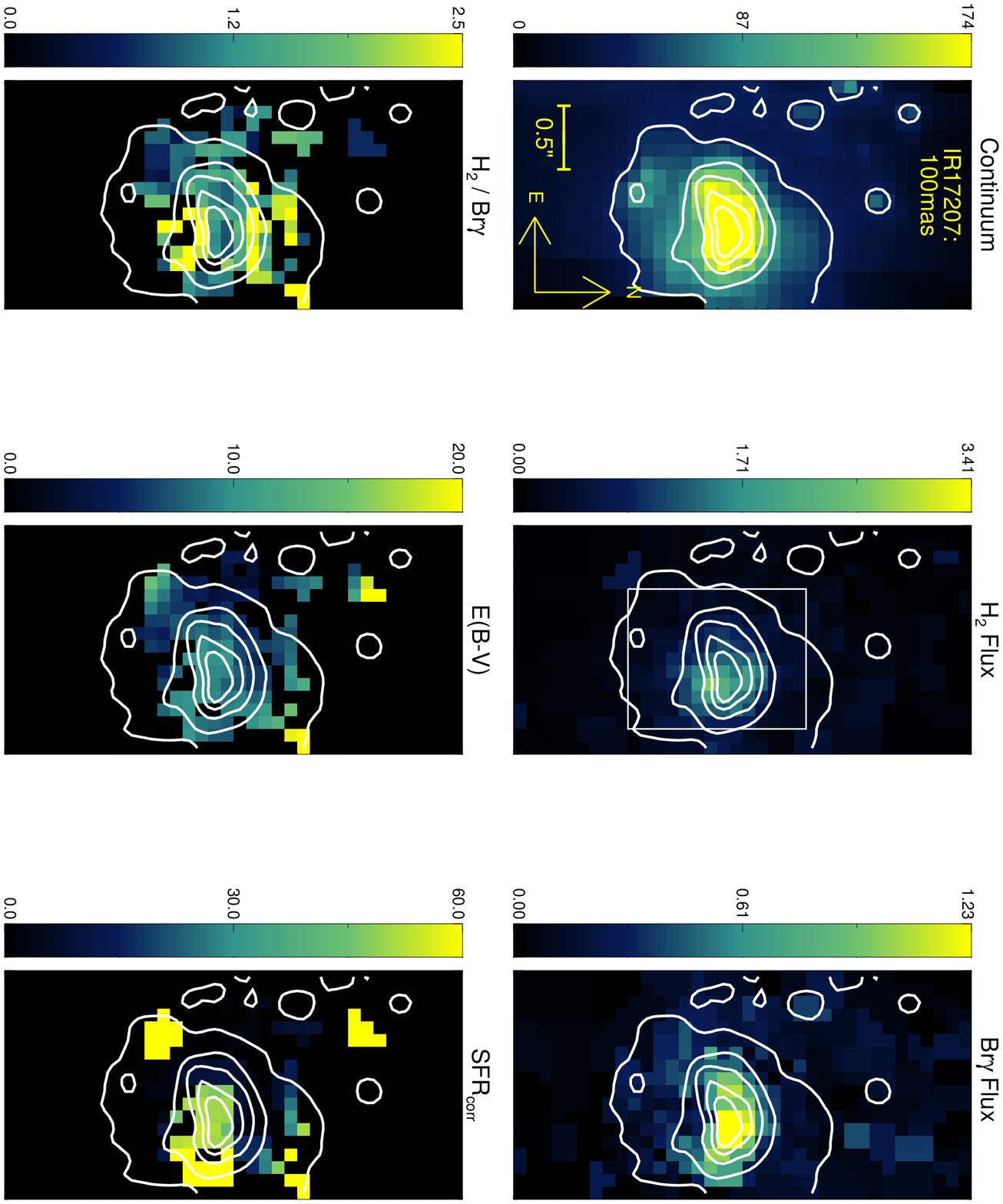}
\figsetgrpnote{Six-panel figure showing, from left to right, top to bottom, 1. $K$-band continuum map in relative flux units and shown with 0.5\arcsec scale bar and compass rose; the continuum contours are shown in all subsequent panels. 2. \molhy~1$-$0 S(1) flux map in 10$^{-16}$ erg s$^{-1}$ cm$^{-2}$, with box highlighting the ``\molhy-dominated region"; 3. \brg~flux map in 10$^{-16}$ erg s$^{-1}$ cm$^{-2}$; 4. \molhy~1$-$0 S(1)/\brg~(\molhy/\brg~hereafter) map showing the reliable spaxels (with S/N $>$ 3 in both \molhy~and \brg); 5. E(B-V) map; and 6. dust-corrected star formation rate map (M$_\odot$ yr$^{-1}$). The
dust extinction and star formation rate maps have been smoothed for
presentation purposes.}
\figsetgrpend

\figsetgrpstart
\figsetgrpnum{2.4}
\figsetgrptitle{UGC 08696}
\figsetplot{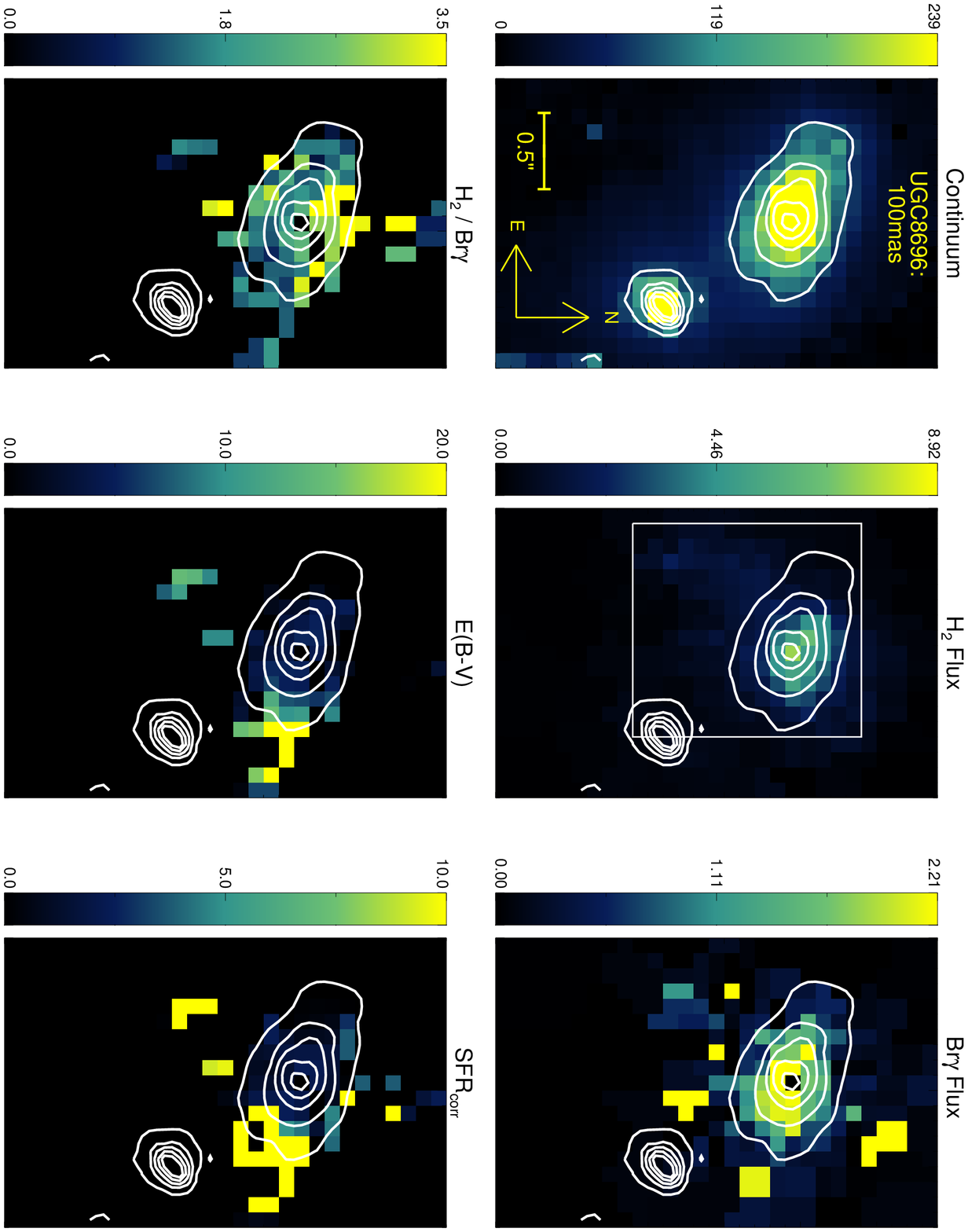}
\figsetgrpnote{Six-panel figure showing, from left to right, top to
  bottom, 1. $K$-band continuum map in relative flux units and shown
  with 0.5\arcsec scale bar and compass rose; the continuum contours
  are shown in all subsequent panels. 2. \molhy~1$-$0 S(1) flux map in
  10$^{-16}$ erg s$^{-1}$ cm$^{-2}$, with box highlighting the ``\molhy-dominated region"; 3. \brg~flux map in 10$^{-16}$
  erg s$^{-1}$ cm$^{-2}$; 4. \molhy~1$-$0
  S(1)/\brg~(\molhy/\brg~hereafter) map showing the reliable spaxels
  (with S/N $>$ 3 in both \molhy~and \brg); 5. E(B-V) map; and
  6. dust-corrected star formation rate map (M$_\odot$ yr$^{-1}$). The
dust extinction and star formation rate maps have been smoothed for
presentation purposes.} 
\figsetgrpend

\figsetgrpstart
\figsetgrpnum{2.5}
\figsetgrptitle{IRAS F22491$-$1808}
\figsetplot{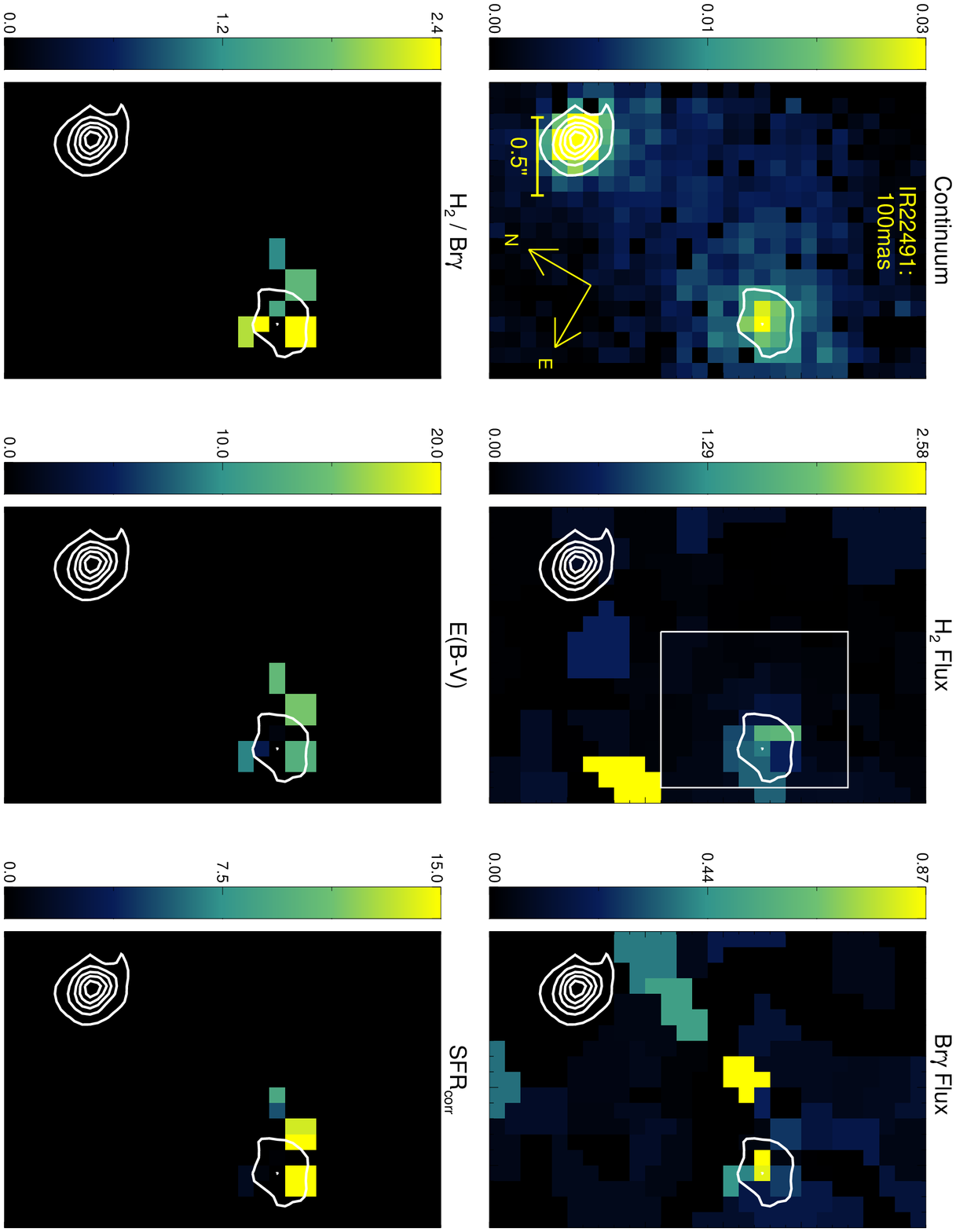}
\figsetgrpnote{Six-panel figure showing, from left to right, top to bottom, 1. $K$-band continuum map in relative flux units and shown with 0.5\arcsec scale bar and compass rose; the continuum contours are shown in all subsequent panels. 2. \molhy~1$-$0 S(1) flux map in 10$^{-16}$ erg s$^{-1}$ cm$^{-2}$, with box highlighting the ``\molhy-dominated region"; 3. \brg~flux map in 10$^{-16}$ erg s$^{-1}$ cm$^{-2}$; 4. \molhy~1$-$0 S(1)/\brg~(\molhy/\brg~hereafter) map; 5. E(B-V) map; and 6. dust-corrected star formation rate map (M$_\odot$ yr$^{-1}$).}
\figsetgrpend

\figsetgrpstart
\figsetgrpnum{2.6}
\figsetgrptitle{IRAS F15250+3608}
\figsetplot{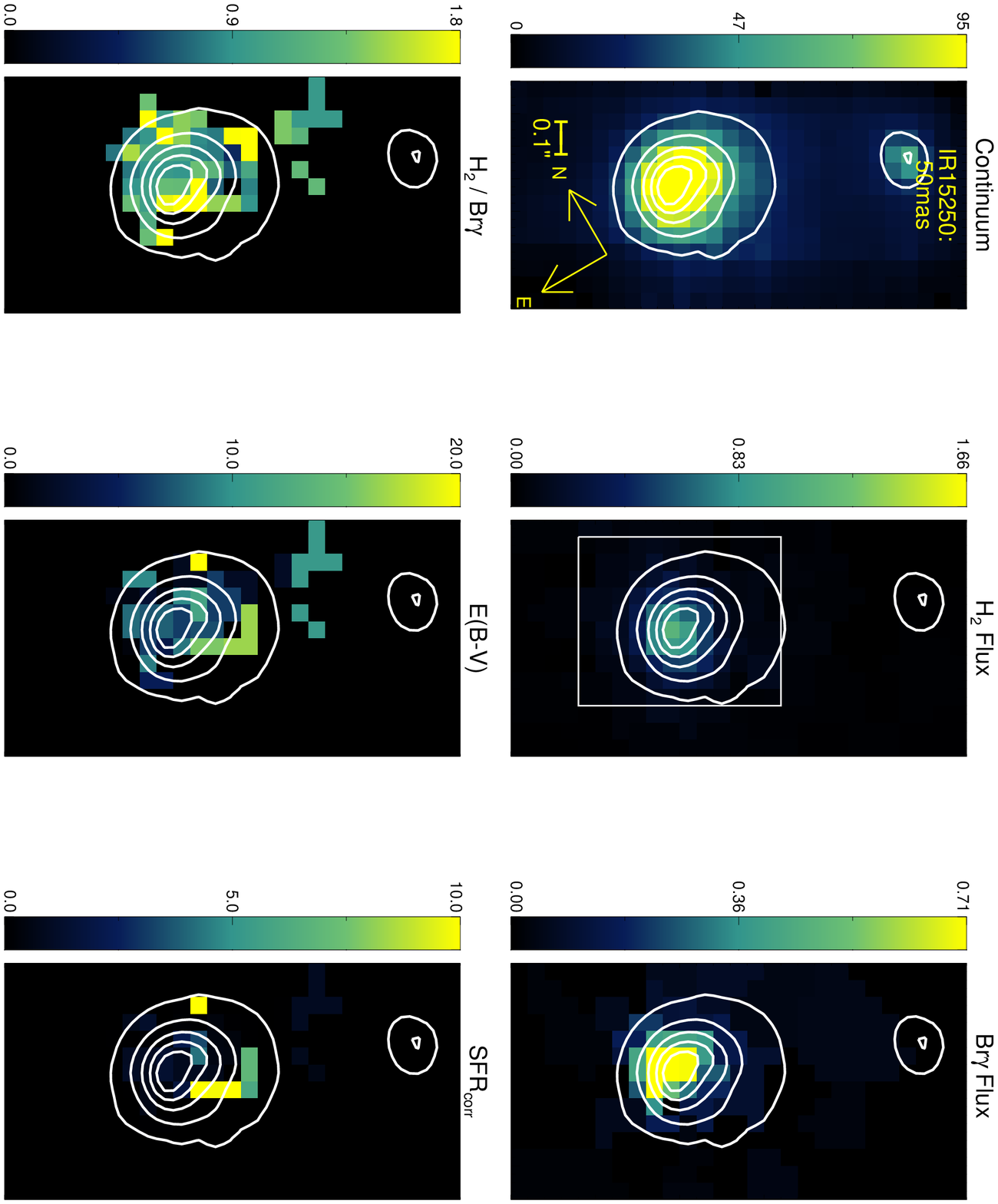}
\figsetgrpnote{Six-panel figure showing, from left to right, top to bottom, 1. $K$-band continuum map in relative flux units and shown with 0.1\arcsec scale bar and compass rose; the continuum contours are shown in all subsequent panels. 2. \molhy~1$-$0 S(1) flux map in 10$^{-16}$ erg s$^{-1}$ cm$^{-2}$, with box highlighting the ``\molhy-dominated region"; 3. \brg~flux map in 10$^{-16}$ erg s$^{-1}$ cm$^{-2}$; 4. \molhy~1$-$0 S(1)/\brg~(\molhy/\brg~hereafter) map showing the reliable spaxels (with S/N $>$ 3 in both \molhy~and \brg); 5. E(B-V) map; and 6. dust-corrected star formation rate map (M$_\odot$ yr$^{-1}$).}
\figsetgrpend

\figsetgrpstart
\figsetgrpnum{2.7}
\figsetgrptitle{UGC 05101}
\figsetplot{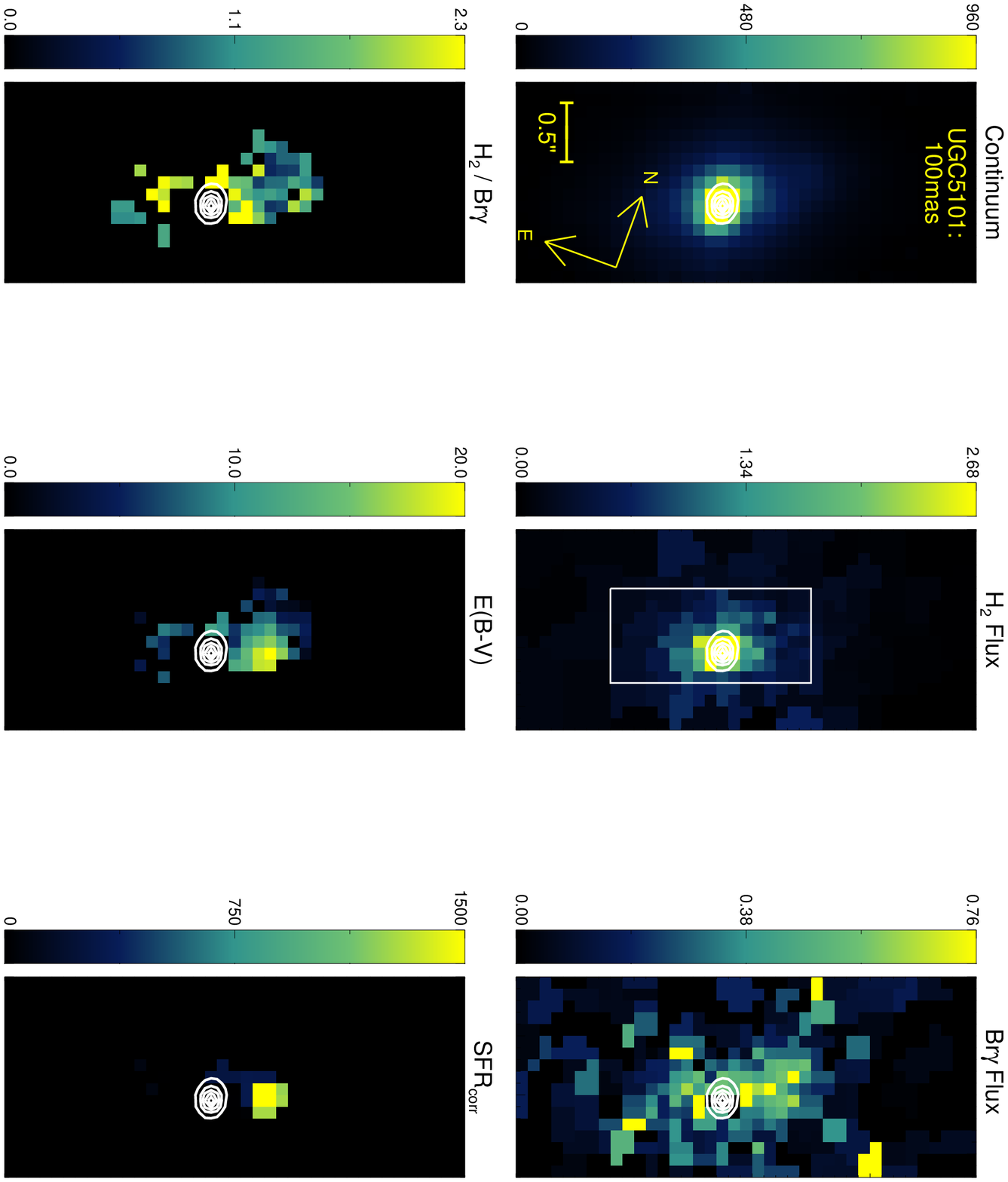}
\figsetgrpnote{Six-panel figure showing, from left to right, top to
  bottom, 1. $K$-band continuum map in relative flux units and shown
  with 0.5\arcsec scale bar and compass rose; the continuum contours
  are shown in all subsequent panels. 2. \molhy~1$-$0 S(1) flux map in
  10$^{-16}$ erg s$^{-1}$ cm$^{-2}$, with box highlighting the ``\molhy-dominated region"; 3. \brg~flux map in 10$^{-16}$
  erg s$^{-1}$ cm$^{-2}$; 4. \molhy~1$-$0
  S(1)/\brg~(\molhy/\brg~hereafter) map showing the reliable spaxels
  (with S/N $>$ 3 in both \molhy~and \brg); 5. E(B-V) map; and
  6. dust-corrected star formation rate map (M$_\odot$ yr$^{-1}$). The
dust extinction and star formation rate maps have been smoothed for
presentation purposes.}
\figsetgrpend

\figsetgrpstart
\figsetgrpnum{2.8}
\figsetgrptitle{VV 340a}
\figsetplot{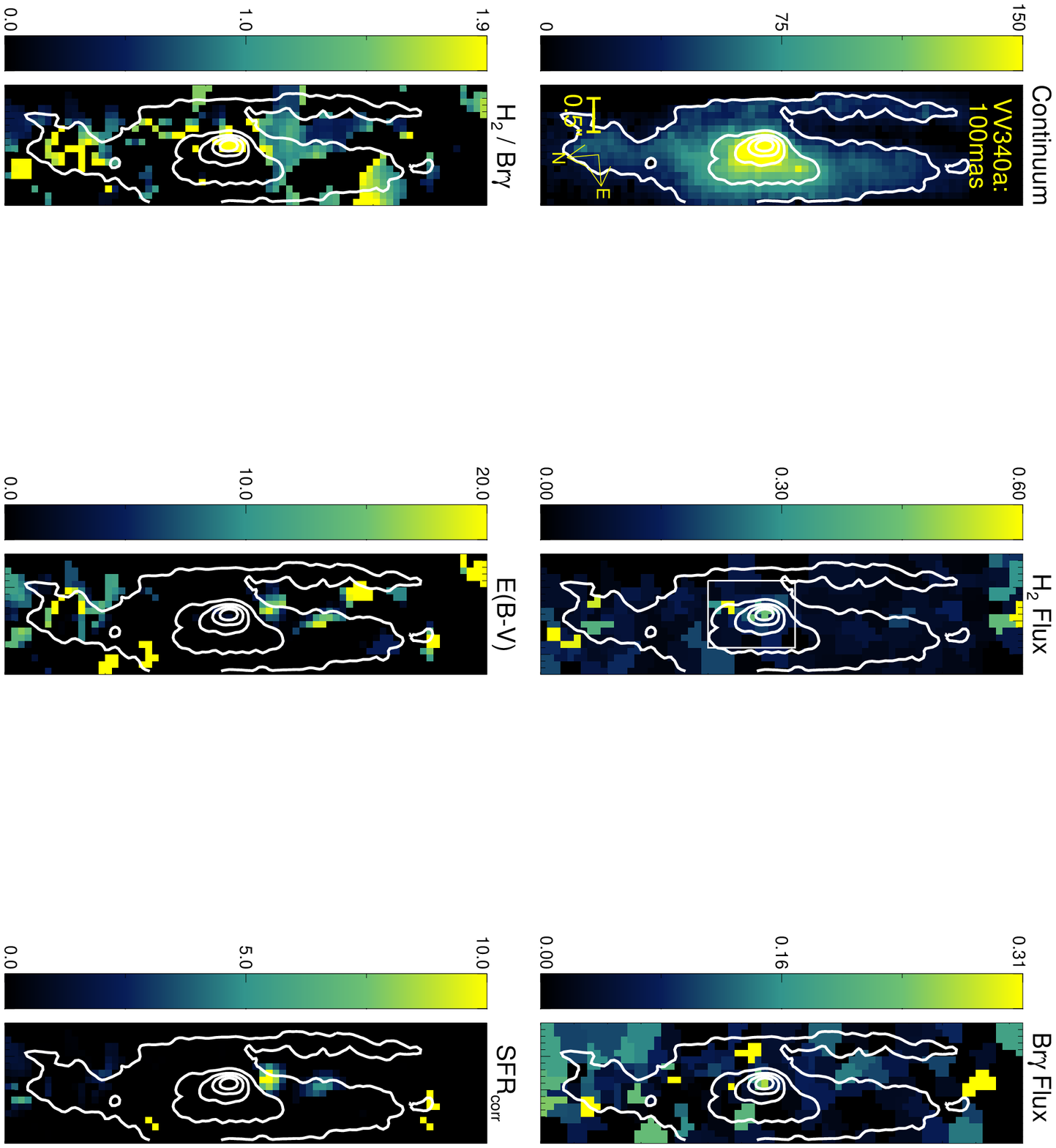}
\figsetgrpnote{Six-panel figure showing, from left to right, top to
  bottom, 1. $K$-band continuum map in relative flux units and shown
  with 0.5\arcsec scale bar and compass rose; the continuum contours
  are shown in all subsequent panels. 2. \molhy~1$-$0 S(1) flux map in
  10$^{-16}$ erg s$^{-1}$ cm$^{-2}$, with box highlighting the ``\molhy-dominated region"; 3. \brg~flux map in 10$^{-16}$
  erg s$^{-1}$ cm$^{-2}$; 4. \molhy~1$-$0
  S(1)/\brg~(\molhy/\brg~hereafter) map; 5. E(B-V) map; and
  6. dust-corrected star formation rate map (M$_\odot$ yr$^{-1}$). The
dust extinction and star formation rate maps have been smoothed for
presentation purposes.}
\figsetgrpend

\figsetgrpstart
\figsetgrpnum{2.9}
\figsetgrptitle{IRAS F01364$-$1042 (100mas)}
\figsetplot{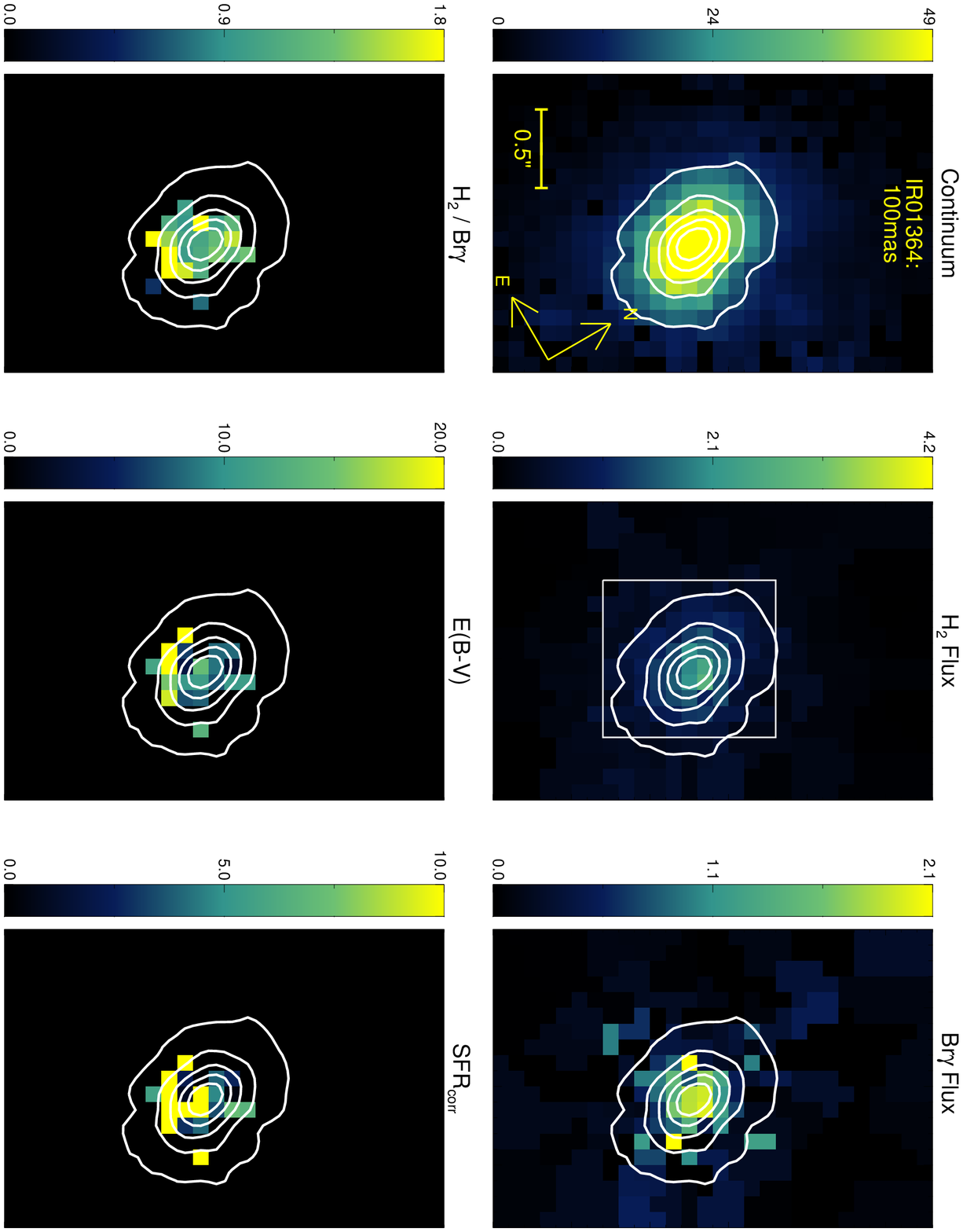}
\figsetgrpnote{Six-panel figure showing, from left to right, top to bottom, 1. $K$-band continuum map in relative flux units and shown with 0.5\arcsec scale bar and compass rose; the continuum contours are shown in all subsequent panels. 2. \molhy~1$-$0 S(1) flux map in 10$^{-16}$ erg s$^{-1}$ cm$^{-2}$, with box highlighting the ``\molhy-dominated region"; 3. \brg~flux map in 10$^{-16}$ erg s$^{-1}$ cm$^{-2}$; 4. \molhy~1$-$0 S(1)/\brg~(\molhy/\brg~hereafter) map showing the reliable spaxels (with S/N $>$ 3 in both \molhy~and \brg); 5. E(B-V) map; and 6. dust-corrected star formation rate map (M$_\odot$ yr$^{-1}$).}
\figsetgrpend

\figsetgrpstart
\figsetgrpnum{2.10}
\figsetgrptitle{IRAS F01364$-$1042 (35mas)}
\figsetplot{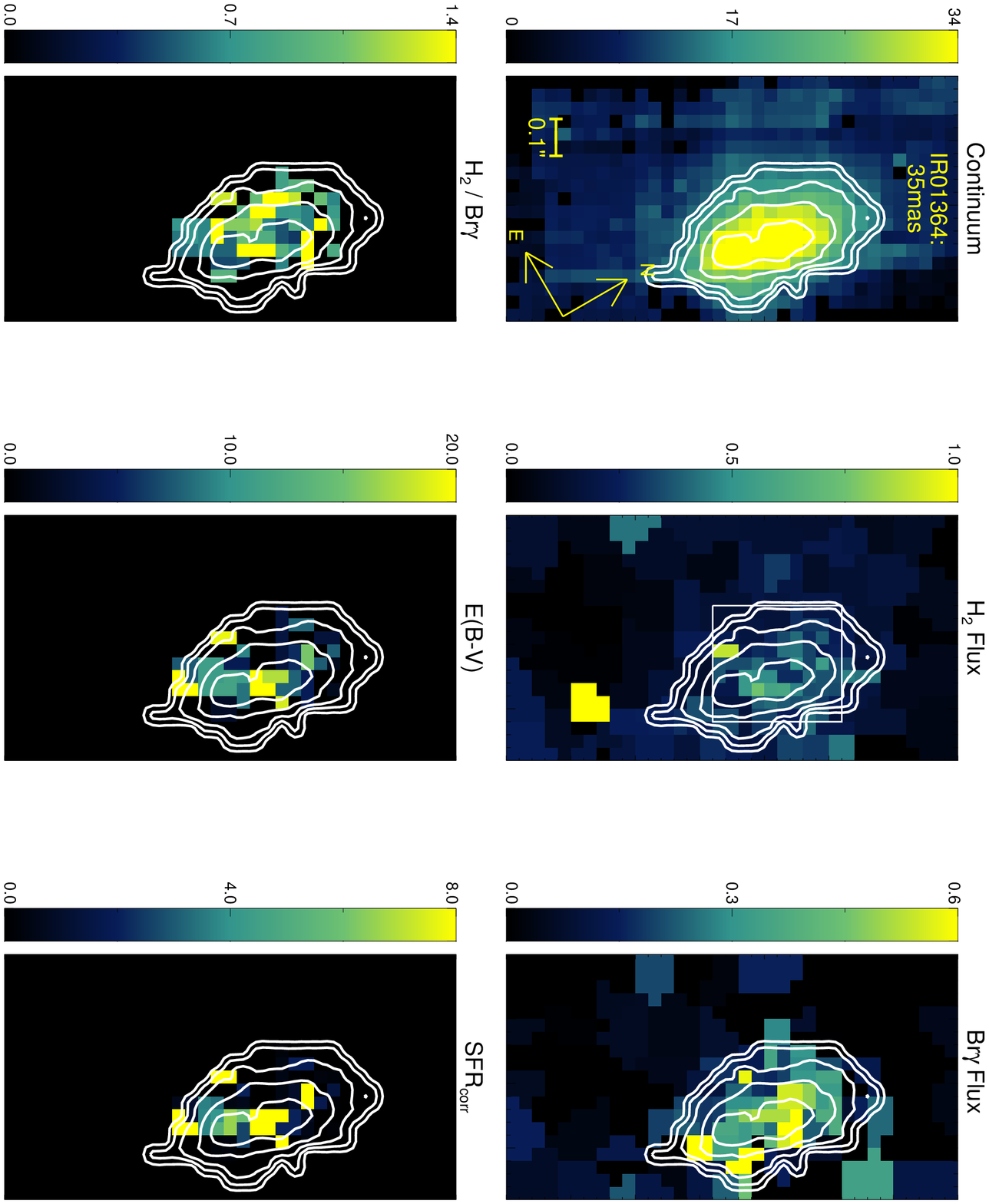}
\figsetgrpnote{Six-panel figure showing, from left to right, top to bottom, 1. $K$-band continuum map in relative flux units and shown with 0.1\arcsec scale bar and compass rose; the continuum contours are shown in all subsequent panels. 2. \molhy~1$-$0 S(1) flux map in 10$^{-16}$ erg s$^{-1}$ cm$^{-2}$, with box highlighting the ``\molhy-dominated region"; 3. \brg~flux map in 10$^{-16}$ erg s$^{-1}$ cm$^{-2}$; 4. \molhy~1$-$0 S(1)/\brg~(\molhy/\brg~hereafter) map showing the reliable spaxels (with S/N $>$ 2.5 in both \molhy~and \brg); 5. E(B-V) map; and 6. dust-corrected star formation rate map (M$_\odot$ yr$^{-1}$).}
\figsetgrpend

\figsetgrpstart
\figsetgrpnum{2.11}
\figsetgrptitle{UGC 08387}
\figsetplot{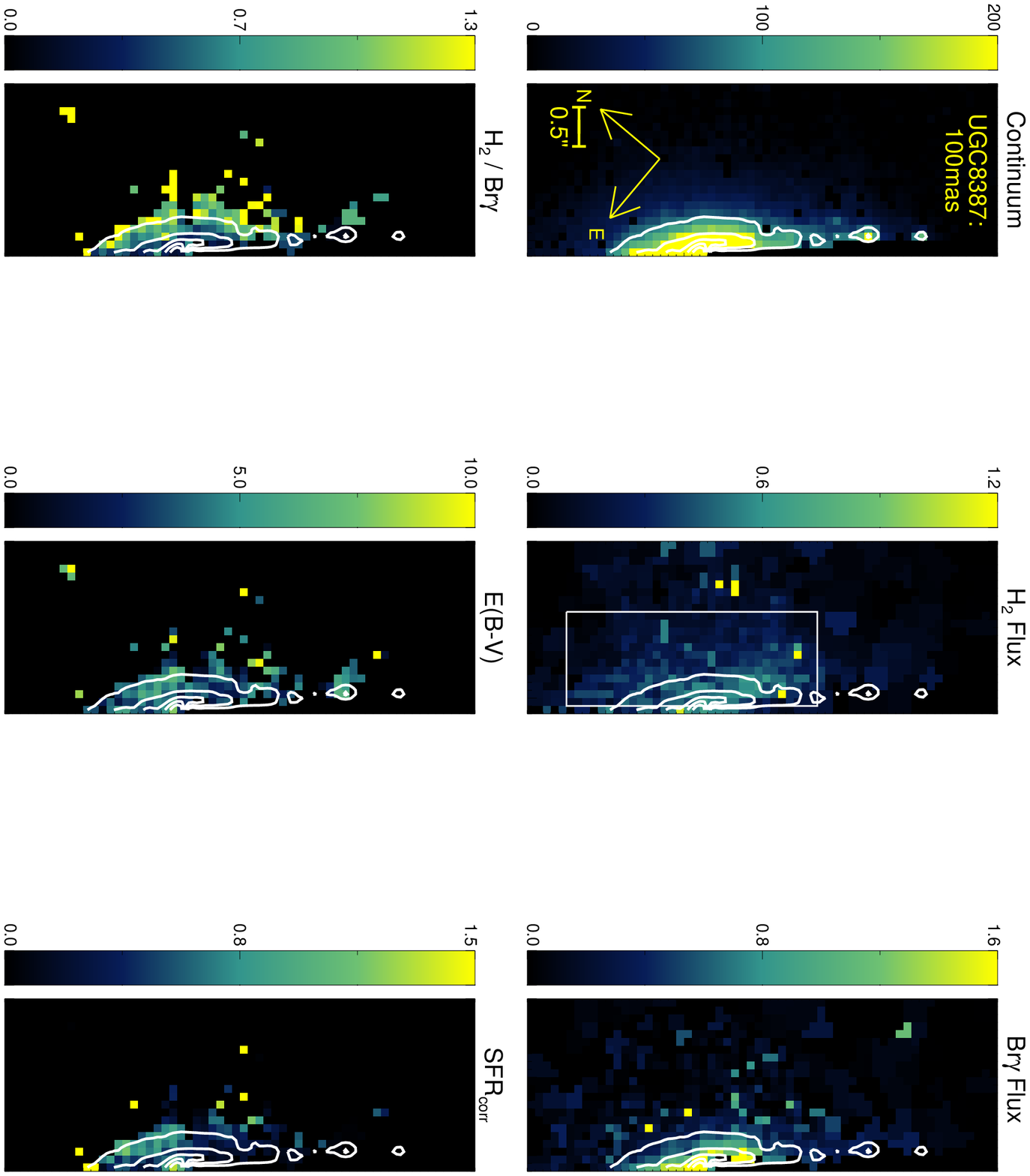}
\figsetgrpnote{Six-panel figure showing, from left to right, top to
  bottom, 1. $K$-band continuum map in relative flux units and shown
  with 0.5\arcsec scale bar and compass rose; the continuum contours
  are shown in all subsequent panels. 2. \molhy~1$-$0 S(1) flux map in
  10$^{-16}$ erg s$^{-1}$ cm$^{-2}$, with box highlighting the ``\molhy-dominated region"; 3. \brg~flux map in 10$^{-16}$
  erg s$^{-1}$ cm$^{-2}$; 4. \molhy~1$-$0
  S(1)/\brg~(\molhy/\brg~hereafter) map showing the reliable spaxels
  (with S/N $>$ 3 in both \molhy~and \brg); 5. E(B-V) map; and
  6. dust-corrected star formation rate map (M$_\odot$ yr$^{-1}$). The
dust extinction and star formation rate maps have been smoothed for
presentation purposes.}
\figsetgrpend

\figsetgrpstart
\figsetgrpnum{2.12}
\figsetgrptitle{CGCG 436$-$030}
\figsetplot{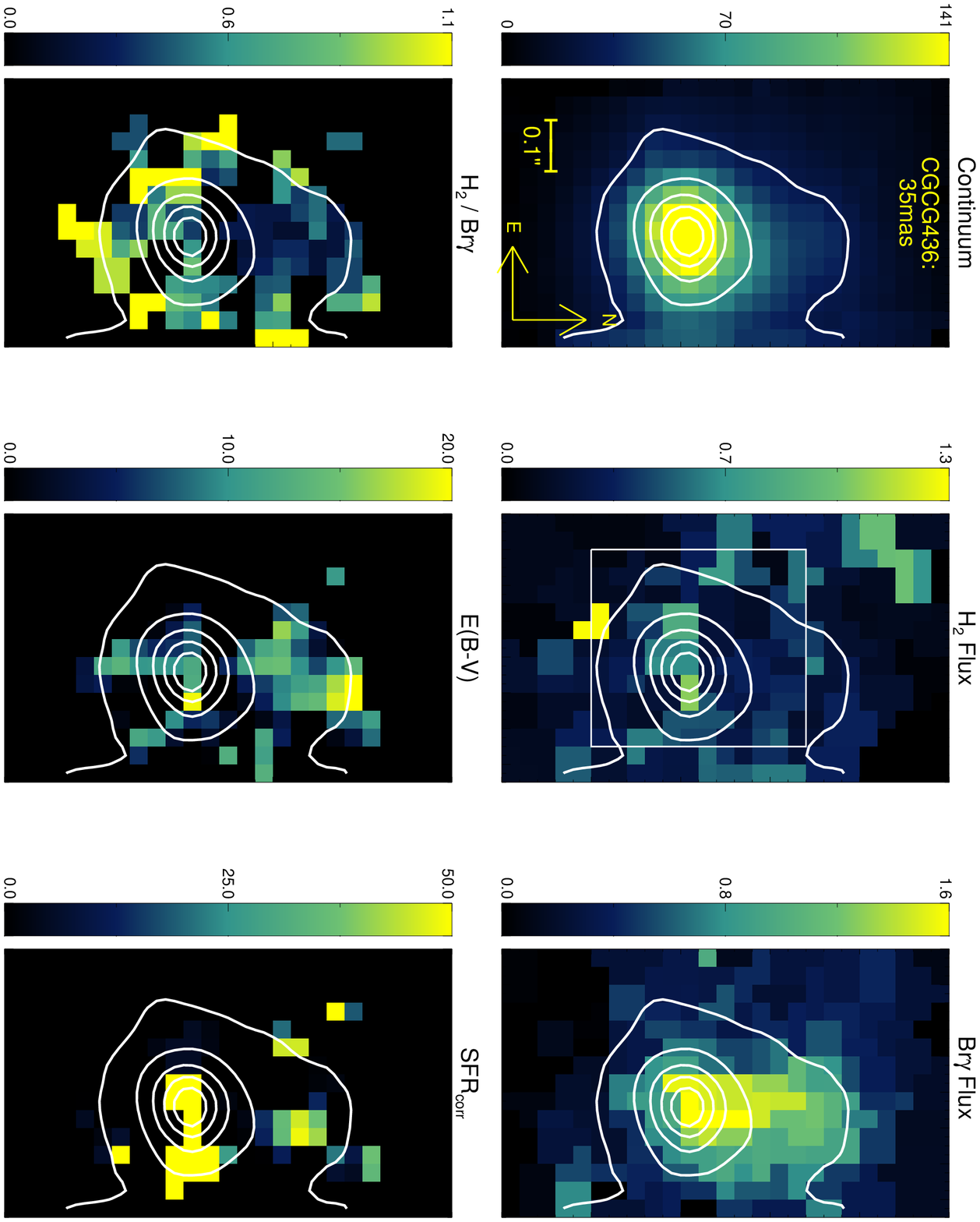}
\figsetgrpnote{Six-panel figure showing, from left to right, top to
  bottom, 1. $K$-band continuum map in relative flux units and shown
  with 0.1\arcsec scale bar and compass rose; the continuum contours
  are shown in all subsequent panels. 2. \molhy~1$-$0 S(1) flux map in
  10$^{-16}$ erg s$^{-1}$ cm$^{-2}$, with box highlighting the ``\molhy-dominated region"; 3. \brg~flux map in 10$^{-16}$
  erg s$^{-1}$ cm$^{-2}$; 4. \molhy~1$-$0
  S(1)/\brg~(\molhy/\brg~hereafter) map showing the reliable spaxels
  (with S/N $>$ 2 in both \molhy~and \brg); 5. E(B-V) map; and
  6. dust-corrected star formation rate map (M$_\odot$ yr$^{-1}$). The
dust extinction and star formation rate maps have been smoothed for
presentation purposes.}
\figsetgrpend

\figsetgrpstart
\figsetgrpnum{2.13}
\figsetgrptitle{NGC 6670E }
\figsetplot{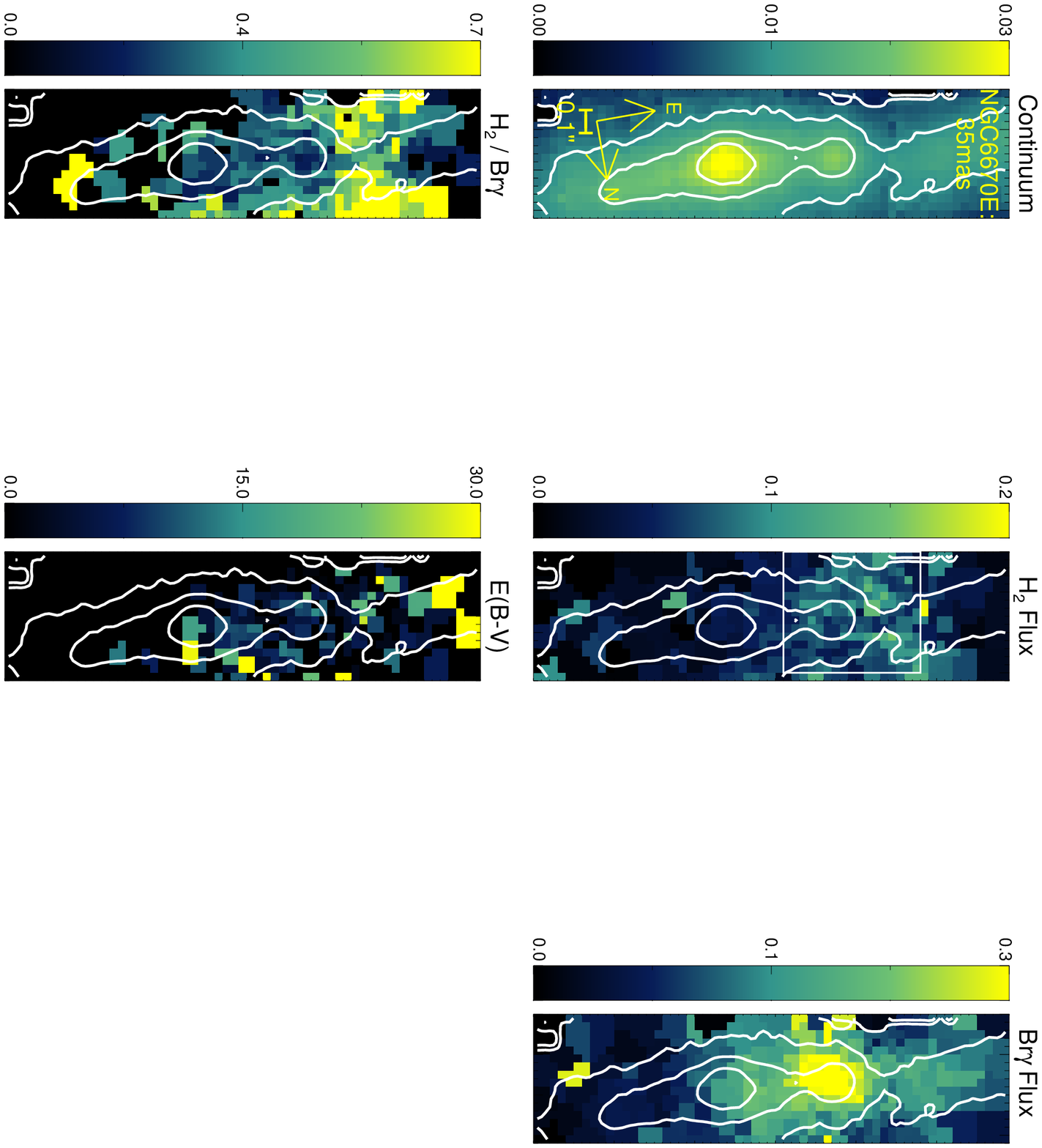}
\figsetgrpnote{Six-panel figure showing, from left to right, top to
  bottom, 1. $K$-band continuum map in relative data count units and shown
  with 0.1\arcsec scale bar and compass rose; the continuum contours
  are shown in all subsequent panels. 2. \molhy~1$-$0 S(1) flux map in
  relative units, with box highlighting the ``\molhy-dominated region"; 3. \brg~flux map in relative units; 4. \molhy~1$-$0
  S(1)/\brg~(\molhy/\brg~hereafter) map showing the reliable spaxels
  (with S/N $>$ 3 in both \molhy~and \brg); and 5. E(B-V) map. No star
  formation rate map is available due to lack of proper calibration
  frames used to flux-calibrate the OSIRIS data.}
\figsetgrpend

\figsetgrpstart
\figsetgrpnum{2.14}
\figsetgrptitle{IRAS F06076$-$2139N}
\figsetplot{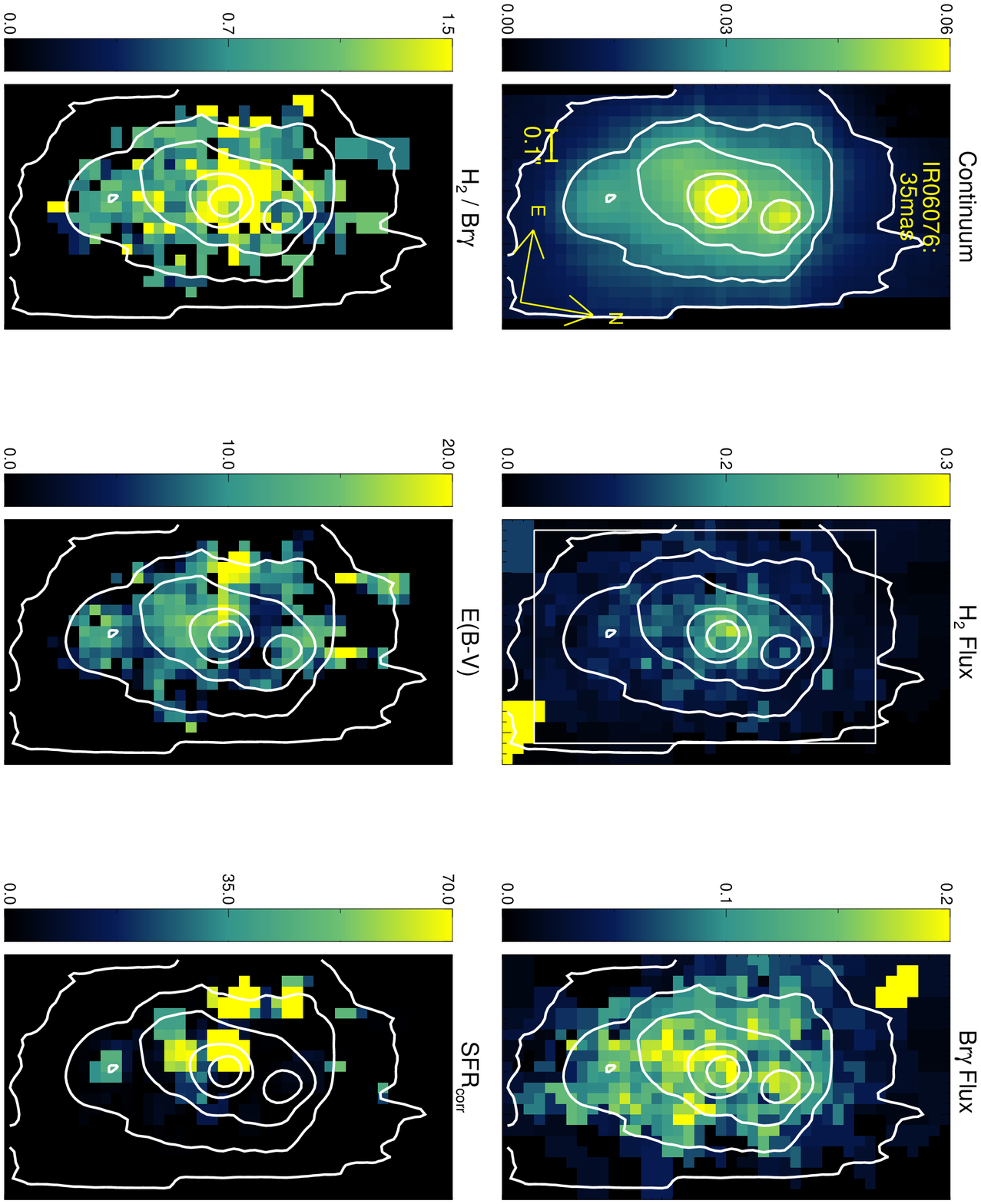}
\figsetgrpnote{Six-panel figure showing, from left to right, top to
  bottom, 1. $K$-band continuum map in relative flux units and shown
  with 0.1\arcsec scale bar and compass rose; the continuum contours
  are shown in all subsequent panels. 2. \molhy~1$-$0 S(1) flux map in
  10$^{-16}$ erg s$^{-1}$ cm$^{-2}$, with box highlighting the ``\molhy-dominated region"; 3. \brg~flux map in 10$^{-16}$
  erg s$^{-1}$ cm$^{-2}$; 4. \molhy~1$-$0
  S(1)/\brg~(\molhy/\brg~hereafter) map showing the reliable spaxels
  (with S/N $>$ 3 in both \molhy~and \brg); 5. E(B-V) map; and
  6. dust-corrected star formation rate map (M$_\odot$ yr$^{-1}$). The
dust extinction and star formation rate maps have been smoothed for
presentation purposes.}
\figsetgrpend

\figsetgrpstart
\figsetgrpnum{2.15}
\figsetgrptitle{IRAS F18090+0130E}
\figsetplot{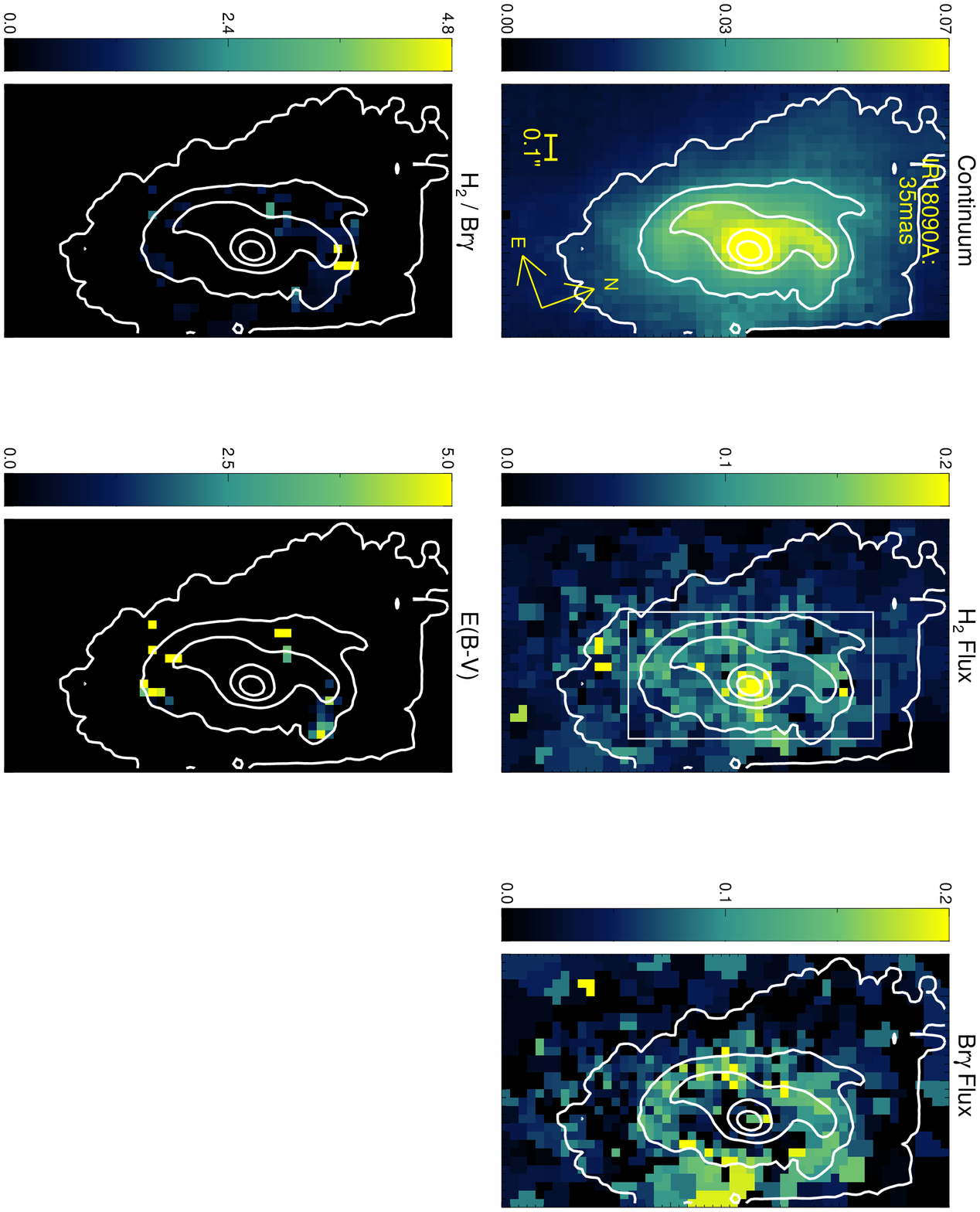}
\figsetgrpnote{Six-panel figure showing, from left to right, top to
  bottom, 1. $K$-band continuum map in relative data count units and shown
  with 0.1\arcsec scale bar and compass rose; the continuum contours
  are shown in all subsequent panels. 2. \molhy~1$-$0 S(1) flux map in
  relative units, with box highlighting the ``\molhy-dominated region"; 3. \brg~flux map in relative units; 4. \molhy~1$-$0
  S(1)/\brg~(\molhy/\brg~hereafter) map showing the reliable spaxels
  (with S/N $>$ 2.5 in both \molhy~and \brg); and 5. E(B-V) map. No star
  formation rate map is available due to lack of proper calibration
  frames used to flux-calibrate the OSIRIS data. The dust extinction
  map has been smoothed for presentation purposes.}
\figsetgrpend

\figsetgrpstart
\figsetgrpnum{2.16}
\figsetgrptitle{IRAS F18090+0130W}
\figsetplot{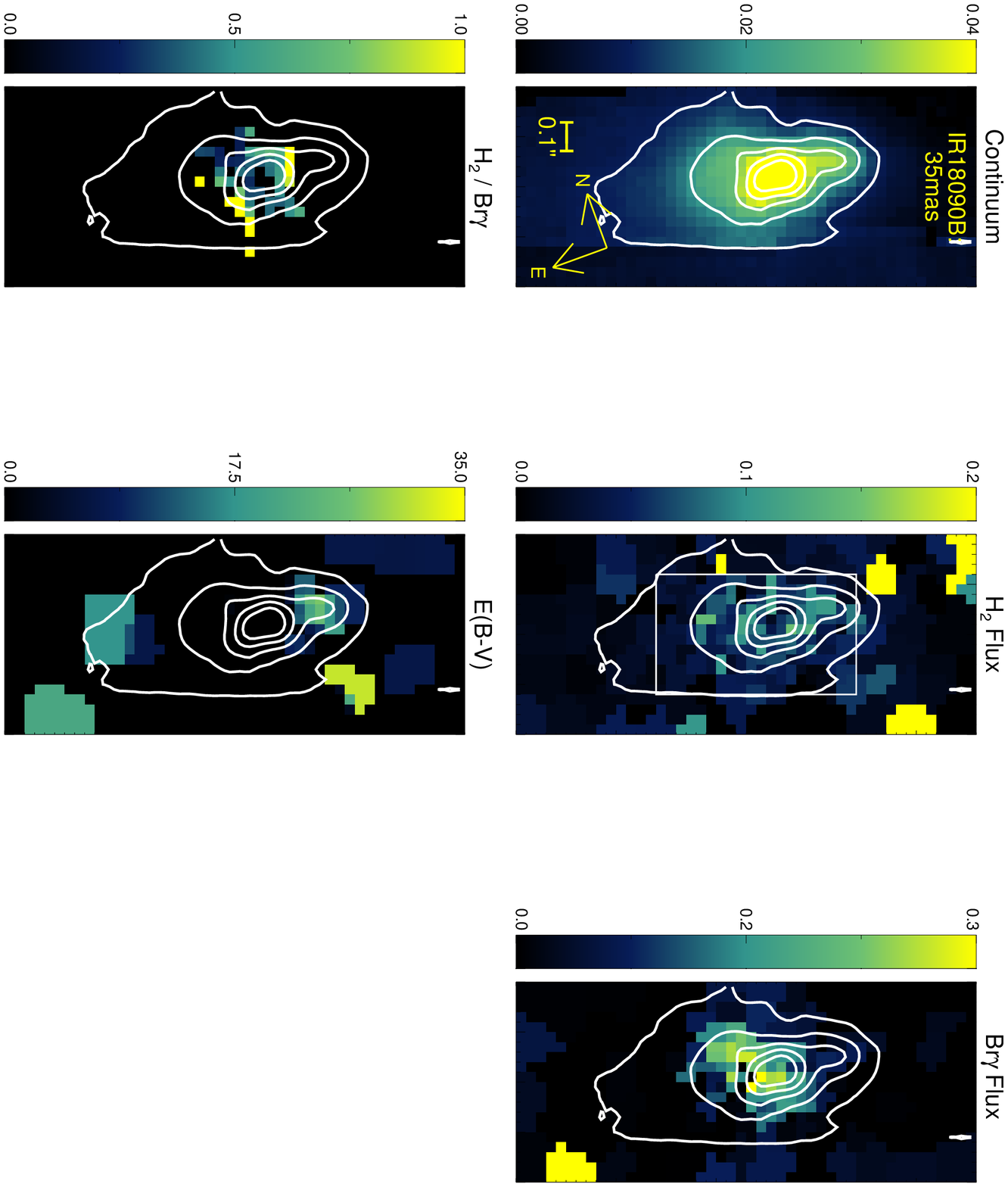}
\figsetgrpnote{Six-panel figure showing, from left to right, top to
  bottom, 1. $K$-band continuum map in relative data count units and shown
  with 0.1\arcsec scale bar and compass rose; the continuum contours
  are shown in all subsequent panels. 2. \molhy~1$-$0 S(1) flux map in
  relative units, with box highlighting the ``\molhy-dominated region"; 3. \brg~flux map in relative units; 4. \molhy~1$-$0
  S(1)/\brg~(\molhy/\brg~hereafter) map showing the reliable spaxels
  (with S/N $>$ 1.5 in both \molhy~and \brg); and 5. E(B-V) map. No star
  formation rate map is available due to lack of proper calibration
  frames used to flux-calibrate the OSIRIS data. The dust extinction
  map has been smoothed for presentation purposes.}
\figsetgrpend

\figsetgrpstart
\figsetgrpnum{2.17}
\figsetgrptitle{III Zw 035}
\figsetplot{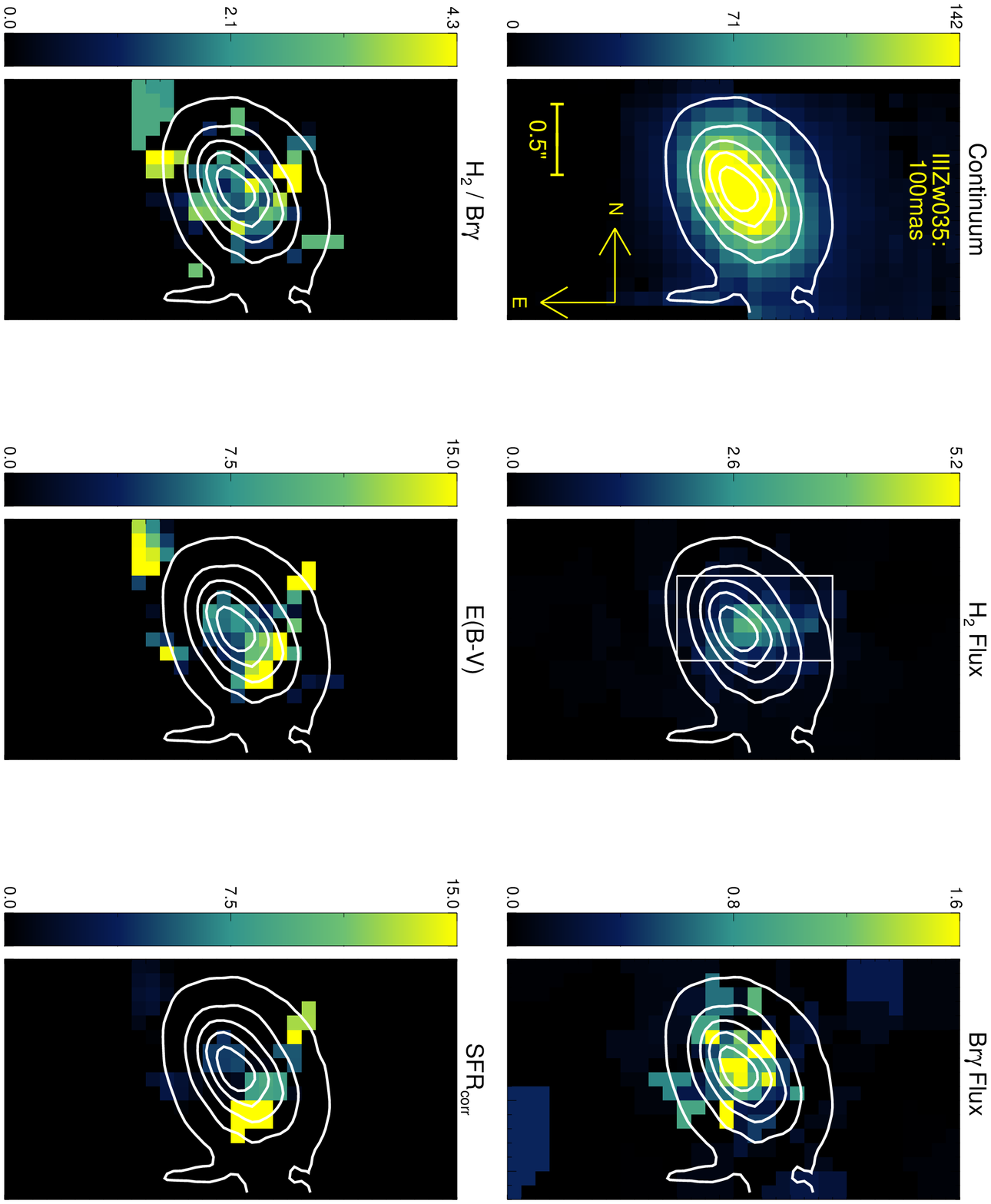}
\figsetgrpnote{Six-panel figure showing, from left to right, top to bottom, 1. $K$-band continuum map in relative flux units and shown with 0.5\arcsec scale bar and compass rose; the continuum contours are shown in all subsequent panels. 2. \molhy~1$-$0 S(1) flux map in 10$^{-16}$ erg s$^{-1}$ cm$^{-2}$, with box highlighting the ``\molhy-dominated region"; 3. \brg~flux map in 10$^{-16}$ erg s$^{-1}$ cm$^{-2}$; 4. \molhy~1$-$0 S(1)/\brg~(\molhy/\brg~hereafter) map showing the reliable spaxels (with S/N $>$ 3 in both \molhy~and \brg); 5. E(B-V) map; and 6. dust-corrected star formation rate map (M$_\odot$ yr$^{-1}$). The
dust extinction and star formation rate maps have been smoothed for
presentation purposes.}
\figsetgrpend

\figsetgrpstart
\figsetgrpnum{2.18}
\figsetgrptitle{IRAS F20351+2521}
\figsetplot{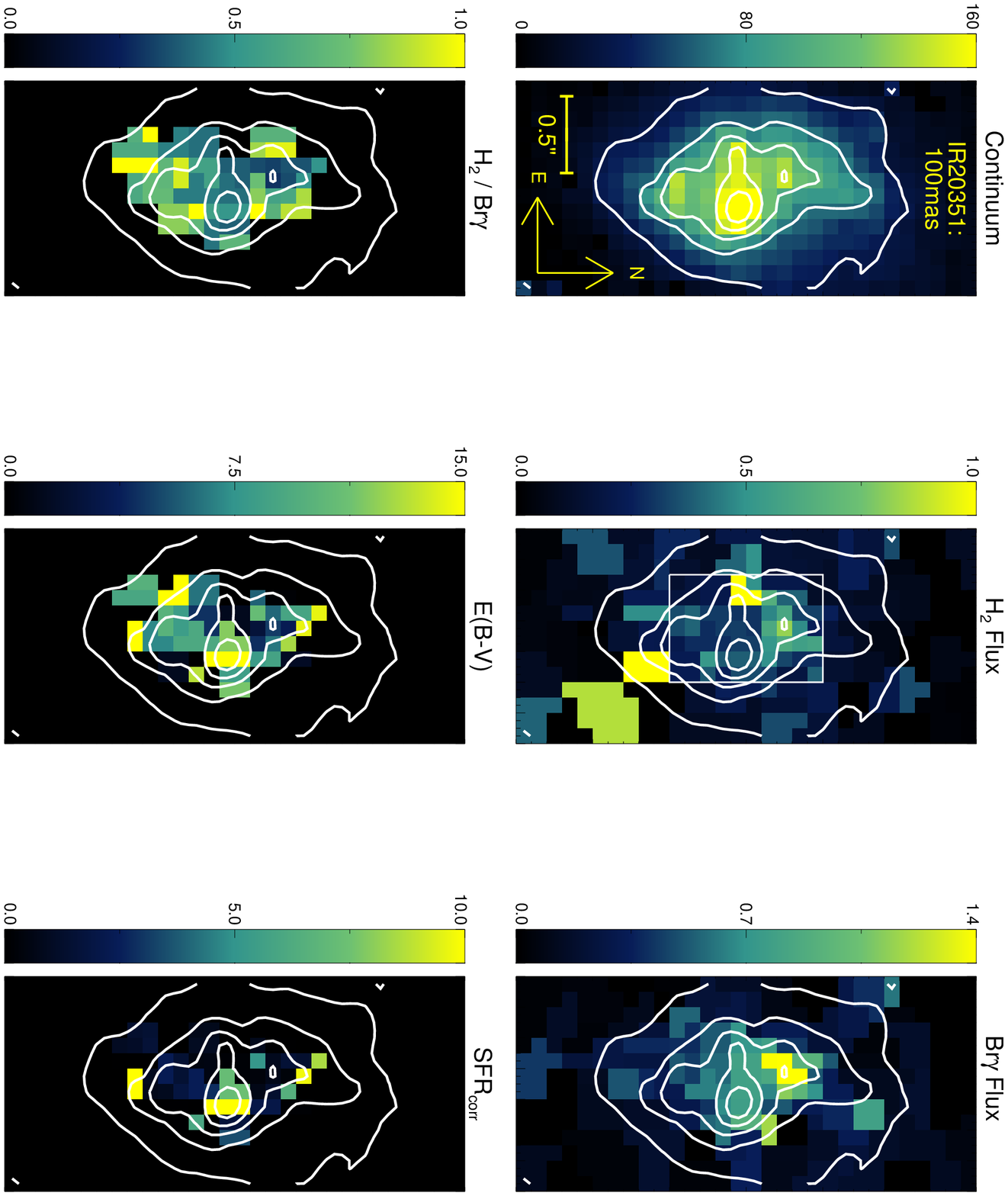}
\figsetgrpnote{Six-panel figure showing, from left to right, top to bottom, 1. $K$-band continuum map in relative flux units and shown with 0.5\arcsec scale bar and compass rose; the continuum contours are shown in all subsequent panels. 2. \molhy~1$-$0 S(1) flux map in 10$^{-16}$ erg s$^{-1}$ cm$^{-2}$, with box highlighting the ``\molhy-dominated region"; 3. \brg~flux map in 10$^{-16}$ erg s$^{-1}$ cm$^{-2}$; 4. \molhy~1$-$0 S(1)/\brg~(\molhy/\brg~hereafter) map showing the reliable spaxels (with S/N $>$ 3 in both \molhy~and \brg); 5. E(B-V) map; and 6. dust-corrected star formation rate map (M$_\odot$ yr$^{-1}$).}
\figsetgrpend

\figsetgrpstart
\figsetgrpnum{2.19}
\figsetgrptitle{NGC 2623}
\figsetplot{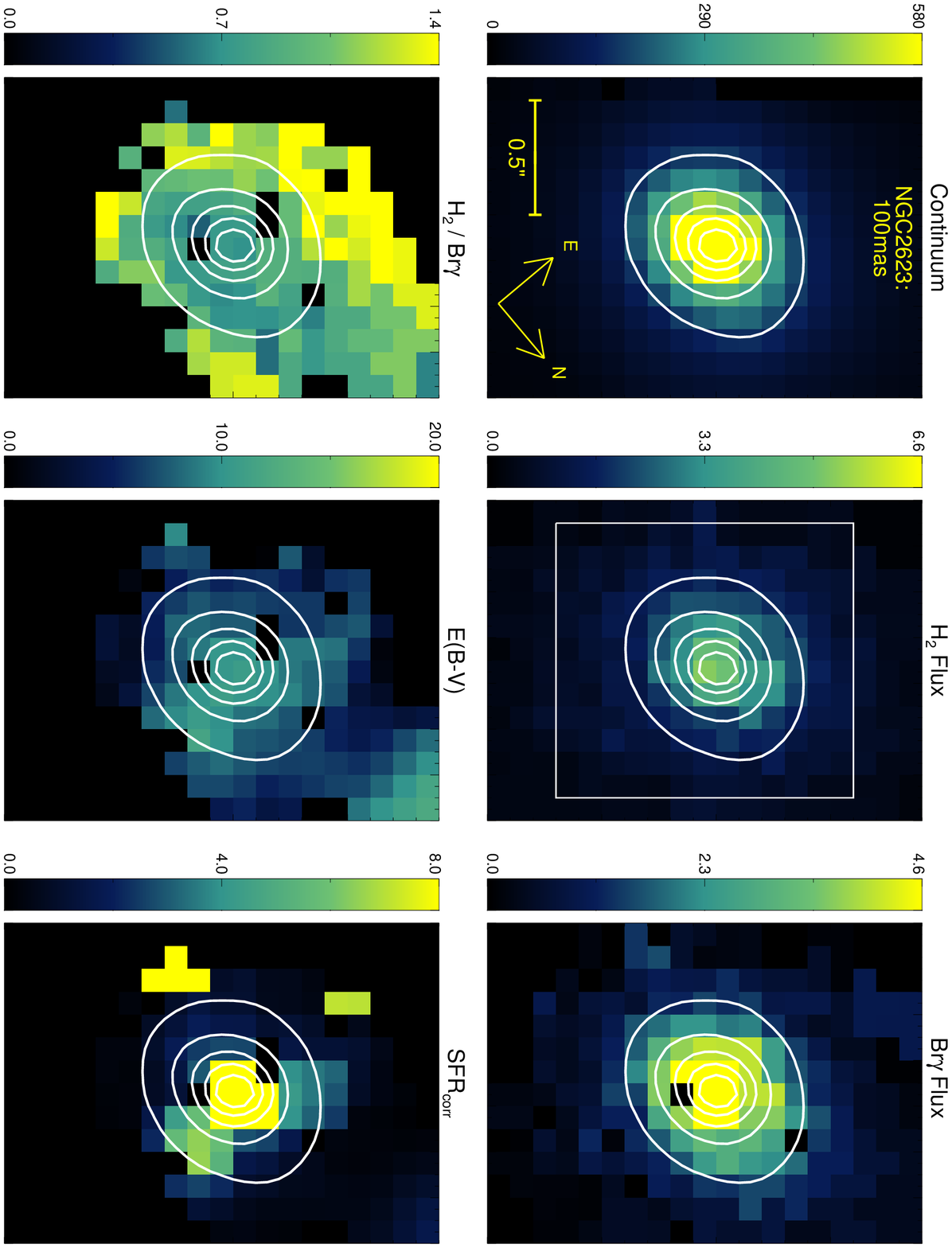}
\figsetgrpnote{Six-panel figure showing, from left to right, top to
  bottom, 1. $K$-band continuum map in relative flux units and shown
  with 0.5\arcsec scale bar and compass rose; the continuum contours
  are shown in all subsequent panels. 2. \molhy~1$-$0 S(1) flux map in
  10$^{-16}$ erg s$^{-1}$ cm$^{-2}$, with box highlighting the ``\molhy-dominated region"; 3. \brg~flux map in 10$^{-16}$
  erg s$^{-1}$ cm$^{-2}$; 4. \molhy~1$-$0
  S(1)/\brg~(\molhy/\brg~hereafter) map showing the reliable spaxels
  (with S/N $>$ 3 in both \molhy~and \brg); 5. E(B-V) map; and
  6. dust-corrected star formation rate map (M$_\odot$ yr$^{-1}$). The
dust extinction and star formation rate maps have been smoothed for
presentation purposes.}
\figsetgrpend

\figsetgrpstart
\figsetgrpnum{2.20}
\figsetgrptitle{NGC 7469N}
\figsetplot{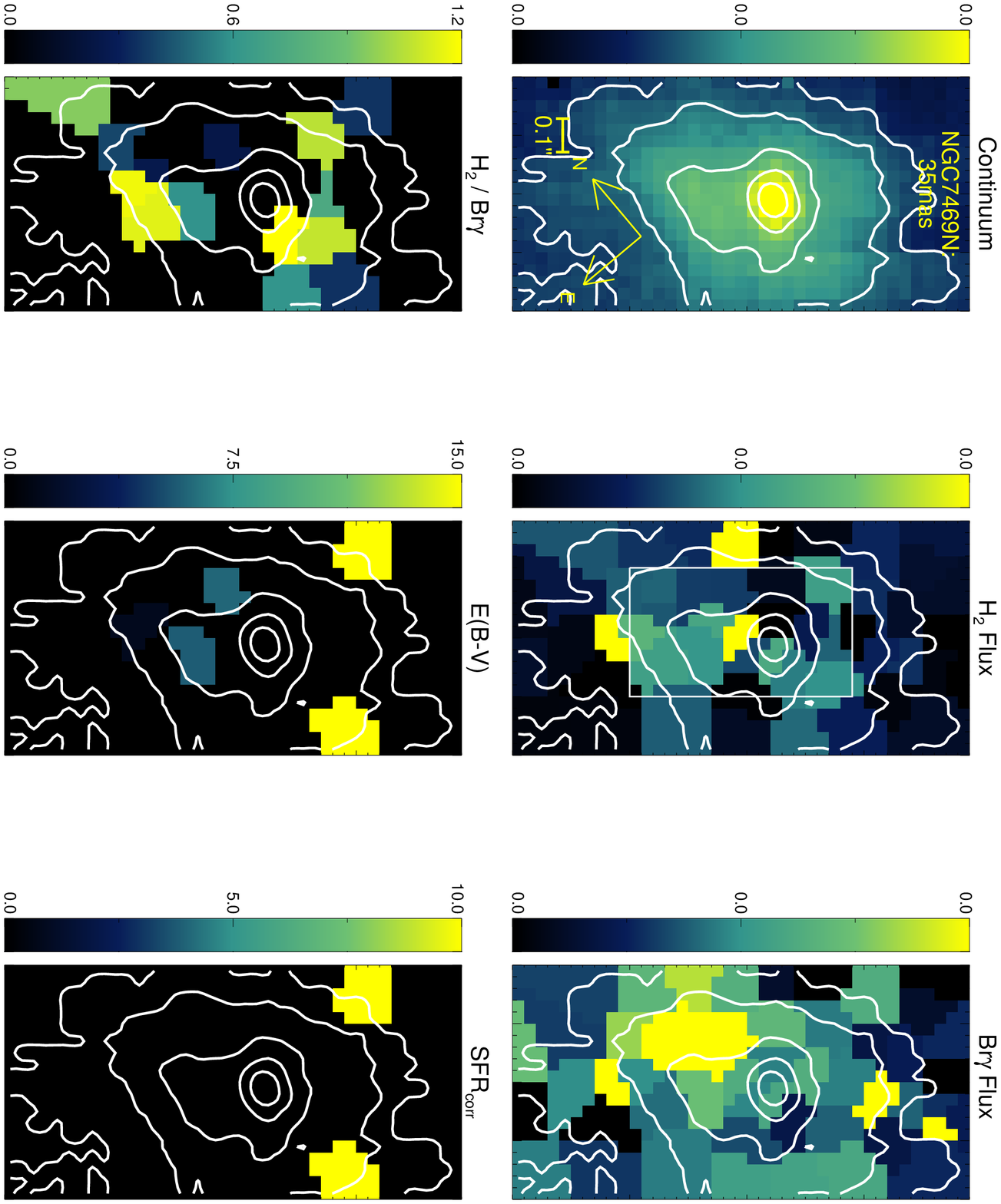}
\figsetgrpnote{Six-panel figure showing, from left to right, top to bottom, 1. $K$-band continuum map in relative flux units and shown with 0.1\arcsec scale bar and compass rose; the continuum contours are shown in all subsequent panels. 2. \molhy~1$-$0 S(1) flux map in 10$^{-16}$ erg s$^{-1}$ cm$^{-2}$, with box highlighting the ``\molhy-dominated region"; 3. \brg~flux map in 10$^{-16}$ erg s$^{-1}$ cm$^{-2}$; 4. \molhy~1$-$0 S(1)/\brg~(\molhy/\brg~hereafter) map showing the reliable spaxels (with S/N $>$ 1.5 in both \molhy~and \brg); 5. E(B-V) map; and 6. dust-corrected star formation rate map (M$_\odot$ yr$^{-1}$).}
\figsetgrpend

\figsetgrpstart
\figsetgrpnum{2.21}
\figsetgrptitle{NGC 6090}
\figsetplot{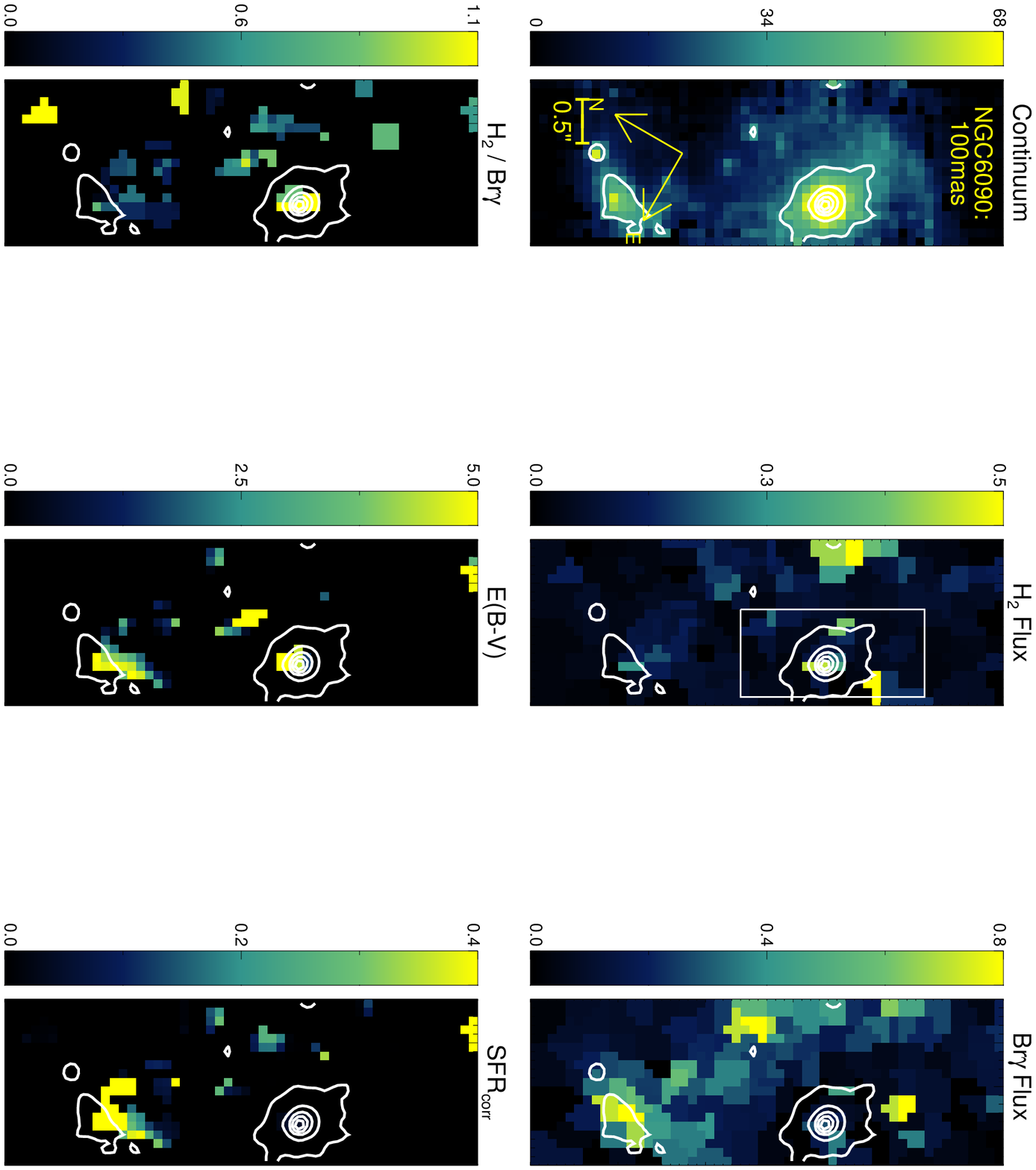}
\figsetgrpnote{Six-panel figure showing, from left to right, top to
  bottom, 1. $K$-band continuum map in relative flux units and shown
  with 0.5\arcsec scale bar and compass rose; the continuum contours
  are shown in all subsequent panels. 2. \molhy~1$-$0 S(1) flux map in
  10$^{-16}$ erg s$^{-1}$ cm$^{-2}$, with box highlighting the ``\molhy-dominated region"; 3. \brg~flux map in 10$^{-16}$
  erg s$^{-1}$ cm$^{-2}$; 4. \molhy~1$-$0
  S(1)/\brg~(\molhy/\brg~hereafter) map showing the reliable spaxels
  (with S/N $>$ 2 in both \molhy~and \brg); 5. E(B-V) map; and
  6. dust-corrected star formation rate map (M$_\odot$ yr$^{-1}$). The
dust extinction and star formation rate maps have been smoothed for
presentation purposes.}
\figsetgrpend

\figsetgrpstart
\figsetgrpnum{2.22}
\figsetgrptitle{NGC 7674W}
\figsetplot{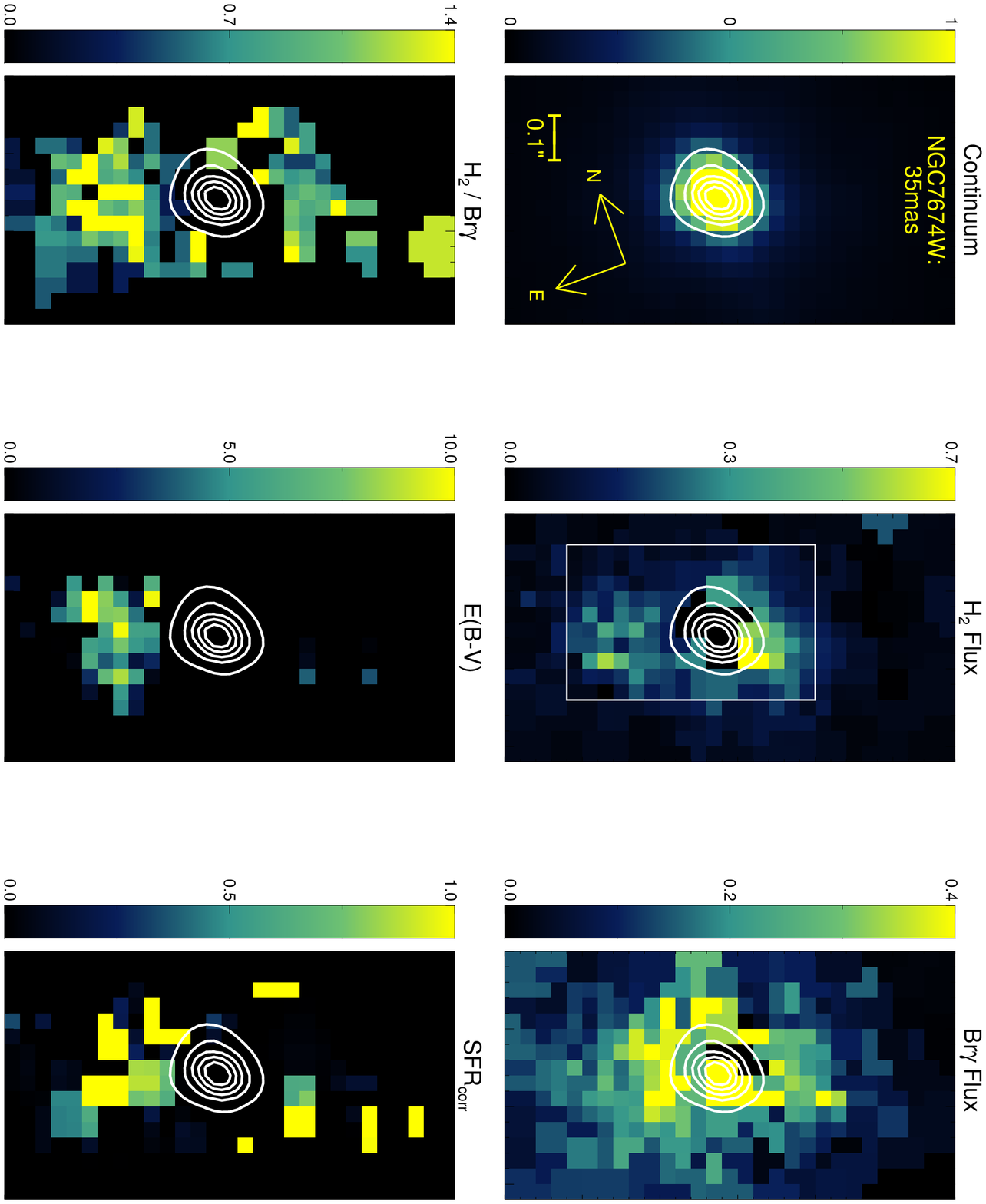}
\figsetgrpnote{Six-panel figure showing, from left to right, top to
  bottom, 1. $K$-band continuum map in relative flux units and shown
  with 0.1\arcsec scale bar and compass rose; the continuum contours
  are shown in all subsequent panels. 2. \molhy~1$-$0 S(1) flux map in
  10$^{-16}$ erg s$^{-1}$ cm$^{-2}$, with box highlighting the ``\molhy-dominated region"; 3. \brg~flux map in 10$^{-16}$
  erg s$^{-1}$ cm$^{-2}$; 4. \molhy~1$-$0
  S(1)/\brg~(\molhy/\brg~hereafter) map showing the reliable spaxels
  (with S/N $>$ 3 in both \molhy~and \brg); 5. E(B-V) map; and
  6. dust-corrected star formation rate map (M$_\odot$ yr$^{-1}$). The
dust extinction and star formation rate maps have been smoothed for
presentation purposes.}
\figsetgrpend

\figsetgrpstart
\figsetgrpnum{2.23}
\figsetgrptitle{IRAS F03359+1523 (100mas)}
\figsetplot{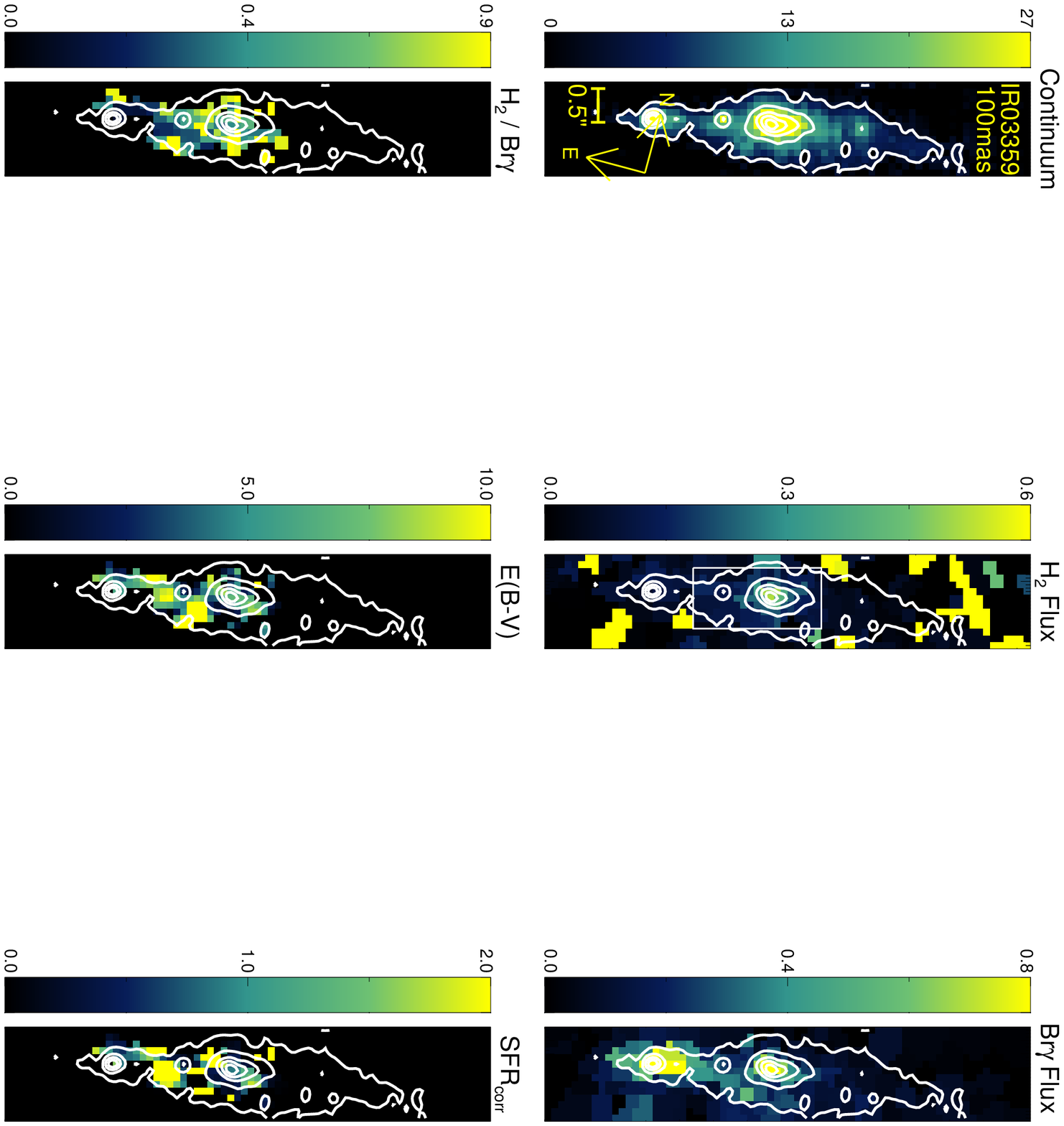}
\figsetgrpnote{Six-panel figure showing, from left to right, top to
  bottom, 1. $K$-band continuum map in relative flux units and shown
  with 0.5\arcsec scale bar and compass rose; the continuum contours
  are shown in all subsequent panels. 2. \molhy~1$-$0 S(1) flux map in
  10$^{-16}$ erg s$^{-1}$ cm$^{-2}$, with box highlighting the ``\molhy-dominated region"; 3. \brg~flux map in 10$^{-16}$
  erg s$^{-1}$ cm$^{-2}$; 4. \molhy~1$-$0
  S(1)/\brg~(\molhy/\brg~hereafter) map showing the reliable spaxels
  (with S/N $>$ 3 in both \molhy~and \brg); 5. E(B-V) map; and
  6. dust-corrected star formation rate map (M$_\odot$ yr$^{-1}$). The
dust extinction and star formation rate maps have been smoothed for
presentation purposes.}
\figsetgrpend

\figsetgrpstart
\figsetgrpnum{2.24}
\figsetgrptitle{IRAS F03359+1523 (35mas)}
\figsetplot{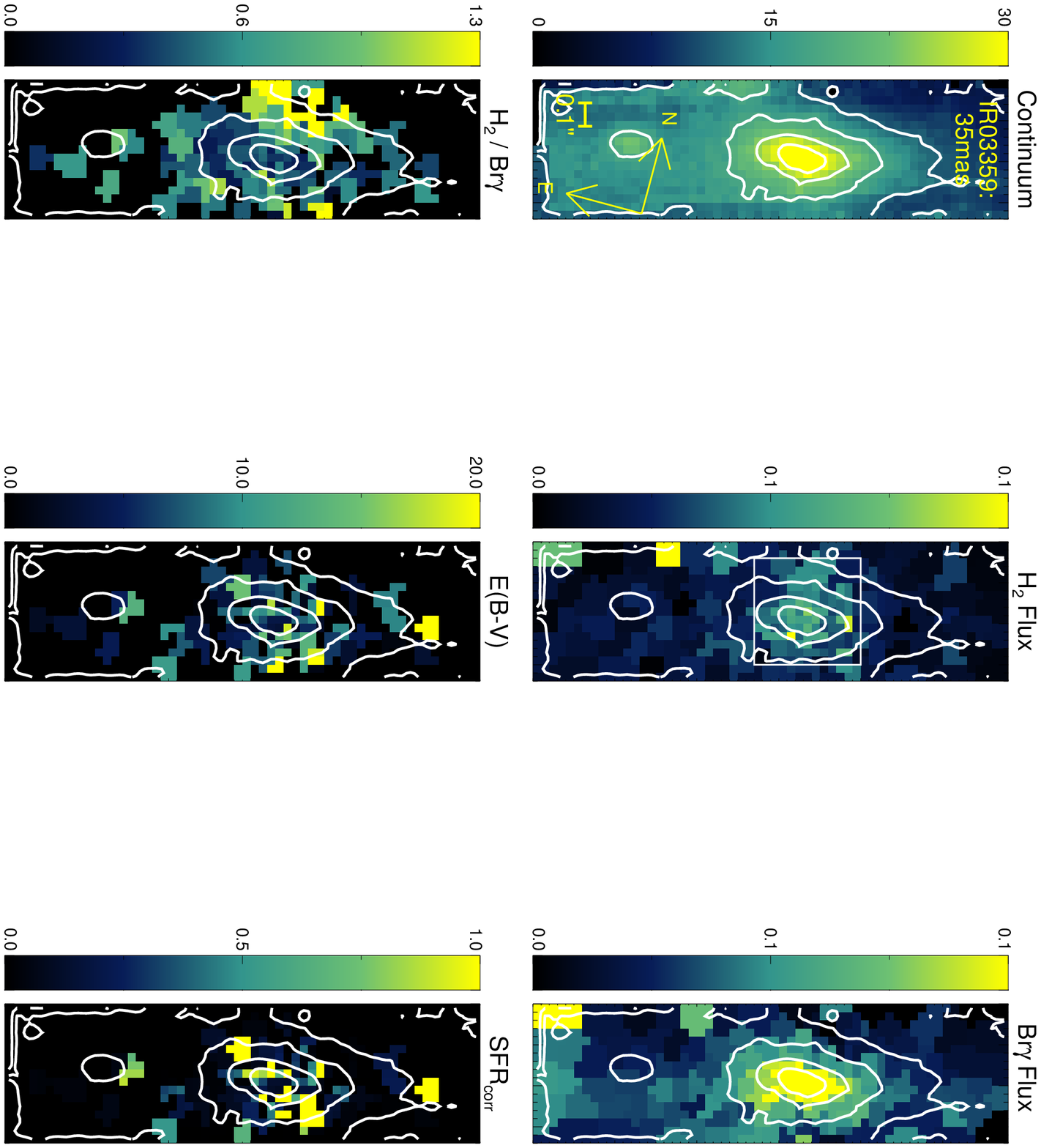}
\figsetgrpnote{Six-panel figure showing, from left to right, top to bottom, 1. $K$-band continuum map in relative flux units and shown with 0.1\arcsec scale bar and compass rose; the continuum contours are shown in all subsequent panels. 2. \molhy~1$-$0 S(1) flux map in 10$^{-16}$ erg s$^{-1}$ cm$^{-2}$, with box highlighting the ``\molhy-dominated region"; 3. \brg~flux map in 10$^{-16}$ erg s$^{-1}$ cm$^{-2}$; 4. \molhy~1$-$0 S(1)/\brg~(\molhy/\brg~hereafter) map showing the reliable spaxels (with S/N $>$ 3 in both \molhy~and \brg); 5. E(B-V) map; and 6. dust-corrected star formation rate map (M$_\odot$ yr$^{-1}$).}
\figsetgrpend

\figsetgrpstart
\figsetgrpnum{2.25}
\figsetgrptitle{MCG +08$-$11$-$002 (100mas)}
\figsetplot{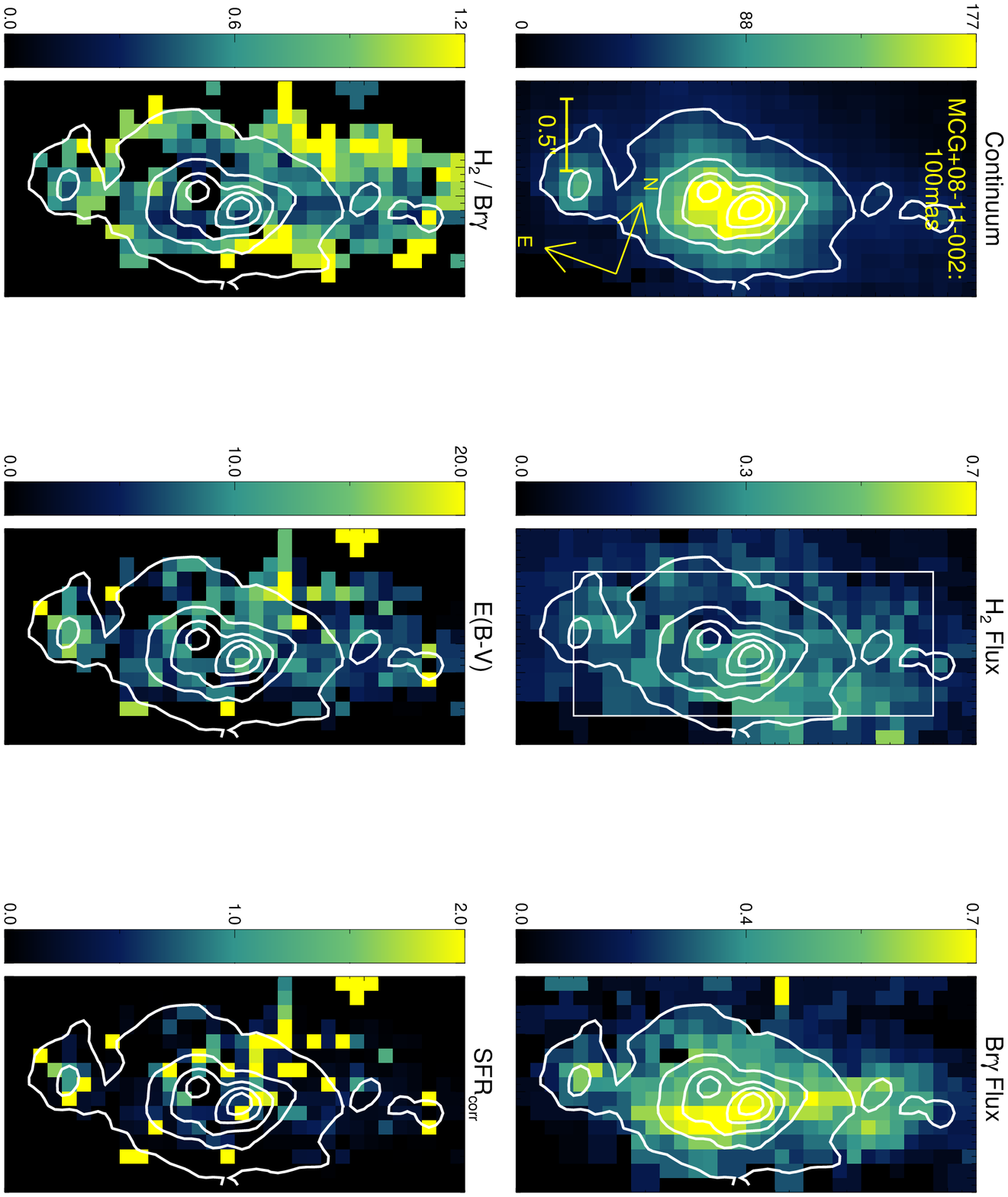}
\figsetgrpnote{Six-panel figure showing, from left to right, top to bottom, 1. $K$-band continuum map in relative flux units and shown with 0.5\arcsec scale bar and compass rose; the continuum contours are shown in all subsequent panels. 2. \molhy~1$-$0 S(1) flux map in 10$^{-16}$ erg s$^{-1}$ cm$^{-2}$, with box highlighting the ``\molhy-dominated region"; 3. \brg~flux map in 10$^{-16}$ erg s$^{-1}$ cm$^{-2}$; 4. \molhy~1$-$0 S(1)/\brg~(\molhy/\brg~hereafter) map showing the reliable spaxels (with S/N $>$ 3 in both \molhy~and \brg); 5. E(B-V) map; and 6. dust-corrected star formation rate map (M$_\odot$ yr$^{-1}$).}
\figsetgrpend

\figsetgrpstart
\figsetgrpnum{2.26}
\figsetgrptitle{MCG +08$-$11$-$002 (35mas)}
\figsetplot{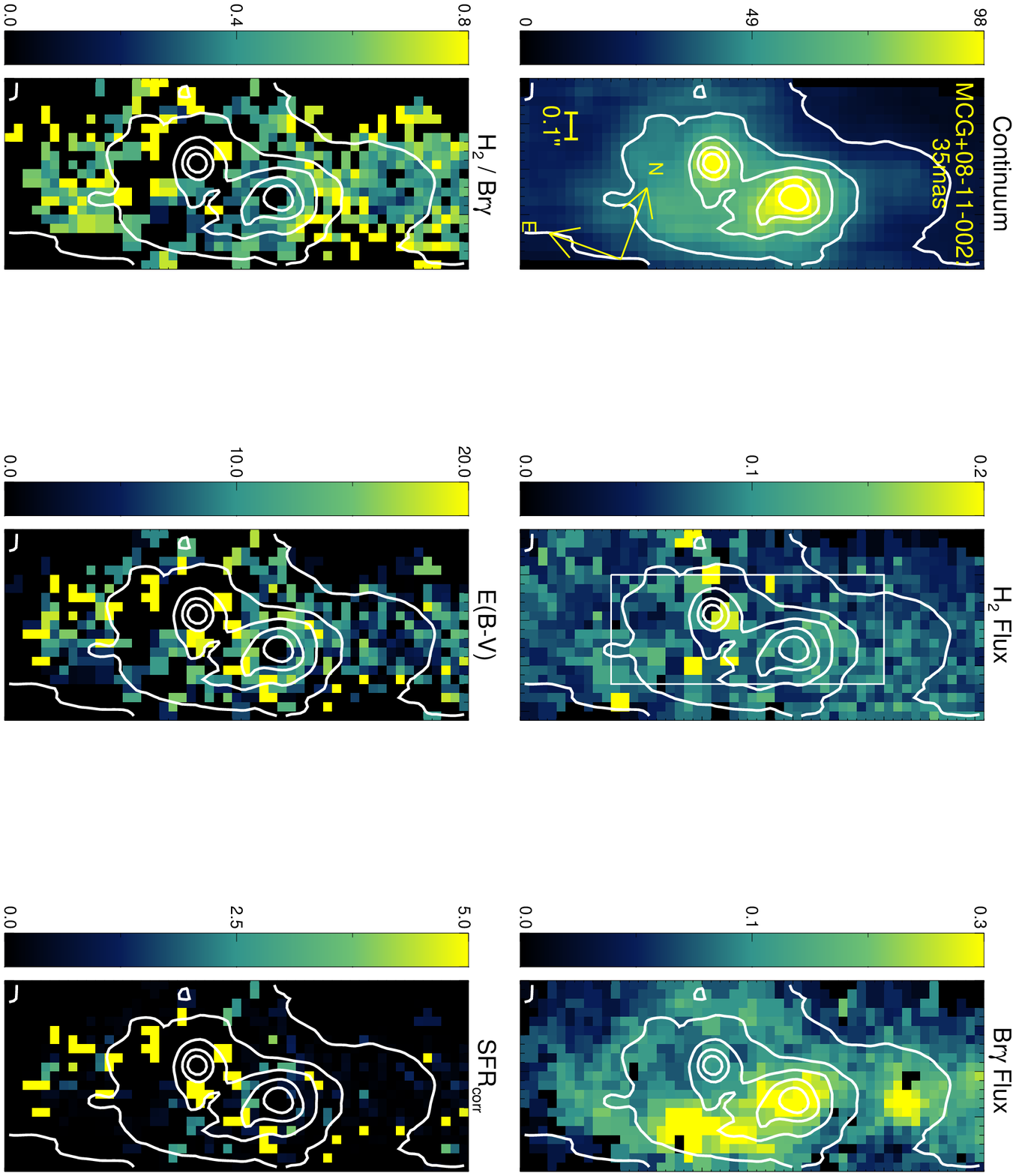}
\figsetgrpnote{Six-panel figure showing, from left to right, top to bottom, 1. $K$-band continuum map in relative flux units and shown with 0.1\arcsec scale bar and compass rose; the continuum contours are shown in all subsequent panels. 2. \molhy~1$-$0 S(1) flux map in 10$^{-16}$ erg s$^{-1}$ cm$^{-2}$, with box highlighting the ``\molhy-dominated region"; 3. \brg~flux map in 10$^{-16}$ erg s$^{-1}$ cm$^{-2}$; 4. \molhy~1$-$0 S(1)/\brg~(\molhy/\brg~hereafter) map showing the reliable spaxels (with S/N $>$ 3 in both \molhy~and \brg); 5. E(B-V) map; and 6. dust-corrected star formation rate map (M$_\odot$ yr$^{-1}$).}
\figsetgrpend

\figsetend

\begin{figure*}[htbp]
%\figurenum{2}
%\plotone{fig_IIIZw035_35scale.eps}
   \centering
   \includegraphics[width=.7\textwidth,angle=90]{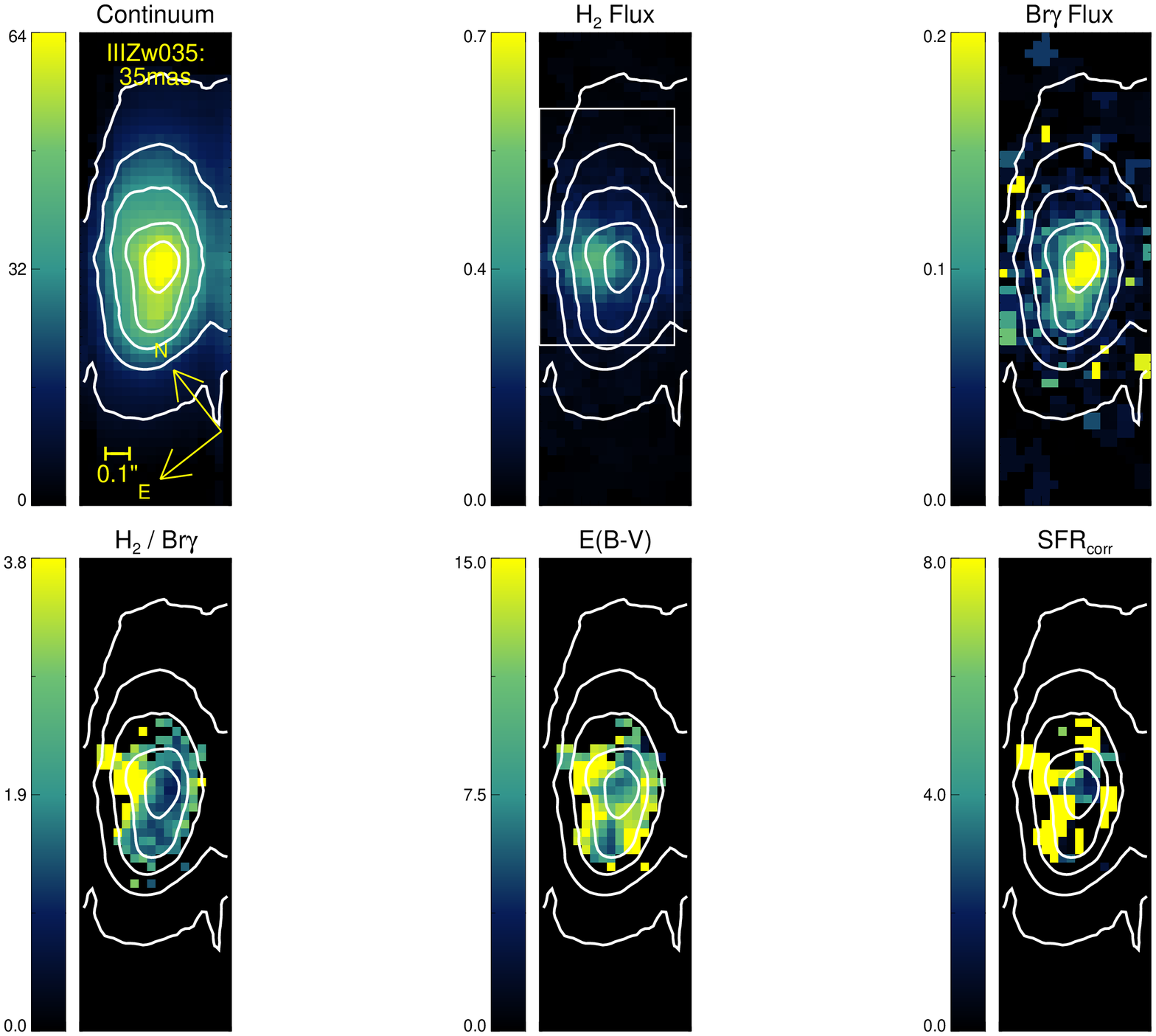}
\caption{Six-panel figure for III Zw 035 (with 35mas sampling)
  showing, from left to right, top to bottom, 1. $K$-band continuum
  map in relative flux units and shown with 0.1\arcsec scale bar and
  compass rose; the continuum contours are shown in all subsequent
  panels. 2. \molhy~1$-$0 S(1) flux map in 10$^{-16}$ erg s$^{-1}$
  cm$^{-2}$, \added{with box highlighting the ``\molhy-dominated region"}; 3. \brg~flux map in 10$^{-16}$ erg s$^{-1}$ cm$^{-2}$;
  4. \molhy~1$-$0 S(1)/\brg~(\molhy/\brg~hereafter) map showing the
  reliable spaxels (with S/N $>$ 3 in both \molhy~and \brg); 5. E(B-V)
  map; and 6. dust-corrected star formation rate map (M$_\odot$
  yr$^{-1}$) \deleted{in logarithmic units}. \added{In several
    galaxies, the SFR exhibits a large dynamic range such that the
    chosen color scale may be limited in showing the full range of the
    derived values. The complete figure set is available in the online journal.}}
  \label{fig:6panel}
\end{figure*}

  \section{Near-Infrared Emission Line Diagnostics}
  \label{results}

  \subsection{\molhy~Excitation Mechanisms and Temperatures}

Five \molhy~emission lines in our spectra trace the warm molecular
gas\deleted{, plausibly a product of both thermal and nonthermal
components}. \replaced{Potential thermal excitation mechanisms refer to those
processes that could cause collisional excitation with the surrounding
gas. Such collisional processes include heating by shocks or turbulence, by UV
photons from O and B stars, or by X-ray ionization}{These lines can be
thermally excited through collisions with the surrounding atomic gas,
itself heated by shocks, turbulence, UV photons from OB stars, or even
X-rays from the central AGN}~\cite[][]{Mouri94}. On
the other hand, non-thermal excitation processes include absorption of
UV-photons (resonance fluorescence), and collisions with
high-energy electrons~\cite[][]{Burton87}. The flux ratios between 
the transitions can partially distinguish between thermal and
nonthermal excitation mechanisms, thanks to the difference in their
efficiencies in populating the various vibrational levels.  
Much effort has been put
into developing models \replaced{that would}{to} predict the values of these line
intensity ratios for the scenarios of nonthermal
excitation~\cite[][]{Black87}, shock-heating~\cite[][]{Brand89}, and
thermal UV excitation~\cite[][]{Sternberg89}. \deleted{Observations of the
\molhy~emission in supernova remnants, photodissociation regions, and
several galaxies have been compared to these models to decipher the
mechanisms that excite the molecular gas.} 
Detailed studies of the near-infrared \molhy~line ratios in
Seyferts~\cite[][]{Muller-Sanchez17}, 
star-forming galaxies~\cite[][]{Riffel13}, 
LIRGs~\cite[][]{Bedregal09,Emonts14,Vaisanen17}, and nearby
spirals~\cite[][]{Mazzalay13,Smajic15,Busch17} have been enabled
within the past decade by a suite of near-infrared integral-field
instruments. Several of the results have indicated that the \replaced{hot}{warm}
molecular gas in these systems is \replaced{close to}{consistent with
  being} thermally excited by a combination of X-ray radiation and 
slow shocks.

\begin{figure}[tbp]
  \centering
  \includegraphics[width=.51\textwidth,trim={2cm 2cm 0 0},clip]{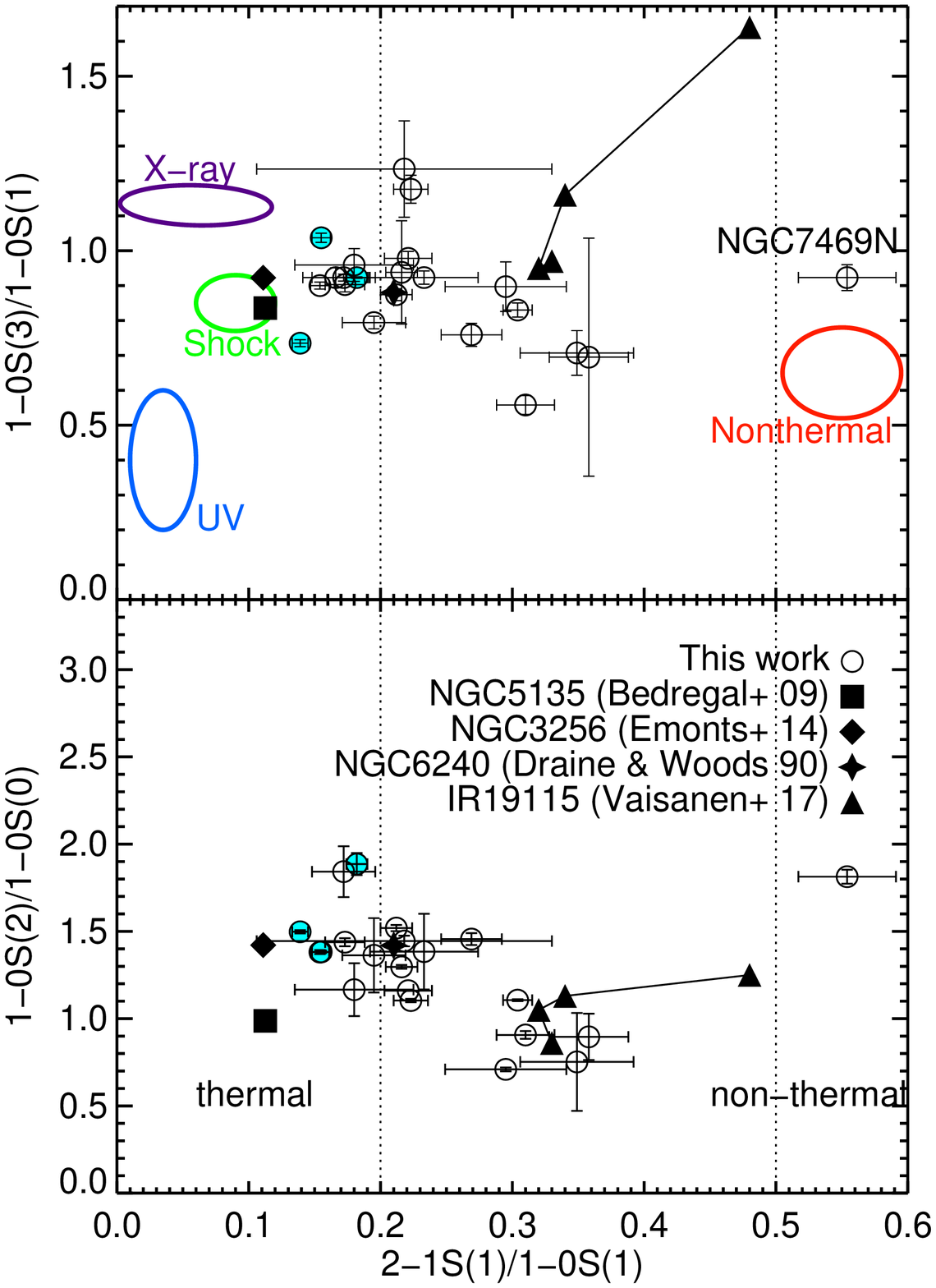}
  \caption{\molhy~ratio diagnostic plots comparing the nuclei of our (U)LIRGs (open
    circles) to other (U)LIRGs from literature (filled symbols). In
    the top plot, the ovals designate models as adopted from~\cite{Draine90}
    (X-rays; purple),~\cite{Brand89} (shock;
    green),~\cite{Sternberg89} (UV; blue),
    and~\cite{Black87} (nonthermal; red). The
    thermal and non-thermal regions are indicated as 
    $\frac{2-1S(1)}{1-0S(1)} < 0.2$ and $\frac{2-1S(1)}{1-0S(1)} >
    0.5$, respectively. Most of the \molhy~line ratios of our 
    merger systems do not align with pure theoretical models,
    indicating that mixing may complicate the identification of the
    main mechanism at
    hand. \added{The filled cyan circles represent Mrk 273, IRAS
    F17207$-$0014, and III Zw 035, which display some of the best cases of
    shocked gas (see \S \ref{sec:h2brg_shocks}) and reside near the
    region predicted by shock models.} The connected filled triangles
  represent spatially-resolved regions 
    within IRAS F19115-2124~\cite[][]{Vaisanen17}, showing that
    different \added{spatial} components within a merger may also host
    distinctly-excited molecular gas. }
  \label{fig:h2ratio}
\end{figure}

 Here we employ two particular \molhy~line ratio diagnostics, $2-1$
 S(1)/$1-0$ S(1) versus $1-0$ S(3)/$1-0$ S(1) and versus $1-0$ S(2)/$1-0$ S(0), to gain
 insight into the excitation mechanisms of our galaxies by extracting
 their integrated values over the \added{high signal-to-noise,}
 \molhy~emission-dominated \added{``nuclear''} region within each galactic nucleus
 \added{(see Figure Set \ref{fig:6panel} and Table \ref{tbl:derivedprop})} and
 comparing them to various theoretical models (Figure
 \ref{fig:h2ratio}). First, we note that the locus of the galaxies lies
 closer to the various thermal models on the left side of the plot,
 suggesting that non-thermal UV fluorescence, though plausibly
 present, is unlikely to contribute to the dominant excitation mechanism. The 
 exception is NGC 7469N, which, interestingly, features an offset ($\sim
 0\farcs2$) \brg~emission relative to the continuum peak.
  We further note that several of the systems where we found shocked
  gas based on enhanced
  \molhy/\brg~ratios
  \replaced{(e.g. Medling et al. 2015b)}{(e.g. Mrk 273 in U et
    al. 2013; IRAS F17207$-$0014 in Medling et al. 2015b; III Zw 035; filled
    cyan circles in the figure)} do reside near the region on the
  diagnostic plot \deleted{as}occupied by predictions from shock
  models. \added{See \S \ref{sec:h2brg_shocks} for further analysis of
    \molhy/\brg~of this sample.}

We also compare our \molhy~line ratios with those of several
LIRGs found in the literature in Figure \ref{fig:h2ratio}. For both
NGC 5135~\cite[][$\log L_{\rm IR}/L_\odot = 11.2$]{Bedregal09} and NGC
3256~\cite[][$\log L_{\rm IR}/L_\odot = 11.6$]{Emonts14}, the 
\molhy~line ratios within their respective nuclei are
represented as single components. In the case of NGC
6240~\cite[][$\log L_{\rm IR}/L_\odot = 11.9$]{Draine90}, the \molhy~line ratios
were extracted from the inner 3\arcsec~region. These sources are
broadly consistent with the (U)LIRGs in our sample, where the
\molhy~gas is likely excited by X-ray irradiation and/or shocks from
supernovae remnants.
We also show the spatially-resolved \molhy~line ratios extracted from
different regions within IRAS F19115-2124, where the aperture
corresponding to the circumnuclear region is shown to be dominated by
non-thermal emission and that corresponding to the strongest
star-forming areas appears to be dominated by thermal
excitation~\cite[][]{Vaisanen17}. 

\begin{figure}[tbp]
  \centering
  \includegraphics[width=.38\textwidth,angle=90]{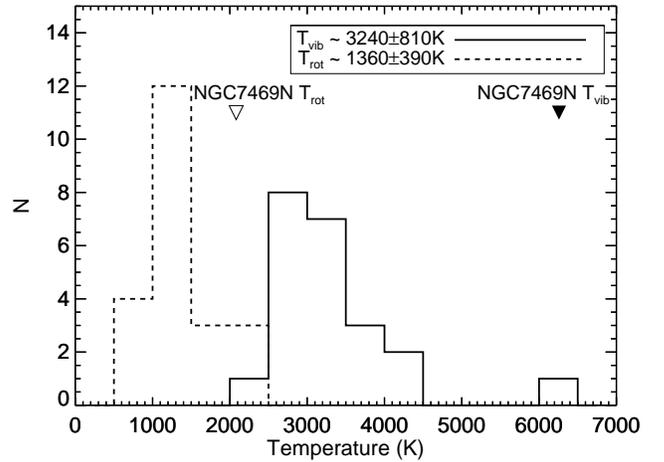}
  \caption{Histograms of the vibrational (solid) and rotational
    (dashed) temperatures computed from \molhy~transitions (see
    Equations in text). The mean values for $T_{\rm vib}$ and $T_{\rm rot}$
  differ appreciably, by $\sim$ a factor of two. The downward
  triangles mark the $T_{\rm vib}$ (filled) and $T_{\rm rot}$ (open)
  for NGC 7469N\added{, which shows the largest discrepancy}. 
}
  \label{fig:h2temp}
\end{figure}

Given the \molhy~line ratios, we make use of the equations
from~\cite{Reunanen02} and~\cite{Rodriguez04,Rodriguez05} to compute
the rotational and vibrational temperatures within the nuclei of our
sample:
\begin{equation}
  T_{\rm vib} \cong 5600 / {\rm ln}(1.355 \times \frac{1-0 S(1)}{2-1 S(1)}) 
\end{equation}
\begin{equation}
\text{and  } \quad T_{\rm rot} \cong -1113 / {\rm ln} (0.323 \times
\frac{1-0 S(2)}{1-0 S(0)}) \text{   .}
\end{equation}
The derived vibrational and rotational temperatures
\replaced{can be found}{are listed} in \tbl \ref{tbl:derivedprop} and
illustrated in Figure \ref{fig:h2temp}. The mean vibrational temperature
for our sample is $\langle T_{\rm vib} \rangle = 3240 \pm 810 K$,
whereas the mean rotational temperature is $\langle T_{\rm rot}
\rangle = 1360 \pm 390 K$. The
vibrational temperatures for our sample tend to be above those
measured for Seyferts~\cite[$T_{\rm vib} \lesssim 2600$
K;][]{Reunanen02}. We note that the mean values for $T_{\rm vib}$ and
$T_{\rm rot}$ differ by a factor of $\sim$2 for most of the sample, but
the difference is particularly appreciable for NGC 7469N. 
\replaced{The}{A large} difference between $T_{\rm rot}$ and $T_{\rm vib}$ is usually
explained by the presence of fluorescently excited \molhy, while
purely thermally excited \molhy~would exhibit much more similar
$T_{\rm rot}$ and $T_{\rm vib}$ values~\cite[\eg][]{Black87,Draine96,Martini99}.

\subsection{Dust Correction and Nuclear Star Formation Rates}

Given the expected level of heavy dust extinction in the nuclei of
(U)LIRGs~\cite[$A_V \sim 3-17$ mag;][]{Piqueras13}, star formation
rates (SFR) as computed from optical or UV 
tracers may not offer the complete picture in this domain. The
hydrogen recombination lines in the near-infrared such as the Brackett
or Paschen series may offer a truer measure of the nuclear SFR as they
trace ionizing photons from young stars. However, if the dust
extinction in the nuclear region were patchy or had an unusual
wavelength dependence, even \added{global} SFRs determined from these
near-infrared lines may be biased. Thus, we compute the
spatially-resolved extinction map for each source and  
determine the nuclear SFR using the dust-corrected \brg~line flux as a
proxy for recent star formation.  Following
e.g.~\cite{Calzetti94,Dominguez13}, we compute the intrinsic
luminosities $L_{\rm int}$ as follows: 
\begin{equation}
L_{\rm int}(\lambda) = L_{\rm obs}(\lambda) 10^{0.4 A_\lambda}
\end{equation}
where $L_{\rm obs}$ are the observed luminosities and $A_\lambda$ is
the extinction at wavelength $\lambda$, $A_\lambda = k(\lambda)
E(B-V)$. We determine the color excess $E(B-V)$ using the relationship
between the nebular emission line color excess and the Brackett
decrement~\cite[\brg~and \brd; \eg][]{Momcheva13}: 
\begin{align*}
E(B-V) &= \frac{E({\rm Br}\delta - {\rm Br}\gamma)}{k({\rm Br}\delta)-k({\rm Br}\gamma)} \\
 &= \frac{2.5}{k({\rm Br}\delta)-k({\rm Br}\gamma)} \log_{10} [ \frac{(Br\gamma/Br\delta)_{\rm
  obs}}{(Br\gamma/Br\delta)_{\rm int}} ] \quad ,
\end{align*}
where $k(Br\delta) = 0.43$ and $k(Br\gamma) = 0.36$ are the reddening
curves evaluated at \brd~and \brg~wavelengths,
respectively~\cite[\eg][Calzetti, \emph{private
  communication}]{Cardelli89,Calzetti01}. We obtain
$(Br\gamma/Br\delta)_{\rm int} = 3.0/2.1 = 1.4 $
from~\cite{Osterbrock89}, which, in combination with the extinction
curves and the observed \brg/\brd~maps, provide the color excess map
for each nucleus and subsequently, the dust-corrected \brg~maps. 
We then convert the luminosity of \brg, $L$(\brg), to SFR using: 
\begin{equation}
{\rm SFR}~({\rm M}_\odot~\rm{yr}^{-1}) = 8.2 \times 10^{-40}~L({\rm
  Br} \gamma)~({\rm ergs~s}^{-1}) 
\end{equation} 
as adopted from~\cite{Kennicutt98}. In order to compare SFRs across
galaxies and data sets covering different physical scales of the
nuclear regions, we further compute the surface density of SFR, 
$\Sigma_{\rm SFR}$ (in M$_\odot$ yr$^{-1}$ kpc$^{-2}$), by taking into
account only the subregion within the OSIRIS FOV where reliable
\molhy~and \brg~fluxes can be measured. 
The observed and extinction-corrected total nuclear
\brg~luminosities along with their corresponding SFR surface densities are listed in
\tbl \ref{tbl:derivedprop}.  

\subsection{Warm molecular gas mass} 
The total warm molecular hydrogen mass in the galactic nuclei can be
computed from the flux-calibrated \molhy~$1-0$ S(1) emission line. Using
the prescriptions from \eg \cite{Mazzalay13},
\begin{equation}
M_{\text{\molhy}} \simeq 5.0875 \times 10^{13}
(\frac{D_L}{\text{Mpc}})^2 (\frac{F_{1-0 S(1)}}{\text{erg s}^{-1} \text{
    cm}^{-2}}) 10^{0.4
  A_{2.2}} \text{  ,}
\end{equation}
where $M_{\text{\molhy}}$ is the mass of the warm \molhy~at $T \simeq
2000$ K in M$_{\odot}$, D$_L$ is the luminosity distance in Mpc,
F$_{1-0 S(1)}$ is the flux of the \molhy~$1-0$ S(1) line, and
$A_{2.2}$ is the extinction at 2.2$\mu$m~\cite[\eg][]{Scoville82}. The
total flux of \molhy~within the same subregion of reliable
measurements and the corresponding derived molecular gas mass are
listed in \tbl \ref{tbl:derivedprop}.

  \subsection{\molhy/\brg~as a Shock Tracer}
  \label{sec:h2brg_shocks}
  Given that the warm (T $\sim$ 2000 K)  molecular gas is most likely
  thermally excited due to shocks or perhaps X-ray radiation (Figure
  \ref{fig:h2ratio}), and that the \brg~line traces UV-ionizing
  radiation from young stars, the \molhy~$1-0$
  S(1)/\brg~(\molhy/\brg~hereafter) ratio quantifies the relative
  contributions from~\replaced{shocks or X-ray radiation versus UV
  radiation}{these energy sources. Specifically, a high \molhy/\brg~ratio signals a
  contribution to the \molhy~heating that is above and beyond that
  contributed by the source of the radiation that produces the H{\sc
    ii} regions. That excess heating may be due to shocks or X-ray
  radiation}. \deleted{Analogous to the use of optical BPT diagram,} The 
  \molhy/\brg~ratio \deleted{, and it alone,} is arguably better at separating
  starbursts, Seyferts, and composite galaxies than \replaced{using
  the}{its} optical \replaced{lines}{counterparts}, \deleted{this is} because
\molhy~traces the \replaced{cooler 
  molecular gas}{warm molecular gas at $\sim$100 K rather than
  10$^4$ K ionized atomic gas} and is therefore less closely coupled with high
  ionization lines (\eg \oiii). The \molhy/\brg~ratio divides
  starburst galaxies and Seyferts at 0.6, while \molhy/\brg~$\gtrsim$
  $2-3$ further indicates shock heating or photoionization by the
  central AGN~\cite[\eg][]{Larkin98,Rodriguez05,Riffel08}.

\begin{figure*}[htb]
  \centering\offinterlineskip
  \includegraphics[width=.13\textwidth,angle=90,trim={16.8cm 0 0 0},clip]{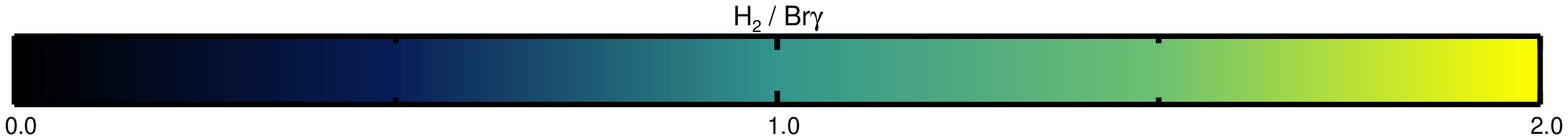}
  \colorbox{black}{\includegraphics[width=.29\textwidth,angle=90,trim={5.2cm 0 6.8cm 0},clip]{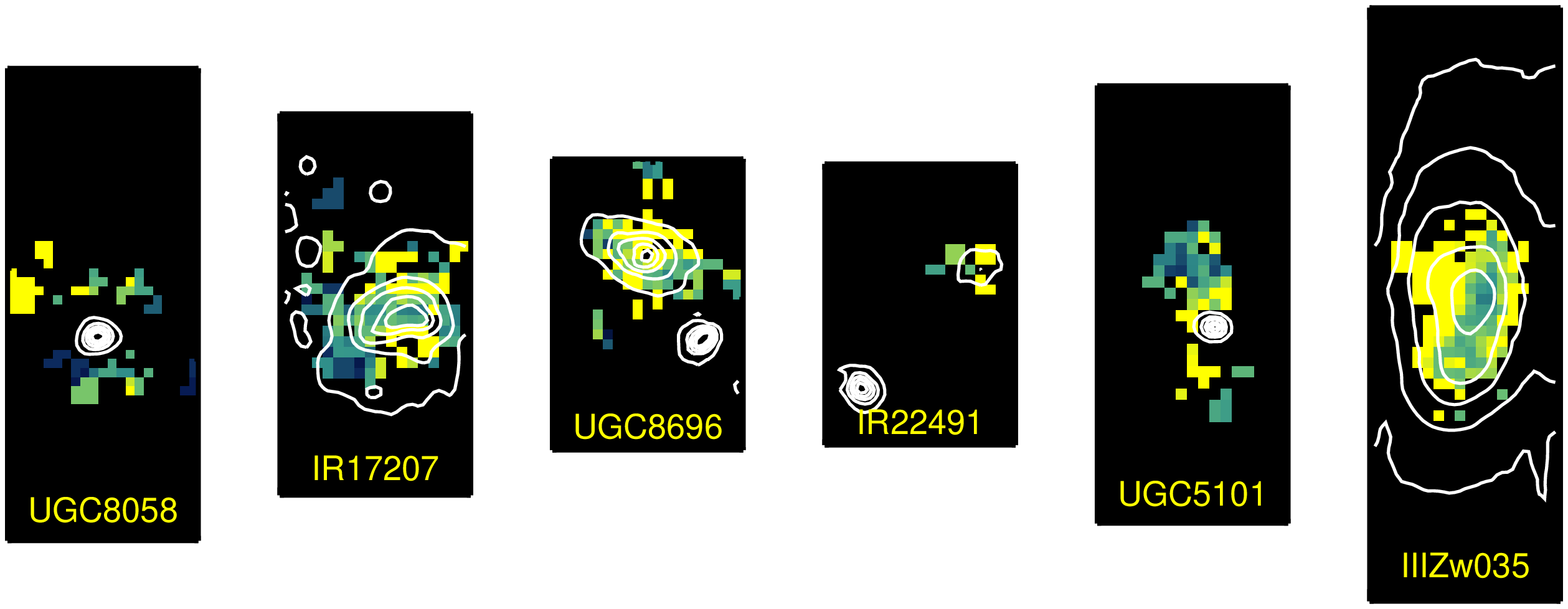}}
   \caption{
Spatial distribution of the \molhy/\brg~maps for potential shock
candidates. These sources exhibit coherent structure with elevated
\molhy/\brg~values by visual inspection. Here, the \replaced{scale}{color} bar is
constant for all the panels. The contours represent the
$K$-band continuum. 
}
  \label{fig:shocked}
\end{figure*}

 \deleted{The \molhy/\brg~maps for all our galaxies are presented in 
  Figure~\ref{fig:6panel} and
%    Figures~\ref{fig:6panel_app1}$-$\ref{fig:6panel_app25}.}  
    Figures A1$-$A25.}  
  In our sample of 21 interacting systems, 6 sources 
  \replaced{feature spaxels}{have regions} with \molhy/\brg~$> 2$ (UGC
  08058, IRAS F17207$-$0014, UGC 
  08696, IRAS F22491$-$1808, UGC 05101, and III Zw 035). These shocked
  candidates consist of \replaced{5}{five} ULIRGs \replaced{plus}{and}
  one LIRG within our sample 
  (see Figure~\ref{fig:shocked}).  
  The shocked nature of the outflowing molecular gas in UGC 08696 and IRAS
  F17207$-$0014 have been discussed extensively in~\cite{U13}
  and~\cite{Medling15_ir17207}, respectively. UGC 08696
  features a dual AGN system where enhanced \molhy/\brg~ratios are
  found both at the site of a hard X-ray AGN in the southwestern
  region as well as near the obscured AGN in the north. Given that the
  northern AGN is very obscured, its surrounding \molhy~gas is likely
  to be primarily heated by shocks despite the presence of an AGN. The
  shocked molecular gas in IRAS F17207$-$0014 is spatially coincident with the
  base of a nuclear superwind. 

  The strongest \added{new} shocked outflow candidate \deleted{that came out of this work}
  is \added{the} OH megamaser host \added{galaxy} III Zw 035, whose \molhy/\brg~line ratio and
  \molhy~kinematic maps point to a clump of shocked, outflowing
  \molhy~just \replaced{left}{south} of the continuum peak (see
  Figure~\ref{fig:6panel}). This shocked gas is seen emanating
  from the minor axis of the disk in a fan shape. There is a hint of
  biconical signature on the opposite side of the disk, but the
  spatial coverage of the OSIRIS FOV is too limiting to
  confirm \replaced{it}{its full extent}. The lack of AGN signature from hard X-ray or [Ne
  \textsc{v}] detection in our multiwavelength
  ancillary data set~\cite[][]{Iwasawa11,Petric11,Inami13} would suggest that
  these shocks~\replaced{have not resulted from AGN ionization, but likely from
  nuclear starbursts}{are not being driven by an AGN}.  IRAS F22491$-$1808 
  hosts two kinematically distinct nuclei that are 2.2 kpc apart (see
  Paper I and the Appendix for details), and a molecular outflow has
  been detected in \molhy~\cite[][]{Emonts17} but not in
  OH~\cite[][]{Veilleux13}. Our OSIRIS data shows that the
  \molhy~outflow entrains shocked gas with \molhy/\brg~$>$ 2 \added{in
    the eastern nucleus}. UGC
  08058 and UGC 05101 both feature a bright AGN in the center, so the
  gas is difficult to see at \deleted{the exact site of} the nucleus
  \added{given the bright continuum source}, but
  excited \molhy~gas is detected in the circumnuclear region
  \replaced{which}{that} may be caused by shocks or AGN photoionization. 

  For the remaining sources, the \molhy/\brg~ratio reaches no
  higher than $\sim$1.5. In several cases, \molhy~is preferentially
  located on the outskirts ($\sim$ 150$-$250 pc) of the nucleus
  relative to \brg~(\eg UGC 08387, NGC 2623, and IRAS
  F03359+1523). Here, the spatial coincidence of \brg~relative to the $K$-band 
  continuum indicates the sites of nuclear star formation, whereas the
  off-nuclear \molhy~gas may be expelled winds from the nearby central
  sources based on kinematics analysis (to be presented in a forthcoming paper).
  For CGCG 436$-$030, IRAS F20351+2521, and NGC 6090, the
  \molhy~flux is weak but cospatial relative to that of \brg, resulting in low
  \molhy/\brg~values across the nuclear region. 

  Statistical results of the \deleted{integrated} \molhy/\brg~line ratio for our
  sample (\added{for} individual galaxies as well as for ULIRGs and
  LIRGs as two populations) are 
  presented in \tbl \ref{tbl:h2brgratios}. We note that the median
  value for our LIRGs, 0.77, is consistent with VLT/SINFONI
  results presented for LIRGs (median = 0.77 with
  5\replaced{\%}{$^{\rm th}$} and 95\replaced{\%}{$^{\rm th}$}
  \replaced{quantile =}{percentiles of} 0.29 and 2.30, respectively) in~\cite{Colina15},
  which consists of a sample of LIRGs \replaced{lower in luminosity within the
  $11.1 < \log L_{\rm IR}/L_\odot < 11.7$ range}{with $\log L_{\rm
    IR}/L_\odot = 11.1-11.7$, reaching a lower luminosity range than
  our LIRGs do}. \replaced{While}{The} VLT/SINFONI
\added{observations} \replaced{has}{span} a \deleted{slightly} larger
FOV and coarser \deleted{spatial} resolution \added{(average FWHM $\sim$
  0\farcs6'')} than \added{our} OSIRIS \added{observations do}. \added{While} 
  the sample in~\cite{Colina15} is \deleted{much} more nearby ($z <
  0.018$), \replaced{such that the IFS
  coverage is comparable for our two sets of statistical
  measurements}{their IFS coverage includes the inner $\sim 2.5-5$ kpc
  regions, more extended than ours in general}.  The statistical results for our
  ULIRGs (median \added{\molhy/\brg}~= 1.43) are similar to those
  obtained from Gemini/NIFS 
  observations of Seyferts (median = 1.41 with 5\replaced{\%}{$^{\rm
      th}$} and 95\replaced{\% quantile}{$^{\rm th}$ percentiles} =
  0.40 and 4.00, respectively) as reported in~\cite{Colina15}, even
  though not all of our ULIRGs host an AGN. The similarity in the high
  \molhy/\brg~ratio between the ULIRGs and the Seyferts suggests that
  the two populations~\replaced{may have comparable excitation mechanism in
  shocks or AGN ionization, or a combination of both}{both have
  significant excesses of warm molecular gas compared to what is seen
  in the larger population of LIRGs}. 

\subsection{Effects of Spatial Resolution}
\explain{New subsection to address the referee's comments on the
  effects of resolution, and on resolved vs. integrated values.} 

The effect of angular resolution on line ratios was considered 
significant in previous studies, particularly in the case of
LIRGs~\cite[e.g.][]{Colina15}. Given that lower-resolution
observations are commonplace, whether due to seeing limitations
or different instrument capabilities, we explored how differences in
spatial resolution might affect our measurements in these exotic
environments within the nuclei of (U)LIRGs. High resolution data, such
as that taken with OSIRIS, present the opportunity to distinguish
resolved sources from diffuse emission at small scales and to quantify
the dilution of the emission line signal, if any. 

First, it is worth reiterating the findings from~\cite{Colina15} that
are relevant to the discussion at hand. %With IFS data taken with
%VLT/SINFONI in seeing-limited mode, the study explored the inner
%$\sim 2.5-5$ kpc region of a sample of 10 LIRGs that are, in general,
%lower luminosity (down to $\log L_{\rm IR}/L_\sun \sim 11.1$) than our
%present sample with average FWHM of $\sim$ 0.6\arcsec, in comparison
%to selected Seyfert and star-forming galaxies. 
In assessing the effect of spatial resolution, they compared the
integrated (e.g. flux-weighted) and median \feii (1.64$\mu$m)/\brg~and
\molhy/\brg~measurements and found that, with the exception
of a few objects, most LIRGs show large differences in the
\feii/\brg~ratio with integrated over median values $\sim 1.5 - 2.2$
times higher. They reasoned that integrated values are more affected
by the locus of the nuclei and might therefore appear as higher
excitation than the median ones, but questioned if this was a general
behavior in LIRGs or galaxies in general if larger volumes of galaxies
with IFS data were present. 

We investigated if a similar effect in line ratios might be seen in
our data. Within our sample, four sources were taken in both the
35 mas and 100 mas plate scales: III Zw 035, IRAS F01364$-$1042, IRAS
F03359$+$1523, and MCG $+$08$-$11$-$002. Since all but the two
observations for MCG $+$08$-$11$-$002 were obtained on different
nights, line ratios provide a more robust test of consistency than
calibrated line fluxes, which may incorporate photometric errors from
different observing conditions. We extract the \molhy/\brg~statistics
from Table~\ref{tbl:h2brgratios} for these objects and compared the
flux-weighted mean over median values at different plate scales. We do
find a slight enhancement in the mean over median values of $\sim 1.1 -
1.3$ times, but we see no systematic differences between those
measured from the 100mas and the 35mas scales. 

There may be several reasons for the apparent discrepancy from
the~\cite{Colina15} claim. The
OSIRIS area coverage is more limited to the central kpc region, which
is dominated by the nuclear source. Thus, a median measurement
representative of the nuclear region is less susceptible to the
diffuse emission on broader physical scales. Even in the case of our
coarser resolution, the nuclear components are resolved. In addition,
\cite{Colina15} illustrated that the disparity between the median and
integrated line ratios is negligible in the Seyferts, but more prominent
in LIRGs and star-forming galaxies. There may be a range of the
expected divergence among different types of galaxies, and our sample
lies in the higher-luminosity end of such a spectrum. Lastly,
the~\cite{Colina15} conclusion was based on the integrated and median
values of the 
\feii/\brg~ratio. They also showed the \molhy/\brg~measurements in
their Figure 7 where the differences appear to be smaller.
%, but it is difficult to quantify. 

%We compared the mean \molhy/\brg~values
%derived from the 100 mas and 35 mas plate-scale data cubes for these
%four objects in Figure \ref{fig:res}. The measurements generally agree
%with each other; the 1-$\sigma$ uncertainties show the range of
%\molhy/\brg~values spanned by the spaxels within the nuclear
%region. In the case of III Zw 035, the exposure time for the 35 mas
%observation was 5 times that of the 100 mas observation; this
%difference accounts for the difference in the quality of
%the detection of the diffuse \brg~emission across the FOV. As one of
%the best examples of having shock-heated molecular gas in our sample,
%the \molhy~appears localized in a region close to the nucleus. Thus,
%one would expect that the mean \molhy/\brg~ratio averaged over the
%smaller FOV of the 35 mas data cube would be higher than that measured
%from the larger FOV of the 100 mas data cube. The diluted line ratio
%from the 100 mas data might lead to an easier overlook of the shocked gas in
%this galaxy. 

To further verify the impact of spatial resolution on our results, we
rebinned all the data to the lowest physical resolution present in the
sample, e.g. that of IRAS F22491$-$1808 at 146.7 pc/spaxel. The most
important effect is the loss of ability to visually identify resolved structures
originally present in the sample, but also to identify potential
shocked gas given how its peak signal is smoothed out. The
quantitative influence from the degradation of angular resolution on
peak \molhy/\brg~values is further shown in Figure \ref{fig:agnfrac} in 
\S \ref{disc:agn}.

%\begin{figure}[htb]
%  \centering
%  \includegraphics[angle=90,width=.49\textwidth]{fig_compres.eps}
%  \caption{Comparison of the mean \molhy/\brg~values measured from 100
%  mas data (orange filled diamond) and 35 mas data (blue open diamond)
%for the four objects in this sample that feature both plate
%scales. The line ratios from different scales agree within a standard
%deviation in each case; small discrepancies are likely due to
%difference in data quality but also to the emission line distributions
%within each source (see text for details).} 
%  \label{fig:res}
%\end{figure}

  \section{Discussion}
  \label{discussion}

\subsection{SFR at nuclear scales}
\label{disc:sfr}
\begin{figure*}[htb]
  \centering
  \includegraphics[angle=90,width=.9\textwidth]{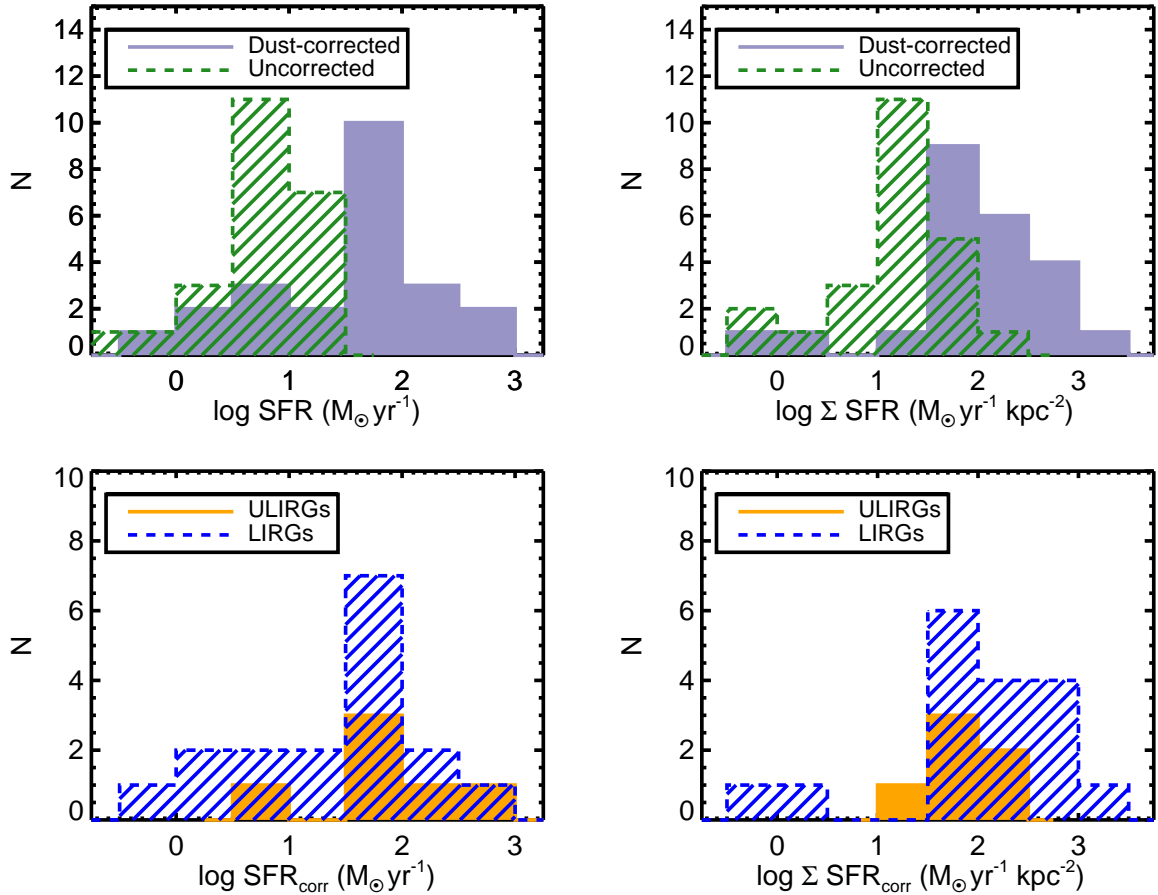}
  \caption{(Top) Histograms of the nuclear SFR (left) and SFR surface density
    (right) on logarithmic scales \deleted{as segregated by} before (green
    hashed) and after (purple solid) dust correction. In both cases,
    dust correction raises the observed SFR by a factor of $\sim
    1-1.5$ \added{dex}. 
    (Bottom) Histograms of the dust-corrected nuclear SFR (left) and SFR surface
    density (right) on logarithmic scales as segregated by ULIRGs
    (orange filled) and LIRGs (blue filled). It can be seen that the
    ULIRGs have SFRs that are comparable to the high end of what the
    LIRGs span, but may be indistinguishable from the LIRGs in terms
    of SFR surface densities. Bin sizes are 0.5 dex \deleted{in the log}.}
  \label{fig:sfr}
\end{figure*}

In Figure~\ref{fig:sfr} we show the histograms of our derived SFR and
$\Sigma_{\rm SFR}$ for the sample both before and after extinction
correction. In both cases, it is clear that there is a $\sim 1-1.5$
dex enhancement in the SFR and SFR surface density that was obscured
by the dust in the nuclear regions of our galaxies. \deleted{Allegedly,} This
\added{correction} corresponds to an extinction up to $\sim 2.5-4$ mag in the $K$ band, or
$\sim 25-40$ in the $V$ band, significantly higher than the $A_V \sim
3-17$ mag as quoted from~\cite{Piqueras13} \added{but consistent with
  the $A_V \sim 5-40$ mag found in other AO studies of LIRG
  nuclei~\cite[][]{Mattila07,Vaisanen17}}. If we take a closer look  
at the VLT/SINFONI results, the $A_V$ value of
individual spaxels do reach upward of 20$-$30 mag. In some cases,
the extinction at the very center of the $K$-band continuum peak is not
reported due to insufficient S/N in the Brackett line fluxes. The lack
of signal at the center, coupled with the integrated nature of the
statistics reported in~\cite{Piqueras13}, could explain the
discrepancy in extinction corrections between this work and their
VLT/SINFONI results.

For comparison, observed SFRs have been derived for both nuclear and
extranuclear star-forming clumps identified in high-resolution
\emph{HST}/WFC3 narrowband images of \paa~and \pab~emission of the
LIRGs in GOALS. The nuclear SFR ranges from 0.02 to 10 M$_\sun$
yr$^{-1}$, while the nuclear $\Sigma_{\rm SFR}$ spans $\sim 4-20$ M$_\sun$
yr$^{-1}$ kpc$^{-2}$ (Larson et al., in prep.). These measurements are
broadly consistent with the median values of our uncorrected SFR and
$\Sigma_{\rm  SFR}$ distributions. \added{Dust corrections computed from 
Paschen lines in the $H$ band and Brackett lines in the $K$ band
should be applied accordingly, as different extinctions would be
determined depending on the lines used and hence, on the depth probed
through the dust~\cite[e.g.][]{Vaisanen17}.} 

Furthermore, we break down the distribution of dust-corrected SFR and
$\Sigma_{\rm SFR}$ by 
infrared luminosity. The bottom panels of Figure~\ref{fig:sfr} show
the histograms of \added{the corrected} SFR and $\Sigma_{\rm SFR}$ for
LIRGs and ULIRGs, respectively. In terms of total SFRs, ULIRGs do, on
average, make more stars than their lower-luminosity counterparts in
the \added{resolved} nuclear regions. However, \added{we see here
  that some LIRGs match their ultraluminous counterparts in SFR. Also,
  the ULIRGs} appear
indistinguishable from the LIRGs in terms of SFR surface densities,
suggesting that they are similarly efficient at producing stars at
these scales \added{from near-infrared star formation indicators. It
  is plausible that the ULIRG nuclei are still optically thick and thus
  under-corrected at these wavelengths; future resolved far-infrared
  observations could confirm these findings. We note that these
observed SFR and SFRSD values within our S/N-defined nuclear regions are not
consistent in physical scales across our sample that spans
a range in redshifts, such that even SFRSD could be affected if a
radial gradient were present}.  The fact that the LIRGs
and ULIRGs exhibit similar distributions in $\Sigma_{\rm SFR}$ is consistent
with what was observed with VLT/SINFONI of the inner regions of
nearby southern (U)LIRGs~\cite[][]{Piqueras16}.

\begin{figure*}[htb]
  \centering
  \includegraphics[width=.7\textwidth,angle=90]{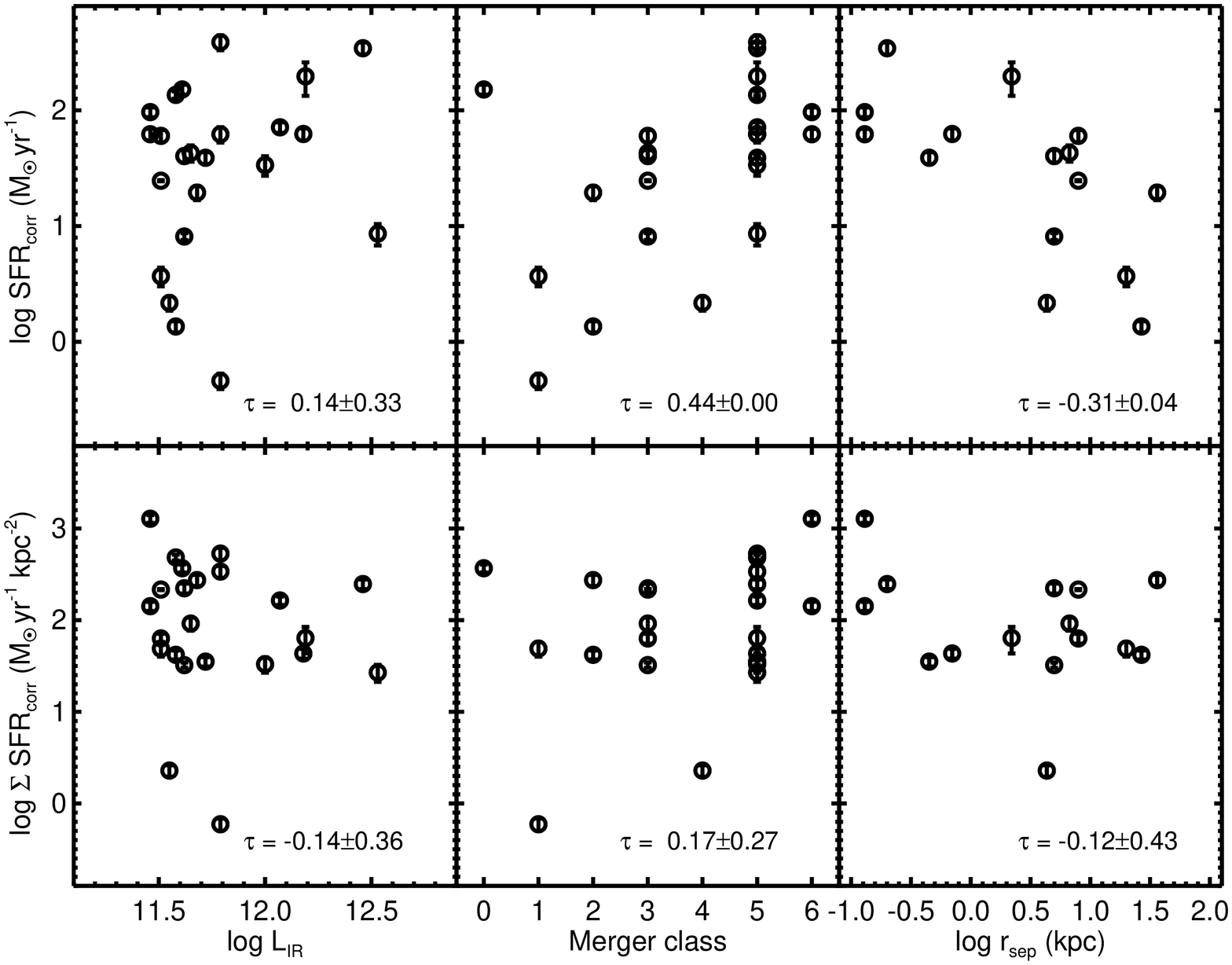}
  \caption{\deleted{(Left to right)} Dust-corrected nuclear SFR
    (\replaced{circle with plus}{top}) and nuclear SFR surface density
    (\replaced{grey filled}{bottom}) on logarithmic scales \replaced{as plotted
    against}{as a function of} galaxy-integrated global infrared luminosity $\log~L_{\rm
      IR}$ (left), merger class (middle), and nuclear separation $\log
    r_{\rm sep}$ (right). \added{%The open circles represent values
%      derived directly from our data at their respective,
%      original resolution. 
%      Dotted lines are linear fits to help guide the eye. 
      Kendall's correlation coefficient $\tau$,
      along with its significance, is given in each panel. %The filled grey circles show values 
%      computed from data that were rebinned to the lowest
%      physical resolution of the sample, 146.7 pc/spaxel. The
%      measurements are at comparable precision levels as those of
%      their finer resolution counterparts, but none of the observed trends with
%      the host property can be replicated with any significance.
}}
  \label{fig:sfrrelation}
\end{figure*}

We further explore the dust-corrected nuclear SFR and SFR surface
density relations with various properties of the (U)LIRG hosts in
Figure~\ref{fig:sfrrelation}. First, we examine how they may correlate
individually with the host's infrared luminosity. Globally, we expect
the infrared luminosity to correlate with SFR, particularly in
(U)LIRGs, \replaced{since}{because} the light from the formation of young stars is
reprocessed through dust in the cooler regime~\cite[c.f.][]{U12}. 
We \deleted{already} saw in Figure~\ref{fig:sfr} that ULIRGs 
occupy the high end of nuclear SFR; \replaced{and indeed a weak correlation may
be seen between SFR and $L_{\rm IR}$ in logarithmic scales}{however,
the relation betwen SFR and $\log L_{\rm IR}$ is difficult to quantify} (see
\added{top} leftmost panel in Figure~\ref{fig:sfrrelation}). The trend
is \replaced{less obvious, if existing at all,}{similarly insignificant} in the
case of SFR surface density \added{(bottom leftmost panel)}. 

Next, we consider the relation of nuclear SFR with merger class of the
(U)LIRG host system. If the most rapid star formation in (U)LIRGs is
triggered by merger activity, we would expect that the nuclear SFR to
be \deleted{the} most enhanced \replaced{in the stage of}{at}
coalescence. \deleted{We adopted the 
merger classification scheme from~\cite{Haan11}: (0) single galaxy
with no obvious major merging companion; (1) separate galaxies
with symmetric disks and no tidal tails; (2) distinguishable
progenitor galaxies with asymmetric disks and/or tidal tails; (3) two
distinct nuclei engulfed in a common envelop within the
merger body; (4) double nuclei with visible tidal tails; (5) single
or obscured nucleus with prominent tails; and (6) single or obscured
nucleus but with disturbed morphology and short faint tails signifying
post-merger remnant.} We have compiled \replaced{this}{the merger
classification for our sample} in our summary table (see
Table~\ref{tbl:summary}). We have also collated, in the same table,
the projected nuclear separation between nuclei, if observed and
resolved, from \added{the} literature, should it offer additional insights into
the accretion of gas onto the galaxy center throughout the merging process.

In the middle panels of Figure~\ref{fig:sfrrelation}, we have plotted
the dust-corrected nuclear SFR \added{(top)} and SFR surface density
\added{(bottom)} on a logarithmic scale as a function of merger class. 
We see that SFR appears to increase with merger class as the
interacting galaxy pair coalesces at merger stages $5-6$ \added{($\tau
  = 0.44\pm0.004$)}. This \added{increase} is
consistent with the picture of merger-induced starburst in the central
kiloparsec region as seen in simulations~\cite[e.g.][]{Moreno15} as
well as in observations~\cite[e.g.][]{Barrera15}.

\deleted{In a related note,} We consider the relation between the nuclear SFR
(and $\Sigma_{\rm SFR}$) with the projected nuclear separation between the nuclei, if seen
and resolved by observations, in the rightmost panels of
Figure~\ref{fig:sfrrelation}. An enhancement in the SFR \deleted{and SFR
surface density} as the projected separation between the nuclei
diminishes is \replaced{depicted}{evident} on this plot \added{($\tau =
-0.31\pm0.04$)}. The scatter in the relation may
\replaced{reduce}{decrease} if the separation was more accurately
measured 
from dynamical modeling of the merger system and did not suffer from
projection effects. We also caution that 
the projected nuclear separation is susceptible to instrument
resolution. For instance, NGC 7674W is listed to have a companion
$\sim$20 pc away~\cite[][]{Haan11}, but~\cite{Kharb17} recently
reported \deleted{incredibly} high-angular resolution observations from the
\emph{VLBA} that resolve the main western nucleus into two radio cores
0.35 pc apart. It is possible that this system has undergone more than
one major merger in the past, and it may be difficult to disentangle the
effects of the various merging events without detailed dynamical
modeling to assess timescales. 

 \subsection{Shocked gas among (U)LIRG hosts}
 From our \molhy/\brg~maps, we have identified six of 22
 galaxy nuclei that host shocked molecular gas.  Five of the
 six ULIRGs plus one LIRG in our sample host shocked molecular gas
 in coherent structures  (with \molhy/\brg~$>$ 2, see Figure \ref{fig:shocked}). 
The fact that shocks appear to be preferentially found in the more
infrared luminous systems may be \replaced{related to the dense gas content in
the nuclei. The denser the gas -- as is the case in the nuclei of ULIRGs -- the
more pronounced effect feedback may have on the ISM, the signature of
which may imprint upon its kinematic properties (to be presented in
a forth coming paper).}{because ULIRGs have more powerful
starbursts and more AGN to heat the dense gas in the nuclear region.}
Wide-field optical IFS studies have found that 
the galaxy-wide shock fraction increases with merger stage~\cite[][]{Rich15},
which also correlates with infrared luminosity. It is thus not
surprising to find more shocked gas among the ULIRG hosts than in
their less-luminous counterparts. However, \deleted{what} our current study
using a high angular resolution instrument such as OSIRIS has
\deleted{uniquely afforded is} the \added{unique} ability to peer
through the dust into the inner 
kiloparsec of the merging system. The shocked gas detected in this
work may represent the smoking gun signature of the more wide-spread
shocks seen in larger IFS surveys~\cite[\eg][]{Rich15,Ho16}, the
connection between which is muddled by drastic differences across
existing instrument resolutions and is yet to be thoroughly investigated. 

The \added{less luminous} exception, and arguably the most unequivocal
case of shocked molecular gas in this sample, was found in \added{the}
LIRG III Zw 035. As the only LIRG that shows \added{spatially coherent,
  definitively} shock-excited \molhy~gas, III Zw 035 has an
unremarkable infrared luminosity of $\log L_{\rm IR}/L_\odot = 11.62$
at a redshift of 0.0278. It is classified as \deleted{being in} merger stage 3,
with a nuclear separation of 5 kpc\deleted{in between the two distinct
nuclei}. The fact that it has a non-detection in the \emph{Swift}/BAT
hard X-ray bands~\cite[][]{Koss13} but exhibits Compton-Thick AGN
qualities in \emph{Spitzer}/IRS data~\cite[][]{Gonzalez15} means that its
center must be heavily obscured. It also has the most compact 33 GHz
continuum emission (with nuclear half light radius of only 30 pc)
among a sample of 22 local (U)LIRGs from the GOALS
sample~\cite[][]{Barcos-Munoz17}. \deleted{More follow-up observations of this
galaxy have been conducted with the Keck/NIRC2 camera, and will be
presented in depth in the future.} 

\subsection{Do AGN shock-heat the molecular gas?}
\label{disc:agn}
Further, we want to investigate whether the detected shocks may be a 
manifestation of AGN- or starburst-driven feedback. We assembled AGN
signatures from across the electromagnetic spectrum in
Table~\ref{tbl:summary}. 
 Among this sample, 12 of 21 systems have been observed by
 \emph{Chandra X-ray observatory} either
 as part of GOALS or by other programs. Four are classified as AGN
 hosts from X-ray observations~\cite[UGC 05101, UGC 08058, UGC 08696, and
 VV 340a;][]{Iwasawa11,Ricci17}. Four additional galaxies are identified as
 AGN hosts via the detection of the \nev~14.3 $\mu$m line
 in the mid-infrared \emph{Spitzer}-IRS spectra~\cite[UGC 08387, NGC
 2623, NGC 7469N, and NGC 7674W;][]{Petric11}.  Among these AGN hosts,
 UGC 05101, UGC 08696, and UGC 08058 have \deleted{been identified to
   feature} shocked molecular gas \replaced{from}{identified in} this
 work. The proximity of the shocked gas \replaced{from}{to}
 the central AGN may indicate AGN feedback as the source for the
 shock-heating \added{in those three cases}. 

On the other hand, VV 340a, UGC 8387, and NGC 7469N are three AGN
hosts that do not show 
any signs of shocks or feedback. VV 340a is at an early stage of
merging~\cite[stage 1;][]{Haan11} and hosts a Compton-Thick AGN as
detected by \emph{Chandra}~\cite[][]{Iwasawa11}. If the presence of
shocks in (U)LIRGs is plausibly induced by galaxy interaction, the
lack of a shock signature in VV 340a~\deleted{, if warranted by the depth of the
data,} could \replaced{place}{suggest} an upper limit on how early
shock-heating \replaced{could}{might} take 
place in merging progenitors. In the case of UGC 8387, the OSIRIS data
is incomplete in the coverage of the nucleus due to unfortunate
observing conditions. A hint of low-velocity, excited \molhy~\replaced{may be
seen}{is present} on the outskirts of the molecular gas disk, so more follow-up
observations at the center of the Seyfert nucleus may be
\replaced{worthy}{worthwhile}. As for NGC 7469, the \deleted{depicted}
northern nucleus \added{we show here} is 26 kpc away from the 
Seyfert 1 source in the system, which \replaced{is}{does} indeed
\deleted{observed to} feature biconical outflows \deleted{as} traced
by \deleted{the} coronal lines~\cite[][]{Muller-Sanchez11}. The
remaining two AGN hosts (NGC 2623 and NGC 7674W) \deleted{may} feature
high dispersion ($\gtrsim 150-200$ km s$^{-1}$) \molhy~gas that may be
heated by AGN photoionization; a detailed study of the kinematics will
\replaced{follow}{be presented in a future paper}. 

\begin{figure*}[htb]
  \centering
  \includegraphics[width=.6\textwidth,angle=90]{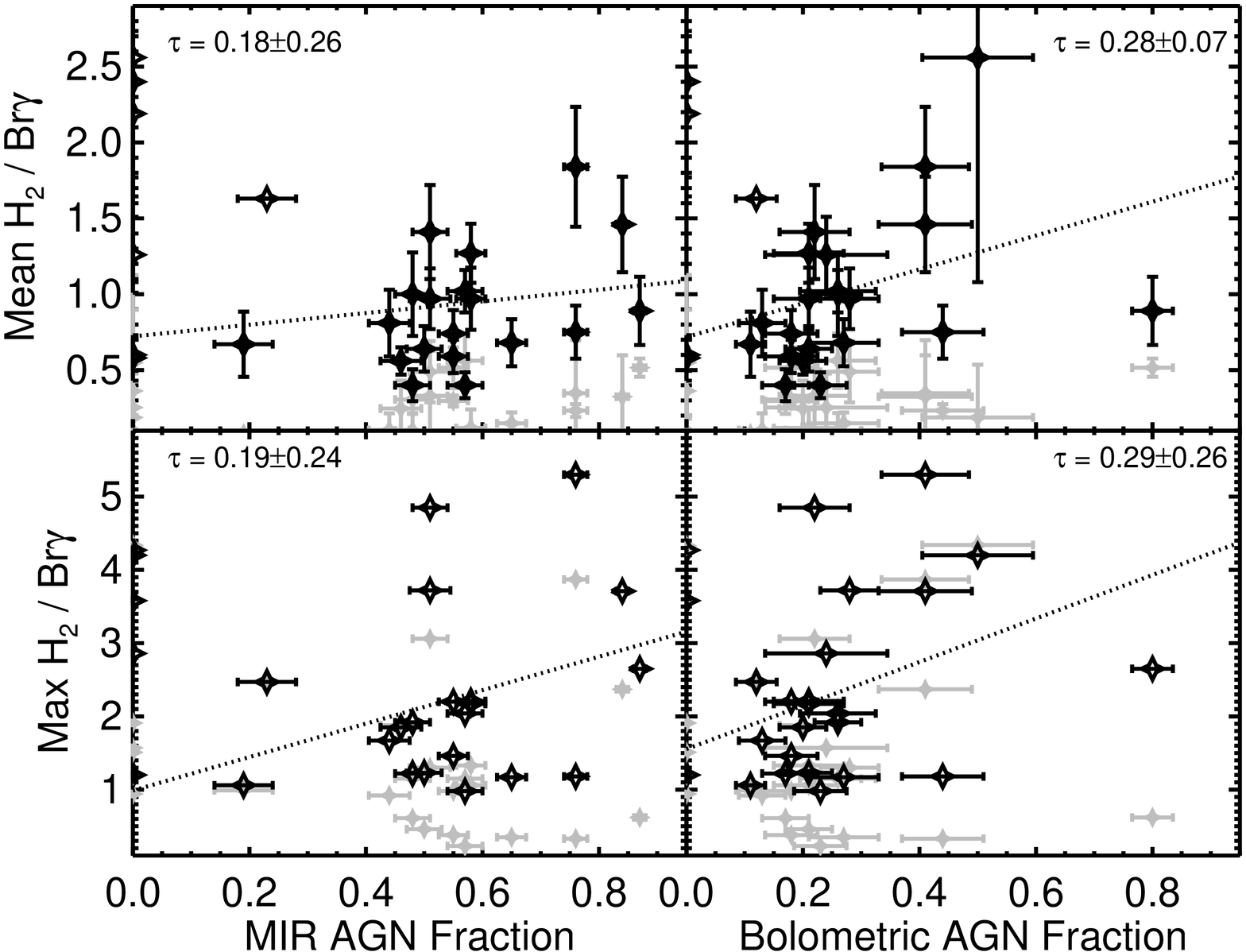}
  \caption{Mean (\replaced{open stars; left axis scale}{top}) and maximum
    \molhy/\brg~(\replaced{filled grey stars; right axis scale}{bottom}) plotted as a
    function of mid-infrared AGN fraction (left) 
    and of bolometric AGN fraction (right) adopted
    from~\cite{Diaz-Santos17}. \replaced{A weak correlation ($\rho =
    0.48\pm0.03$) can be seen between the mean \molhy/\brg~and
    mid-infrared AGN fraction; see text for details.}{Dotted lines are linear fits
      through the various sets of points excluding those at zero AGN
      fraction. Kendall's correlation coefficient $\tau$, along with
      its significance, is given in each panel. The filled grey stars
      represent the mean and maximum \molhy/\brg~values, respectively,
      as determined from data that were rebinned to the lowest physical
      resolution of the sample, and show an overall dilution as
      expected from lower-resolution data.}}
  \label{fig:agnfrac}
\end{figure*}

We explore the relation between nuclear shocks and AGN strength more
quantitatively in Figure~\ref{fig:agnfrac}. We adopted the average
mid-infrared and bolometric AGN fractions computed from various
\emph{Spitzer}/IRS and other diagnostics in~\cite{Diaz-Santos17}; see
references therein. \replaced{A weak correlation ($\rho = 0.48\pm0.03$,
excluding all the sources with zero AGN fraction) can be seen between
the mean \molhy/\brg~and mid-infrared AGN fraction}{While the signature
from shocked \molhy~is diluted in the mean \molhy/\brg~ratio
relative to the corresponding maximum value, weak correlations between
shocked gas and AGN strength measured in the mid-infrared and bolometrically
are seen (Kendall's correlation coefficient $\tau \sim 0.3$, though
generally insignificantly)}, suggesting that some of the observed 
shocks may be powered, at least in part, by AGN in
these nuclei. \replaced{But its scatter, and its non-existing correlation with the
bolometric AGN fraction ($\rho = 0.12\pm0.60$, excluding all the
sources with zero AGN fraction),}{The scatter in these weak
correlations} can be attributed to patchy dust obscuration in the
near-infrared wavelengths or the presence of other mechanisms at work
to drive shock excitation. 

Our finding that nearly half of the AGN hosts 
exhibit shocked \molhy~suggests that AGN may shock-heat the molecular gas,
but it is not ubiquitous and photoionization may also be
important. Further, we cannot rule out the possibility 
that other mechanisms such as cloud-cloud collisions or mechanical
perturbations of the ISM might also contribute to the shocks in these
AGN hosts.  \added{We also find shocked gas in four non-AGN hosts that
  must have been excited by other ionizing sources.} We will 
further discuss the impact of AGN-driven and starburst-driven winds
on the ISM in a future paper. 

\subsection{Comparison with OH Gas}
In Table~\ref{tbl:summary}, we record the detection of feedback
signatures in other wavelength regimes \replaced{as compiled from}{in the} literature. 
Molecular outflows in galaxy mergers have previously been identified
\added{in the longer far-infrared and millimeter wavelengths, for instance}
using the OH 119$\mu$m feature from observations taken with the
\emph{Herschel Space Observatory}~\cite[][]{Veilleux13}. Due to differences in the target
selections, there are only six ULIRGs in our overlapping sample. Of
these, two sources (UGC 08696 and UGC 08058) exhibit OH
outflows with median velocities of $\sim-$200 km s$^{-1}$. These
velocities agree very well with those of our molecular outflows \replaced{based
on}{seen in} the near-infrared \molhy~transitions~\cite[][]{U13,Medling15_ir17207}.
We do see outflowing gas in UGC 08696, though the signal-to-noise of
our data cubes render the case of UGC 08058 less conclusive. 

Of the remaining four sources in our overlapping sample with \emph{Herschel}
observations, three have detected inflows based on redshifted OH
absorption feature with median velocity $v_{50}$(abs) $\geq$ 50 km
s$^{-1}$: IRAS F22491$-$1808, IRAS F15250+3608, and IRAS F17207$-$0014. 
However, the spectral resolution of OSIRIS ($\Delta v \sim$
80 km s$^{-1}$) hinders our ability to detect \deleted{the} slow, inflowing
gas. As for IRAS F17207$-$0014, our in-depth analysis of the OSIRIS data set
along with large-scale optical IFS data has revealed molecular
gas dynamics that are more consistent with
outflows~\cite[][]{Medling15_ir17207}. The other discrepancy
between our analysis and that of~\cite{Veilleux13} rests with UGC
05101, where we see shocked, highly turbulent molecular gas well
in excess of 200 km s$^{-1}$ emanating from the
Seyfert nucleus, but only a $v_{50}$(abs) $= -9$ km s$^{-1}$ was
detected of the OH 119 $\mu$m feature. It is reasonable to explain
this difference in terms of physical scales and the multiphase nature
of outflows: it would be normal to have fast, warm outflows closer to the
nuclei, and slower, cool outflows spatially-averaged over kiloparsec
scales, which is where most of the outflowing mass would be. We note, however, that a
maximum velocity $v_{max}$(abs) of $-1200$ km s$^{-1}$ was reported in
the {\it Herschel} work, which is among the fastest outflow velocities in
that analysis. 

\subsection{Gas, Dust, and Star Formation in (U)LIRGs}
\begin{figure*}[htb]
  \centering
  \includegraphics[width=.52\textwidth,angle=90]{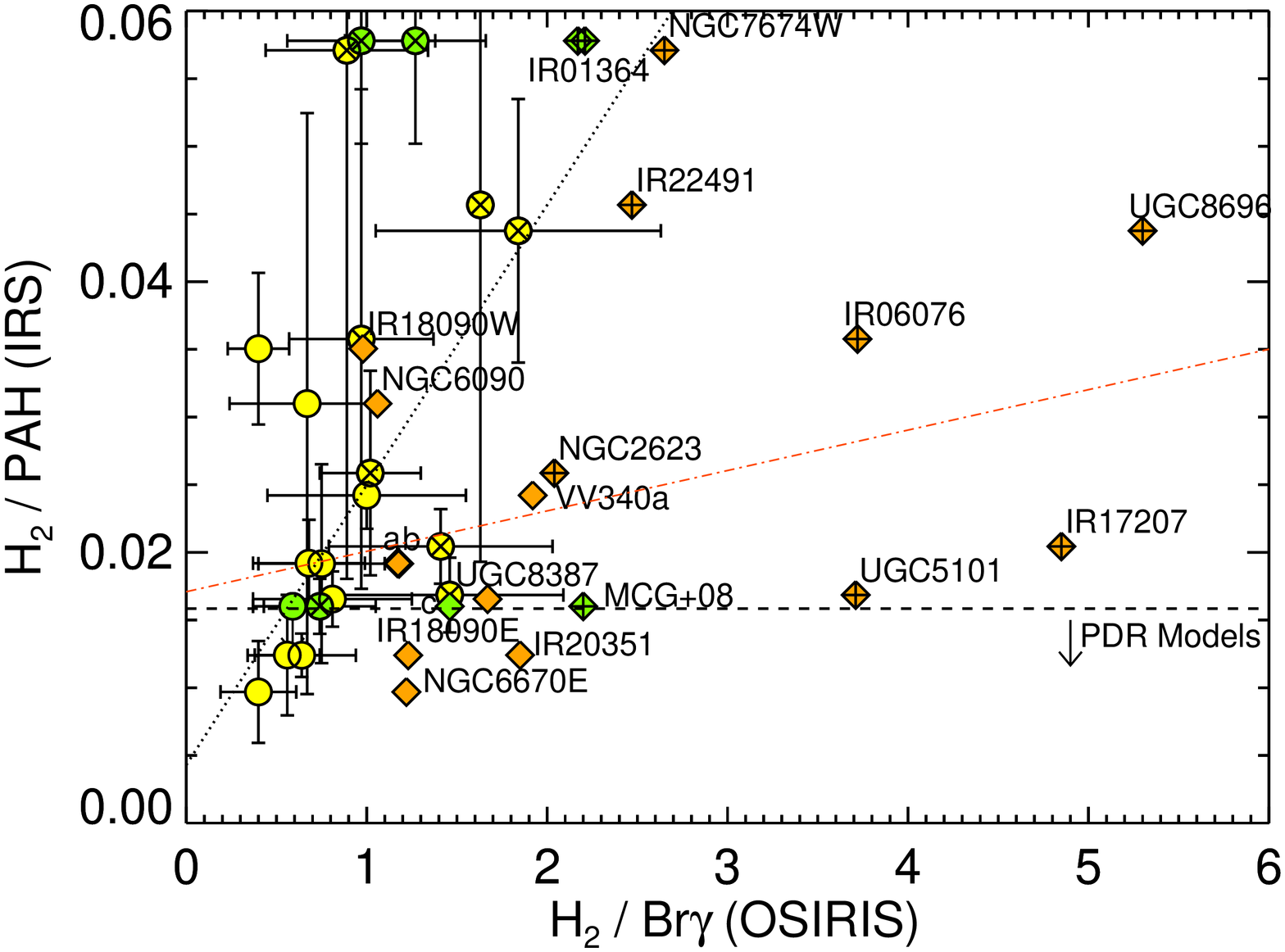}
  \caption{Comparison of \added{the} near-infrared diagnostic \molhy/\brg~from
    OSIRIS to mid-infrared line ratio \molhy/PAH from
    \emph{Spitzer}-IRS from~\cite{Stierwalt14}. \replaced{Nineteen}{Twenty}
    systems in the overlap sample that have the 7.7$\mu$m PAH feature
    detected are represented in this plot; \added{each galaxy is
      plotted twice to illustrate the difference between two
      derivation methods}. The filled circles indicate the 
    3-$\sigma$ clipped mean \molhy/\brg~values with 1-$\sigma$
    horizontal error bars; \replaced{upper limits are imposed by the
      sensitivity of IRS detection in}{the large vertical error bars on 4 of the
      data points are due to the limited detection for} one or more of
    the mid-IR \molhy~lines \added{in the IRS spectra}. In 
    contrast, the filled diamonds represent the maximum
    \molhy/\brg~values indicating the most highly-excited spaxels in
    the resolved nuclear regions for the same galaxies. The crossed
    symbols signify those objects identified \replaced{to be}{as} shock candidates
    from the 2D \molhy/\brg~morphology. The \replaced{linear
      regression}{nonparametric linear fit} for the
    maximum-value \replaced{points}{diamonds} (dot-dashed line) is 
    flatter with larger scatter than that for the mean value \replaced{points}{circles}
    (dotted line), indicating that while excited \molhy~can be seen in
    both the near- and mid-infrared diagnostics, high angular
    observations is necessary to pick out the shock-excited regions.
    The dashed line reflects the upper limit in \molhy/PAH \replaced{as set}{explained} by
    PDR models, above which another type of excitation, most likely
    shocks, will play a role.  The green points indicate the two instances where an
    object is plotted twice for their different plate scales: IRAS
    F01364$-$1042 and MCG +08$-$11$-$002. Sources a, \deleted{b,} and
    \replaced{c}{b} correspond to CGCG 436$-$030 (35mas) \added{and}
    NGC 7469N \deleted{, and MCG +08$-$11$-$002 (35mas)},
    respectively, \added{to clarify the particularly crowded points}.}
  \label{fig:fig_compIRS}
\end{figure*}

In order to get a sense of \added{how} the ionized gas and \deleted{PAH}
\added{dust content might correlate with the warm molecular gas and
  star formation} in these dusty 
galaxies, we also compare the near-infrared \deleted{shock} diagnostic
\molhy/\brg~with the mid-infrared line ratio \molhy/PAH \deleted{as}
extracted from \emph{Spitzer}-IRS~\cite[][]{Stierwalt14} in Figure
\ref{fig:fig_compIRS} for the \replaced{19}{18} galaxies in our overlapping
sample. (All of the galaxies in our sample have been observed with
\emph{Spitzer}-IRS, but only \replaced{19}{18} of them\added{, two
  with two separate observations at different plate scales,} have
detected the 7.7 $\mu$m PAH feature enabling the ensuing
analysis. \added{The mid-infrared numerator \molhy~incorporates the total line flux
  summed from the mid-infrared S0-S2 transitions. In cases where one
  of the \molhy~lines is a mere upper-limit detection, its error has been
  conservatively estimated at a level equal to its corresponding flux
  measurement and subsequently propagated into the error of the total
  \molhy~flux.}) The \molhy/\brg~ratios are 
the clipped mean and maximum values (excluding outliers) within the
central $\lesssim$ 400 pc region. The adopted \molhy/PAH values
incorporate the \molhy (S0-S2) transitions and the 7.7 $\mu$m PAH
feature in the larger IRS slit ($\lesssim$ 2 kpc).  

Overall, the sample \replaced{is seen to correlate}{shows a
  correlation} between the two diagnostic 
line ratios (as indicated by the dotted \deleted{regression} line in Figure
\ref{fig:fig_compIRS}, with correlation coefficient \replaced{$\rho =
  0.57\pm0.01$}{$\tau = 0.40$ and $p$-value of 0.0135});
\replaced{since the \molhy~is expected to be largely related 
across the two wavelength regimes as being excited by the same
processes}{a correlation is expected because the \molhy~is likely to
be excited by the same processes across the two wavelength
regimes}. \added{The nonparametric generalized Kendall's $\tau$ and 
  the Theil-Sen median-based linear model (implemented in the {\tt
    mblm} R statistical package) were used for the correlation
  measurement and linear fitting given their robust treatment of
  potential outliers.} The scatter about the
\replaced{regression}{linear fit} may be due to differences in patchy extinction in the
nuclear regions and in the mismatched physical
scales. \replaced{It}{Some scatter} is also likely due to
\brg~and PAH spatial variations, since the ionized gas and the dust
emission regions need not
correlate \added{on small scales,} as was seen
in~\cite{Diaz-Santos08,Diaz-Santos10a} using 
high-resolution (0\farcs4) mid-infrared Gemini/T-ReCS data. \deleted{The
near-infrared ro-vibrational transitions also correlate with higher
excitation temperatures, which implies that non-thermal excitation may
contribute to the warm \molhy~with a larger extent.} 

Also shown on the plot are the outlier-excluded maximum 
\molhy/\brg~values for each galaxy -- the variance in how much they
are offset with respect to their average counterpart is challenging
to predict based on integrated properties, but it \replaced{lends to the}{highlights}
importance of high angular resolution observations. The flatter
fit with a negligible slope (\replaced{$\rho = 0.34\pm0.15$}{$\tau = 0.21$ and
  $p$-value of 0.194}) between the \molhy/PAH values from
integrated light through the IRS slit and the maximally-enhanced
excited OSIRIS spaxels suggests that shocked regions may be small and
unresolved at 10\arcsec ($\sim$ 10 kpc) scales. Spatially
resolving and quantifying the cooler gas and dust 
will require mid-infrared imaging or integral field spectroscopy
observations using the upcoming \replaced{James Webb Space
  Telescope}{\emph{JWST}} \deleted{in our future work}. 

\replaced{The}{Our OSIRIS} shock candidates \deleted{based on OSIRIS
  analysis, \ie, hosting spatially coherent excited spaxels with
  \molhy/\brg~$>$ 2,} span a range in 
\molhy/PAH ratios, likely due to the fact that significant star formation, which could
drive shocks and feedback, contributes to the PAH emission. We note
that all the identified shock candidates \replaced{rest}{lie} above the upper 
limit \replaced{in}{of} \molhy/PAH \deleted{as} expected
from photodissociation region (PDR) models~\cite[see][for the adoption
of this value in these \molhy~and PAH transitions]{Stierwalt14}. 

  \section{Summary}
  \label{summary}

We present high resolution (0$\farcs$035$-$0$\farcs$100/spaxel,
\added{$\sim$ 40pc/spaxel})
near-infrared AO-assisted integral-field observations taken with
Keck/OSIRIS for a sample of 21 nearby ($z < 0.05$) (U)LIRG systems (22
nuclei) from the \replaced{GOALS-KOALA}{KOALA-GOALS} Survey. In
particular, we focus on 
examining the nuclear SFR and identifying shocks using
\molhy/\brg~as a tracer. We summarize our findings as follows:

\begin{itemize}

\item Using \molhy~line ratio diagnostics, we find that the molecular
  gas in our sample of (U)LIRGs is more likely excited by thermal emission
  (from X-ray irradiation or shocks, or a combination of both). 
 Systems where we identified shocked gas based on excess
 \molhy~relative to \brg~gas \deleted{do indeed} reside on the diagnostic
 diagram nearest to that occupied by shock predictions.

\item %\added{High extinction levels (up to $A_V \sim 40$ mag) were
%    measured, leading to a factor of $10-30$ enhancement in the
 %   corrected SFR in these nuclear regions.} 
We compute dust-corrected
  nuclear SFR and SFR surface density 
  based on calibrated \brg~fluxes, and find that the nuclear SFR
  correlates \deleted{, albeit weakly and not without scatter, with
  global infrared luminosity,} with  
  merger class and diminishing nuclear separation. These trends are
  largely consistent with the picture of merger-induced starbursts
  within the central kiloparsec region of galaxy mergers.

\item Six of our sources (5 ULIRGs and 1 LIRG) feature shocked
  molecular gas as identified by near-infrared diagnostic
  \molhy/\brg~$> 2$. Considering the infrared 
  luminosity of these shock hosts, it appears that shocks are
  preferentially found in the ultraluminous systems,
  but may also be triggered at an earlier merging stage. 

\item Given that nearly half of the AGN hosts exhibit shocked \molhy,
  \replaced{AGN may shock-heat the molecular gas in some systems and drive
  photoionization in others}{it is clear that on circumnuclear scales,
  AGN have a strong effect on heating the surrounding molecular
  gas}. However, the weak correlation between nuclear 
  shocks and AGN strength~\replaced{may be due to patchy dust obscuration,
  mistmatched physical scales probed, or
  other contributing factors to shock excitation}{indicates that it is
  not simply the relative strength of the AGN that drives the excess
  in warm \molhy. The coupling of the AGN power to the dense molecular
  gas is likely to be complicated and depends on orientation, dust
  shielding, density, and other factors}. Shocks may also be triggered by
  cloud-cloud collisions or mechanical perturbations of the ISM.  

\item The near-infrared and mid-infrared diagnostics largely agree,
  since the warm and cooler \molhy~is expected to be excited by the
  same processes. Scatter about the relation may be due to
  differences in patchy extinction within the nuclei, and also likely
  due to \brg~and PAH since the ionized gas and the dust need not
  correlate \added{at scales comparable to H\textsc{II} regions}. Detailed
    understanding of molecular gas physics will 
  require spatially resolving and quantifying the cooler gas and dust
  components as enabled by our upcoming \replaced{James Webb Space
    Telescope}{\emph{JWST}} Early Release Science observations (PI:
  Armus, ID 1328) in the future.

\end{itemize}

\acknowledgements

\added{We thank the anonymous referee  and the statisical editor for
  thoughtful suggestions that significantly improved our manuscript.}
We appreciate informative discussions with B. Groves and P. Creasey 
regarding photoionization models, and wish to acknowledge helpful
communication with D. Calzetti regarding dust extinction curves.
We thank all the Keck staff for help with carrying out the
observations. The data presented herein were obtained at the W.M. Keck Observatory,
which is operated as a scientific partnership among the California
Institute of Technology, the University of California and the National
Aeronautics and Space Administration. The Observatory was made
possible by the generous financial support of the W. M. Keck
Foundation. The authors wish to recognize and acknowledge the very
significant cultural role and reverence that the summit of Mauna Kea
has always had within the indigenous Hawaiian community. We are most
fortunate to have the opportunity to conduct observations from this
mountain. We also acknowledge the Evans Remote Observing Room at UC
Irvine for a number of the remote observing sessions carried out there.

V.U acknowledges funding support from the University of
California Chancellor's Postdoctoral Fellowship, JPL Contract/IRAC GTO
Grant No. 1256790, and NSF grant AST-1412693.  
\replaced{A.M. acknowledges funding support from the Hubble Fellowship.}{Support for A.M.M. is provided by NASA through Hubble Fellowship grant
\#HST-HF2-51377 awarded by the Space Telescope Science Institute,
which is operated by the Association of Universities for Research in
Astronomy, Inc., for NASA, under contract NAS5-26555.} 
T.D-S. acknowledges support from ALMA-CONICYT project 31130005 and
FONDECYT project 1151239. 
G.C.P. acknowledges support from the University of Florida. 
This work was conducted in part at the Aspen
Center for Physics, which is supported by NSF
grant PHY-1607611; we thank the Center for its
hospitality during the Astrophysics of Massive Black Holes Merger
workshop in June and July 2018.
This research has made use of the NASA/IPAC Extragalactic Database
 (NED) which is operated by the Jet Propulsion Laboratory, California
 Institute of Technology, under contract with the National Aeronautics
 and Space Administration.

\facility{Keck:I/II (OSIRIS, AO)}

\bibliographystyle{aasjournal}
\bibliography{ulirgs}

\begin{thebibliography}{}
\expandafter\ifx\csname natexlab\endcsname\relax\def\natexlab#1{#1}\fi

\bibitem[{{Aladro} {et~al.}(2018){Aladro}, {K{\"o}nig}, {Aalto},
  {Gonz{\'a}lez-Alfonso}, {Falstad}, {Mart{\'{\i}}n}, {Muller},
  {Garc{\'{\i}}a-Burillo}, {Henkel}, {van der Werf}, {Mills}, {Fischer},
  {Costagliola}, {Krips}, \& {.}}]{Aladro18}
{Aladro}, R., {K{\"o}nig}, S., {Aalto}, S., {et~al.} 2018, ArXiv e-prints,
  arXiv:1805.11582

\bibitem[{{Alatalo}(2015)}]{Alatalo15}
{Alatalo}, K. 2015, \apjl, 801, L17

\bibitem[{{Allen} {et~al.}(2008){Allen}, {Groves}, {Dopita}, {Sutherland}, \&
  {Kewley}}]{Allen08}
{Allen}, M.~G., {Groves}, B.~A., {Dopita}, M.~A., {Sutherland}, R.~S., \&
  {Kewley}, L.~J. 2008, ApJS, 178, 20

\bibitem[{{Alonso-Herrero} {et~al.}(2012){Alonso-Herrero}, {Pereira-Santaella},
  {Rieke}, \& {Rigopoulou}}]{Alonso12}
{Alonso-Herrero}, A., {Pereira-Santaella}, M., {Rieke}, G.~H., \& {Rigopoulou},
  D. 2012, \apj, 744, 2

\bibitem[{{Armus} {et~al.}(1987){Armus}, {Heckman}, \& {Miley}}]{Armus87}
{Armus}, L., {Heckman}, T., \& {Miley}, G. 1987, \aj, 94, 831

\bibitem[{{Armus} {et~al.}(2004){Armus}, {Charmandaris}, {Spoon}, {Houck},
  {Soifer}, {Brandl}, {Appleton}, {Teplitz}, {Higdon}, {Weedman}, {Devost},
  {Morris}, {Uchida}, {van Cleve}, {Barry}, {Sloan}, {Grillmair}, {Burgdorf},
  {Fajardo-Acosta}, {Ingalls}, {Higdon}, {Hao}, {Bernard-Salas}, {Herter},
  {Troeltzsch}, {Unruh}, \& {Winghart}}]{Armus04}
{Armus}, L., {Charmandaris}, V., {Spoon}, H.~W.~W., {et~al.} 2004, \apjs, 154,
  178

\bibitem[{{Armus} {et~al.}(2007){Armus}, {Charmandaris}, {Bernard-Salas},
  {Spoon}, {Marshall}, {Higdon}, {Desai}, {Teplitz}, {Hao}, {Devost}, {Brandl},
  {Wu}, {Sloan}, {Soifer}, {Houck}, \& {Herter}}]{Armus07}
{Armus}, L., {Charmandaris}, V., {Bernard-Salas}, J., {et~al.} 2007, \apj, 656,
  148

\bibitem[{{Armus} {et~al.}(2009){Armus}, {Mazzarella}, {Evans}, {Surace},
  {Sanders}, {Iwasawa}, {Frayer}, {Howell}, {Chan}, {Petric}, {Vavilkin},
  {Kim}, {Haan}, {Inami}, {Murphy}, {Appleton}, {Barnes}, {Bothun}, {Bridge},
  {Charmandaris}, {Jensen}, {Kewley}, {Lord}, {Madore}, {Marshall},
  {Melbourne}, {Rich}, {Satyapal}, {Schulz}, {Spoon}, {Sturm}, {U}, {Veilleux},
  \& {Xu}}]{Armus09}
{Armus}, L., {Mazzarella}, J.~M., {Evans}, A.~S., {et~al.} 2009, \pasp, 121,
  559

\bibitem[{{Arribas} {et~al.}(2008){Arribas}, {Colina}, {Monreal-Ibero},
  {Alfonso}, {Garc{\'{\i}}a-Mar{\'{\i}}n}, \& {Alonso-Herrero}}]{Arribas08}
{Arribas}, S., {Colina}, L., {Monreal-Ibero}, A., {et~al.} 2008, \aap, 479, 687

\bibitem[{{Baldwin} {et~al.}(1981){Baldwin}, {Phillips}, \&
  {Terlevich}}]{Baldwin81}
{Baldwin}, J.~A., {Phillips}, M.~M., \& {Terlevich}, R. 1981, PASP, 93, 5

\bibitem[{{Barcos-Mu{\~n}oz} {et~al.}(2017){Barcos-Mu{\~n}oz}, {Leroy},
  {Evans}, {Condon}, {Privon}, {Thompson}, {Armus}, {D{\'{\i}}az-Santos},
  {Mazzarella}, {Meier}, {Momjian}, {Murphy}, {Ott}, {Sanders}, {Schinnerer},
  {Stierwalt}, {Surace}, \& {Walter}}]{Barcos-Munoz17}
{Barcos-Mu{\~n}oz}, L., {Leroy}, A.~K., {Evans}, A.~S., {et~al.} 2017, \apj,
  843, 117

\bibitem[{{Barrera-Ballesteros} {et~al.}(2015){Barrera-Ballesteros},
  {S{\'a}nchez}, {Garc{\'{\i}}a-Lorenzo}, {Falc{\'o}n-Barroso}, {Mast},
  {Garc{\'{\i}}a-Benito}, {Husemann}, {van de Ven}, {Iglesias-P{\'a}ramo},
  {Rosales-Ortega}, {P{\'e}rez-Torres}, {M{\'a}rquez}, {Kehrig}, {Marino},
  {Vilchez}, {Galbany}, {L{\'o}pez-S{\'a}nchez}, {Walcher}, \& {Califa
  Collaboration}}]{Barrera15}
{Barrera-Ballesteros}, J.~K., {S{\'a}nchez}, S.~F., {Garc{\'{\i}}a-Lorenzo},
  B., {et~al.} 2015, \aap, 579, A45

\bibitem[{{Bedregal} {et~al.}(2009){Bedregal}, {Colina}, {Alonso-Herrero}, \&
  {Arribas}}]{Bedregal09}
{Bedregal}, A.~G., {Colina}, L., {Alonso-Herrero}, A., \& {Arribas}, S. 2009,
  \apj, 698, 1852

\bibitem[{{Black} \& {van Dishoeck}(1987)}]{Black87}
{Black}, J.~H., \& {van Dishoeck}, E.~F. 1987, \apj, 322, 412

\bibitem[{{Brand} {et~al.}(1989){Brand}, {Toner}, {Geballe}, {Webster},
  {Williams}, \& {Burton}}]{Brand89}
{Brand}, P.~W.~J.~L., {Toner}, M.~P., {Geballe}, T.~R., {et~al.} 1989, \mnras,
  236, 929

\bibitem[{{Burton}(1987)}]{Burton87}
{Burton}, M.~G. 1987, PhD thesis, University of Edinburgh

\bibitem[{{Busch} {et~al.}(2017){Busch}, {Eckart}, {Valencia-S.}, {Fazeli},
  {Scharw{\"a}chter}, {Combes}, \& {Garc{\'{\i}}a-Burillo}}]{Busch17}
{Busch}, G., {Eckart}, A., {Valencia-S.}, M., {et~al.} 2017, A\&A, 598, A55

\bibitem[{{Calzetti}(2001)}]{Calzetti01}
{Calzetti}, D. 2001, \pasp, 113, 1449

\bibitem[{{Calzetti} {et~al.}(1994){Calzetti}, {Kinney}, \&
  {Storchi-Bergmann}}]{Calzetti94}
{Calzetti}, D., {Kinney}, A.~L., \& {Storchi-Bergmann}, T. 1994, \apj, 429, 582

\bibitem[{{Cappellari} \& {Copin}(2003)}]{Cappellari03}
{Cappellari}, M., \& {Copin}, Y. 2003, \mnras, 342, 345

\bibitem[{{Cardelli} {et~al.}(1989){Cardelli}, {Clayton}, \&
  {Mathis}}]{Cardelli89}
{Cardelli}, J.~A., {Clayton}, G.~C., \& {Mathis}, J.~S. 1989, \apj, 345, 245

\bibitem[{{Chisholm} {et~al.}(2016){Chisholm}, {Tremonti Christy}, {Leitherer},
  \& {Chen}}]{Chisholm16}
{Chisholm}, J., {Tremonti Christy}, A., {Leitherer}, C., \& {Chen}, Y. 2016,
  \mnras, 463, 541

\bibitem[{{Cicone} {et~al.}(2014){Cicone}, {Maiolino}, {Sturm},
  {Graci{\'a}-Carpio}, {Feruglio}, {Neri}, {Aalto}, {Davies}, {Fiore},
  {Fischer}, {Garc{\'{\i}}a-Burillo}, {Gonz{\'a}lez-Alfonso},
  {Hailey-Dunsheath}, {Piconcelli}, \& {Veilleux}}]{Cicone14}
{Cicone}, C., {Maiolino}, R., {Sturm}, E., {et~al.} 2014, \aap, 562, A21

\bibitem[{{Colina} {et~al.}(2015){Colina}, {Piqueras L{\'o}pez}, {Arribas},
  {Riffel}, {Riffel}, {Rodriguez-Ardila}, {Pastoriza}, {Storchi-Bergmann},
  {Alonso-Herrero}, \& {Sales}}]{Colina15}
{Colina}, L., {Piqueras L{\'o}pez}, J., {Arribas}, S., {et~al.} 2015, \aap,
  578, A48

\bibitem[{{Cortijo-Ferrero} {et~al.}(2017){Cortijo-Ferrero}, {Gonz{\'a}lez
  Delgado}, {P{\'e}rez}, {Cid Fernandes}, {S{\'a}nchez}, {de Amorim}, {Di
  Matteo}, {Garc{\'{\i}}a-Benito}, {Lacerda}, {L{\'o}pez Fern{\'a}ndez}, \&
  {Tadhunter}}]{Cortijo17}
{Cortijo-Ferrero}, C., {Gonz{\'a}lez Delgado}, R.~M., {P{\'e}rez}, E., {et~al.}
  2017, \mnras, 467, 3898

\bibitem[{{Dale} {et~al.}(2009){Dale}, {Smith}, {Schlawin}, {Armus},
  {Buckalew}, {Cohen}, {Helou}, {Jarrett}, {Johnson}, {Moustakas}, {Murphy},
  {Roussel}, {Sheth}, {Staudaher}, {Bot}, {Calzetti}, {Engelbracht}, {Gordon},
  {Hollenbach}, {Kennicutt}, \& {Malhotra}}]{Dale09}
{Dale}, D.~A., {Smith}, J.~D.~T., {Schlawin}, E.~A., {et~al.} 2009, \apj, 693,
  1821

\bibitem[{{Dasyra} \& {Combes}(2011)}]{Dasyra11b}
{Dasyra}, K.~M., \& {Combes}, F. 2011, \aap, 533, L10

\bibitem[{{Dasyra} {et~al.}(2011){Dasyra}, {Ho}, {Netzer}, {Combes},
  {Trakhtenbrot}, {Sturm}, {Armus}, \& {Elbaz}}]{Dasyra11a}
{Dasyra}, K.~M., {Ho}, L.~C., {Netzer}, H., {et~al.} 2011, \apj, 740, 94

\bibitem[{{Davies} {et~al.}(2016){Davies}, {Medling}, {U}, {Max}, {Sanders}, \&
  {Kewley}}]{Davies16}
{Davies}, R.~L., {Medling}, A.~M., {U}, V., {et~al.} 2016, \mnras, 458, 158

\bibitem[{{D{\'{\i}}az-Santos} {et~al.}(2010){D{\'{\i}}az-Santos},
  {Alonso-Herrero}, {Colina}, {Packham}, {Levenson}, {Pereira-Santaella},
  {Roche}, \& {Telesco}}]{Diaz-Santos10a}
{D{\'{\i}}az-Santos}, T., {Alonso-Herrero}, A., {Colina}, L., {et~al.} 2010,
  \apj, 711, 328

\bibitem[{{D{\'{\i}}az-Santos} {et~al.}(2008){D{\'{\i}}az-Santos},
  {Alonso-Herrero}, {Colina}, {Packham}, {Radomski}, \&
  {Telesco}}]{Diaz-Santos08}
---. 2008, \apj, 685, 211

\bibitem[{{D{\'{\i}}az-Santos} {et~al.}(2007){D{\'{\i}}az-Santos},
  {Alonso-Herrero}, {Colina}, {Ryder}, \& {Knapen}}]{Diaz-Santos07}
{D{\'{\i}}az-Santos}, T., {Alonso-Herrero}, A., {Colina}, L., {Ryder}, S.~D.,
  \& {Knapen}, J.~H. 2007, ApJ, 661, 149

\bibitem[{{D{\'{\i}}az-Santos} {et~al.}(2017){D{\'{\i}}az-Santos}, {Armus},
  {Charmandaris}, {Lu}, {Stierwalt}, {Stacey}, {Malhotra}, {van der Werf},
  {Howell}, {Privon}, {Mazzarella}, {Goldsmith}, {Murphy}, {Barcos-Mu{\~n}oz},
  {Linden}, {Inami}, {Larson}, {Evans}, {Appleton}, {Iwasawa}, {Lord},
  {Sanders}, \& {Surace}}]{Diaz-Santos17}
{D{\'{\i}}az-Santos}, T., {Armus}, L., {Charmandaris}, V., {et~al.} 2017, \apj,
  846, 32

\bibitem[{{Dom{\'{\i}}nguez} {et~al.}(2013){Dom{\'{\i}}nguez}, {Siana},
  {Henry}, {Scarlata}, {Bedregal}, {Malkan}, {Atek}, {Ross}, {Colbert},
  {Teplitz}, {Rafelski}, {McCarthy}, {Bunker}, {Hathi}, {Dressler}, {Martin},
  \& {Masters}}]{Dominguez13}
{Dom{\'{\i}}nguez}, A., {Siana}, B., {Henry}, A.~L., {et~al.} 2013, \apj, 763,
  145

\bibitem[{{Draine} \& {Bertoldi}(1996)}]{Draine96}
{Draine}, B.~T., \& {Bertoldi}, F. 1996, \apj, 468, 269

\bibitem[{{Draine} \& {Woods}(1990)}]{Draine90}
{Draine}, B.~T., \& {Woods}, D.~T. 1990, \apj, 363, 464

\bibitem[{{Emonts} {et~al.}(2017){Emonts}, {Colina}, {Piqueras-L{\'o}pez},
  {Garcia-Burillo}, {Pereira-Santaella}, {Arribas}, {Labiano}, \&
  {Alonso-Herrero}}]{Emonts17}
{Emonts}, B.~H.~C., {Colina}, L., {Piqueras-L{\'o}pez}, J., {et~al.} 2017,
  \aap, 607, A116

\bibitem[{{Emonts} {et~al.}(2014){Emonts}, {Piqueras-L{\'o}pez}, {Colina},
  {Arribas}, {Villar-Mart{\'{\i}}n}, {Pereira-Santaella}, {Garcia-Burillo}, \&
  {Alonso-Herrero}}]{Emonts14}
{Emonts}, B.~H.~C., {Piqueras-L{\'o}pez}, J., {Colina}, L., {et~al.} 2014,
  \aap, 572, A40

\bibitem[{{Evans} {et~al.}(2008){Evans}, {Vavilkin}, {Pizagno}, {Modica},
  {Mazzarella}, {Iwasawa}, {Howell}, {Surace}, {Armus}, {Petric}, {Spoon},
  {Barnes}, {Suer}, {Sanders}, {Chan}, \& {Lord}}]{Evans08}
{Evans}, A.~S., {Vavilkin}, T., {Pizagno}, J., {et~al.} 2008, \apjl, 675, L69

\bibitem[{{Farage} {et~al.}(2010){Farage}, {McGregor}, {Dopita}, \&
  {Bicknell}}]{Farage10}
{Farage}, C.~L., {McGregor}, P.~J., {Dopita}, M.~A., \& {Bicknell}, G.~V. 2010,
  \apj, 724, 267

\bibitem[{{Garc{\'{\i}}a-Burillo} {et~al.}(2015){Garc{\'{\i}}a-Burillo},
  {Combes}, {Usero}, {Aalto}, {Colina}, {Alonso-Herrero}, {Hunt}, {Arribas},
  {Costagliola}, {Labiano}, {Neri}, {Pereira-Santaella}, {Tacconi}, \& {van der
  Werf}}]{GarciaBurillo15}
{Garc{\'{\i}}a-Burillo}, S., {Combes}, F., {Usero}, A., {et~al.} 2015, \aap,
  580, A35

\bibitem[{{Genzel} {et~al.}(1998){Genzel}, {Lutz}, {Sturm}, {Egami}, {Kunze},
  {Moorwood}, {Rigopoulou}, {Spoon}, {Sternberg}, {Tacconi-Garman}, {Tacconi},
  \& {Thatte}}]{Genzel98}
{Genzel}, R., {Lutz}, D., {Sturm}, E., {et~al.} 1998, \apj, 498, 579

\bibitem[{{Gonz{\'a}lez-Alfonso} {et~al.}(2017){Gonz{\'a}lez-Alfonso},
  {Fischer}, {Spoon}, {Stewart}, {Ashby}, {Veilleux}, {Smith}, {Sturm},
  {Farrah}, {Falstad}, {Mel{\'e}ndez}, {Graci{\'a}-Carpio}, {Janssen}, \&
  {Lebouteiller}}]{Gonzalez17}
{Gonz{\'a}lez-Alfonso}, E., {Fischer}, J., {Spoon}, H.~W.~W., {et~al.} 2017,
  \apj, 836, 11

\bibitem[{{Gonz{\'a}lez-Mart{\'{\i}}n}
  {et~al.}(2009){Gonz{\'a}lez-Mart{\'{\i}}n}, {Masegosa}, {M{\'a}rquez}, \&
  {Guainazzi}}]{Gonzalez09}
{Gonz{\'a}lez-Mart{\'{\i}}n}, O., {Masegosa}, J., {M{\'a}rquez}, I., \&
  {Guainazzi}, M. 2009, \apj, 704, 1570

\bibitem[{{Gonz{\'a}lez-Mart{\'{\i}}n}
  {et~al.}(2015){Gonz{\'a}lez-Mart{\'{\i}}n}, {Masegosa}, {M{\'a}rquez},
  {Rodr{\'{\i}}guez-Espinosa}, {Acosta-Pulido}, {Ramos Almeida}, {Dultzin},
  {Hern{\'a}ndez-Garc{\'{\i}}a}, {Ruschel-Dutra}, \&
  {Alonso-Herrero}}]{Gonzalez15}
{Gonz{\'a}lez-Mart{\'{\i}}n}, O., {Masegosa}, J., {M{\'a}rquez}, I., {et~al.}
  2015, \aap, 578, A74

\bibitem[{{Groves} {et~al.}(2008){Groves}, {Dopita}, {Sutherland}, {Kewley},
  {Fischera}, {Leitherer}, {Brandl}, \& {van Breugel}}]{Groves08}
{Groves}, B., {Dopita}, M.~A., {Sutherland}, R.~S., {et~al.} 2008, \apjs, 176,
  438

\bibitem[{{Haan} {et~al.}(2011){Haan}, {Surace}, {Armus}, {Evans}, {Howell},
  {Mazzarella}, {Kim}, {Vavilkin}, {Inami}, {Sanders}, {Petric}, {Bridge},
  {Melbourne}, {Charmandaris}, {Diaz-Santos}, {Murphy}, {U}, {Stierwalt}, \&
  {Marshall}}]{Haan11}
{Haan}, S., {Surace}, J.~A., {Armus}, L., {et~al.} 2011, \aj, 141, 100

\bibitem[{{Heckman} {et~al.}(1990){Heckman}, {Armus}, \& {Miley}}]{Heckman90}
{Heckman}, T.~M., {Armus}, L., \& {Miley}, G.~K. 1990, The Astrophysical
  Journal Supplement Series, 74, 833

\bibitem[{{Hinshaw} {et~al.}(2009){Hinshaw}, {Weiland}, {Hill}, {Odegard},
  {Larson}, {Bennett}, {Dunkley}, {Gold}, {Greason}, {Jarosik}, {Komatsu},
  {Nolta}, {Page}, {Spergel}, {Wollack}, {Halpern}, {Kogut}, {Limon}, {Meyer},
  {Tucker}, \& {Wright}}]{Hinshaw09}
{Hinshaw}, G., {Weiland}, J.~L., {Hill}, R.~S., {et~al.} 2009, \apjs, 180, 225

\bibitem[{{Ho} {et~al.}(2016){Ho}, {Medling}, {Bland-Hawthorn}, {Groves},
  {Kewley}, {Kobayashi}, {Dopita}, {Leslie}, {Sharp}, {Allen}, {Bourne},
  {Bryant}, {Cortese}, {Croom}, {Dunne}, {Fogarty}, {Goodwin}, {Green},
  {Konstantopoulos}, {Lawrence}, {Lorente}, {Owers}, {Richards}, {Sweet},
  {Tescari}, \& {Valiante}}]{Ho16}
{Ho}, I.-T., {Medling}, A.~M., {Bland-Hawthorn}, J., {et~al.} 2016, \mnras,
  457, 1257

\bibitem[{{Hopkins} {et~al.}(2005){Hopkins}, {Hernquist}, {Cox}, {Di Matteo},
  {Martini}, {Robertson}, \& {Springel}}]{Hopkins05}
{Hopkins}, P.~F., {Hernquist}, L., {Cox}, T.~J., {et~al.} 2005, \apj, 630, 705

\bibitem[{{Imanishi} {et~al.}(2016){Imanishi}, {Nakanishi}, \&
  {Izumi}}]{Imanishi16}
{Imanishi}, M., {Nakanishi}, K., \& {Izumi}, T. 2016, \aj, 152, 218

\bibitem[{{Imanishi} {et~al.}(2003){Imanishi}, {Terashima}, {Anabuki}, \&
  {Nakagawa}}]{Imanishi03}
{Imanishi}, M., {Terashima}, Y., {Anabuki}, N., \& {Nakagawa}, T. 2003, \apjl,
  596, L167

\bibitem[{{Inami} {et~al.}(2013){Inami}, {Armus}, {Charmandaris}, {Groves},
  {Kewley}, {Petric}, {Stierwalt}, {D{\'{\i}}az-Santos}, {Surace}, {Rich},
  {Haan}, {Howell}, {Evans}, {Mazzarella}, {Marshall}, {Appleton}, {Lord},
  {Spoon}, {Frayer}, {Matsuhara}, \& {Veilleux}}]{Inami13}
{Inami}, H., {Armus}, L., {Charmandaris}, V., {et~al.} 2013, \apj, 777, 156

\bibitem[{{Iwasawa} {et~al.}(2011){Iwasawa}, {Sanders}, {Teng}, {U}, {Armus},
  {Evans}, {Howell}, {Komossa}, {Mazzarella}, {Petric}, {Surace}, {Vavilkin},
  {Veilleux}, \& {Trentham}}]{Iwasawa11}
{Iwasawa}, K., {Sanders}, D.~B., {Teng}, S.~H., {et~al.} 2011, \aap, 529, A106

\bibitem[{{Kennicutt}(1998)}]{Kennicutt98}
{Kennicutt}, Jr., R.~C. 1998, \araa, 36, 189

\bibitem[{{Kennicutt} {et~al.}(2003){Kennicutt}, {Armus}, {Bendo}, {Calzetti},
  {Dale}, {Draine}, {Engelbracht}, {Gordon}, {Grauer}, {Helou}, {Hollenbach},
  {Jarrett}, {Kewley}, {Leitherer}, {Li}, {Malhotra}, {Regan}, {Rieke},
  {Rieke}, {Roussel}, {Smith}, {Thornley}, \& {Walter}}]{Kennicutt03}
{Kennicutt}, Jr., R.~C., {Armus}, L., {Bendo}, G., {et~al.} 2003, \pasp, 115,
  928

\bibitem[{{Kewley} {et~al.}(2001){Kewley}, {Dopita}, {Sutherland}, {Heisler},
  \& {Trevena}}]{Kewley01}
{Kewley}, L.~J., {Dopita}, M.~A., {Sutherland}, R.~S., {Heisler}, C.~A., \&
  {Trevena}, J. 2001, \apj, 556, 121

\bibitem[{{Kharb} {et~al.}(2017){Kharb}, {Lal}, \& {Merritt}}]{Kharb17}
{Kharb}, P., {Lal}, D.~V., \& {Merritt}, D. 2017, Nature Astronomy, 1, 727

\bibitem[{{Kim} {et~al.}(2013){Kim}, {Evans}, {Vavilkin}, {Armus},
  {Mazzarella}, {Sheth}, {Surace}, {Haan}, {Howell}, {D{\'{\i}}az-Santos},
  {Petric}, {Iwasawa}, {Privon}, \& {Sanders}}]{Kim13}
{Kim}, D.-C., {Evans}, A.~S., {Vavilkin}, T., {et~al.} 2013, \apj, 768, 102

\bibitem[{{Koss} {et~al.}(2013){Koss}, {Mushotzky}, {Baumgartner}, {Veilleux},
  {Tueller}, {Markwardt}, \& {Casey}}]{Koss13}
{Koss}, M., {Mushotzky}, R., {Baumgartner}, W., {et~al.} 2013, \apjl, 765, L26

\bibitem[{{Krabbe} {et~al.}(2004){Krabbe}, {Gasaway}, {Song}, {Iserlohe},
  {Weiss}, {Larkin}, {Barczys}, \& {Lafreniere}}]{Krabbe04}
{Krabbe}, A., {Gasaway}, T., {Song}, I., {et~al.} 2004, in Society of
  Photo-Optical Instrumentation Engineers (SPIE) Conference Series, Vol. 5492,
  Society of Photo-Optical Instrumentation Engineers (SPIE) Conference Series,
  ed. {A.~F.~M.~Moorwood \& M.~Iye}, 1403--1410

\bibitem[{{Larkin} {et~al.}(2006){Larkin}, {Barczys}, {Krabbe}, {Adkins},
  {Aliado}, {Amico}, {Brims}, {Campbell}, {Canfield}, {Gasaway}, {Honey},
  {Iserlohe}, {Johnson}, {Kress}, {LaFreniere}, {Lyke}, {Magnone}, {Magnone},
  {McElwain}, {Moon}, {Quirrenbach}, {Skulason}, {Song}, {Spencer}, {Weiss}, \&
  {Wright}}]{Larkin06}
{Larkin}, J., {Barczys}, M., {Krabbe}, A., {et~al.} 2006, in \procspie, Vol.
  6269, Society of Photo-Optical Instrumentation Engineers (SPIE) Conference
  Series, 62691A

\bibitem[{{Larkin} {et~al.}(1998){Larkin}, {Armus}, {Knop}, {Soifer}, \&
  {Matthews}}]{Larkin98}
{Larkin}, J.~E., {Armus}, L., {Knop}, R.~A., {Soifer}, B.~T., \& {Matthews}, K.
  1998, \apjs, 114, 59

\bibitem[{{Larson} {et~al.}(2016){Larson}, {Sanders}, {Barnes}, {Ishida},
  {Evans}, {U}, {Mazzarella}, {Kim}, {Privon}, {Mirabel}, \&
  {Flewelling}}]{Larson16}
{Larson}, K.~L., {Sanders}, D.~B., {Barnes}, J.~E., {et~al.} 2016, \apj, 825,
  128

\bibitem[{{Laurent} {et~al.}(2000){Laurent}, {Mirabel}, {Charmandaris},
  {Gallais}, {Madden}, {Sauvage}, {Vigroux}, \& {Cesarsky}}]{Laurent00}
{Laurent}, O., {Mirabel}, I.~F., {Charmandaris}, V., {et~al.} 2000, \aap, 359,
  887

\bibitem[{{Martin}(2006)}]{Martin06}
{Martin}, C.~L. 2006, \apj, 647, 222

\bibitem[{{Martini} {et~al.}(1999){Martini}, {Sellgren}, \&
  {DePoy}}]{Martini99}
{Martini}, P., {Sellgren}, K., \& {DePoy}, D.~L. 1999, \apj, 526, 772

\bibitem[{{Mattila} {et~al.}(2007){Mattila}, {V{\"a}is{\"a}nen}, {Farrah},
  {Efstathiou}, {Meikle}, {Dahlen}, {Fransson}, {Lira}, {Lundqvist},
  {{\"O}stlin}, {Ryder}, \& {Sollerman}}]{Mattila07}
{Mattila}, S., {V{\"a}is{\"a}nen}, P., {Farrah}, D., {et~al.} 2007, \apjl, 659,
  L9

\bibitem[{{Mazzalay} {et~al.}(2013){Mazzalay}, {Saglia}, {Erwin}, {Fabricius},
  {Rusli}, {Thomas}, {Bender}, {Opitsch}, {Nowak}, \& {Williams}}]{Mazzalay13}
{Mazzalay}, X., {Saglia}, R.~P., {Erwin}, P., {et~al.} 2013, \mnras, 428, 2389

\bibitem[{{Medling} {et~al.}(2014){Medling}, {U}, {Guedes}, {Max}, {Mayer},
  {Armus}, {Holden}, {Ro{\v s}kar}, \& {Sanders}}]{Medling14}
{Medling}, A.~M., {U}, V., {Guedes}, J., {et~al.} 2014, \apj, 784, 70

\bibitem[{{Medling} {et~al.}(2015{\natexlab{a}}){Medling}, {U}, {Max},
  {Sanders}, {Armus}, {Holden}, {Mieda}, {Wright}, \& {Larkin}}]{Medling15_bh}
{Medling}, A.~M., {U}, V., {Max}, C.~E., {et~al.} 2015{\natexlab{a}}, \apj,
  803, 61

\bibitem[{{Medling} {et~al.}(2015{\natexlab{b}}){Medling}, {U}, {Rich},
  {Kewley}, {Armus}, {Dopita}, {Max}, {Sanders}, \&
  {Sutherland}}]{Medling15_ir17207}
{Medling}, A.~M., {U}, V., {Rich}, J.~A., {et~al.} 2015{\natexlab{b}}, \mnras,
  448, 2301

\bibitem[{{Mihos} \& {Hernquist}(1994)}]{Mihos94}
{Mihos}, J.~C., \& {Hernquist}, L. 1994, \apjl, 431, L9

\bibitem[{{Momcheva} {et~al.}(2013){Momcheva}, {Lee}, {Ly}, {Salim}, {Dale},
  {Ouchi}, {Finn}, \& {Ono}}]{Momcheva13}
{Momcheva}, I.~G., {Lee}, J.~C., {Ly}, C., {et~al.} 2013, \aj, 145, 47

\bibitem[{{Moran} {et~al.}(1999){Moran}, {Lehnert}, \& {Helfand}}]{Moran99}
{Moran}, E.~C., {Lehnert}, M.~D., \& {Helfand}, D.~J. 1999, \apj, 526, 649

\bibitem[{{Moreno} {et~al.}(2015){Moreno}, {Torrey}, {Ellison}, {Patton},
  {Bluck}, {Bansal}, \& {Hernquist}}]{Moreno15}
{Moreno}, J., {Torrey}, P., {Ellison}, S.~L., {et~al.} 2015, \mnras, 448, 1107

\bibitem[{{Mouri}(1994)}]{Mouri94}
{Mouri}, H. 1994, \apj, 427, 777

\bibitem[{{Mudd} {et~al.}(2014){Mudd}, {Mathur}, {Guainazzi}, {Piconcelli},
  {Bianchi}, {Komossa}, {Vignali}, {Lanzuisi}, {Nicastro}, {Fiore}, \&
  {Maiolino}}]{Mudd14}
{Mudd}, D., {Mathur}, S., {Guainazzi}, M., {et~al.} 2014, \apj, 787, 40

\bibitem[{{M{\"u}ller-S{\'a}nchez} {et~al.}(2017){M{\"u}ller-S{\'a}nchez},
  {Hicks}, {Malkan}, {Davies}, {Yu}, {Shaver}, \& {Davis}}]{Muller-Sanchez17}
{M{\"u}ller-S{\'a}nchez}, F., {Hicks}, E.~K.~S., {Malkan}, M., {et~al.} 2017,
  ArXiv e-prints, arXiv:1705.06678

\bibitem[{{M{\"u}ller-S{\'a}nchez} {et~al.}(2011){M{\"u}ller-S{\'a}nchez},
  {Prieto}, {Hicks}, {Vives-Arias}, {Davies}, {Malkan}, {Tacconi}, \&
  {Genzel}}]{Muller-Sanchez11}
{M{\"u}ller-S{\'a}nchez}, F., {Prieto}, M.~A., {Hicks}, E.~K.~S., {et~al.}
  2011, \apj, 739, 69

\bibitem[{{Muratov} {et~al.}(2015){Muratov}, {Kere{\v s}},
  {Faucher-Gigu{\`e}re}, {Hopkins}, {Quataert}, \& {Murray}}]{Muratov15}
{Muratov}, A.~L., {Kere{\v s}}, D., {Faucher-Gigu{\`e}re}, C.-A., {et~al.}
  2015, MNRAS, 454, 2691

\bibitem[{{Narayanan} {et~al.}(2006){Narayanan}, {Cox}, {Robertson},
  {Dav{\'e}}, {Di Matteo}, {Hernquist}, {Hopkins}, {Kulesa}, \&
  {Walker}}]{Narayanan06}
{Narayanan}, D., {Cox}, T.~J., {Robertson}, B., {et~al.} 2006, \apjl, 642, L107

\bibitem[{{Narayanan} {et~al.}(2008){Narayanan}, {Cox}, {Kelly}, {Dav{\'e}},
  {Hernquist}, {Di Matteo}, {Hopkins}, {Kulesa}, {Robertson}, \&
  {Walker}}]{Narayanan08}
{Narayanan}, D., {Cox}, T.~J., {Kelly}, B., {et~al.} 2008, The Astrophysical
  Journal Supplement Series, 176, 331

\bibitem[{{Nims} {et~al.}(2015){Nims}, {Quataert}, \&
  {Faucher-Gigu{\`e}re}}]{Nims15}
{Nims}, J., {Quataert}, E., \& {Faucher-Gigu{\`e}re}, C.-A. 2015, MNRAS, 447,
  3612

\bibitem[{{Osterbrock}(1989)}]{Osterbrock89}
{Osterbrock}, D.~E. 1989, {Astrophysics of gaseous nebulae and active galactic
  nuclei}

\bibitem[{{Peterson} {et~al.}(2014){Peterson}, {Grier}, {Horne}, {Pogge},
  {Bentz}, {De Rosa}, {Denney}, {Martini}, {Sergeev}, {Kaspi}, {Minezaki},
  {Zu}, {Kochanek}, {Siverd}, {Shappee}, {Araya Salvo}, {Beatty}, {Bird},
  {Bord}, {Borman}, {Che}, {Chen}, {Cohen}, {Dietrich}, {Doroshenko}, {Drake},
  {Efimov}, {Free}, {Ginsburg}, {Henderson}, {King}, {Koshida}, {Mogren},
  {Molina}, {Mosquera}, {Motohara}, {Nazarov}, {Okhmat}, {Pejcha}, {Rafter},
  {Shields}, {Skowron}, {Skowron}, {Valluri}, {van Saders}, \&
  {Yoshii}}]{Peterson14}
{Peterson}, B.~M., {Grier}, C.~J., {Horne}, K., {et~al.} 2014, ApJ, 795, 149

\bibitem[{{Petric} {et~al.}(2011){Petric}, {Armus}, {Howell}, {Chan},
  {Mazzarella}, {Evans}, {Surace}, {Sanders}, {Appleton}, {Charmandaris},
  {D{\'{\i}}az-Santos}, {Frayer}, {Haan}, {Inami}, {Iwasawa}, {Kim}, {Madore},
  {Marshall}, {Spoon}, {Stierwalt}, {Sturm}, {U}, {Vavilkin}, \&
  {Veilleux}}]{Petric11}
{Petric}, A.~O., {Armus}, L., {Howell}, J., {et~al.} 2011, ApJ, 730, 28

\bibitem[{{Piqueras L{\'o}pez} {et~al.}(2013){Piqueras L{\'o}pez}, {Colina},
  {Arribas}, \& {Alonso-Herrero}}]{Piqueras13}
{Piqueras L{\'o}pez}, J., {Colina}, L., {Arribas}, S., \& {Alonso-Herrero}, A.
  2013, \aap, 553, A85

\bibitem[{{Piqueras L{\'o}pez} {et~al.}(2016){Piqueras L{\'o}pez}, {Colina},
  {Arribas}, {Pereira-Santaella}, \& {Alonso-Herrero}}]{Piqueras16}
{Piqueras L{\'o}pez}, J., {Colina}, L., {Arribas}, S., {Pereira-Santaella}, M.,
  \& {Alonso-Herrero}, A. 2016, \aap, 590, A67

\bibitem[{{Ptak} {et~al.}(2003){Ptak}, {Heckman}, {Levenson}, {Weaver}, \&
  {Strickland}}]{Ptak03}
{Ptak}, A., {Heckman}, T., {Levenson}, N.~A., {Weaver}, K., \& {Strickland}, D.
  2003, \apj, 592, 782

\bibitem[{{Reunanen} {et~al.}(2002){Reunanen}, {Kotilainen}, \&
  {Prieto}}]{Reunanen02}
{Reunanen}, J., {Kotilainen}, J.~K., \& {Prieto}, M.~A. 2002, \mnras, 331, 154

\bibitem[{{Ricci} {et~al.}(2017){Ricci}, {Bauer}, {Treister}, {Schawinski},
  {Privon}, {Blecha}, {Arevalo}, {Armus}, {Harrison}, {Ho}, {Iwasawa},
  {Sanders}, \& {Stern}}]{Ricci17}
{Ricci}, C., {Bauer}, F.~E., {Treister}, E., {et~al.} 2017, \mnras, 468, 1273

\bibitem[{{Rich} {et~al.}(2010){Rich}, {Dopita}, {Kewley}, \& {Rupke}}]{Rich10}
{Rich}, J.~A., {Dopita}, M.~A., {Kewley}, L.~J., \& {Rupke}, D.~S.~N. 2010,
  ApJ, 721, 505

\bibitem[{{Rich} {et~al.}(2011){Rich}, {Kewley}, \& {Dopita}}]{Rich11}
{Rich}, J.~A., {Kewley}, L.~J., \& {Dopita}, M.~A. 2011, \apj, 734, 87

\bibitem[{{Rich} {et~al.}(2015){Rich}, {Kewley}, \& {Dopita}}]{Rich15}
---. 2015, \apjs, 221, 28

\bibitem[{{Riffel} {et~al.}(2013){Riffel}, {Rodr{\'{\i}}guez-Ardila}, {Aleman},
  {Brotherton}, {Pastoriza}, {Bonatto}, \& {Dors}}]{Riffel13}
{Riffel}, R., {Rodr{\'{\i}}guez-Ardila}, A., {Aleman}, I., {et~al.} 2013,
  \mnras, 430, 2002

\bibitem[{{Riffel} {et~al.}(2006){Riffel}, {Rodr{\'{\i}}guez-Ardila}, \&
  {Pastoriza}}]{Riffel06}
{Riffel}, R., {Rodr{\'{\i}}guez-Ardila}, A., \& {Pastoriza}, M.~G. 2006, \aap,
  457, 61

\bibitem[{{Riffel} {et~al.}(2008){Riffel}, {Storchi-Bergmann}, {Winge},
  {McGregor}, {Beck}, \& {Schmitt}}]{Riffel08}
{Riffel}, R.~A., {Storchi-Bergmann}, T., {Winge}, C., {et~al.} 2008, \mnras,
  385, 1129

\bibitem[{{Rodr{\'{\i}}guez-Ardila} {et~al.}(2004){Rodr{\'{\i}}guez-Ardila},
  {Pastoriza}, {Viegas}, {Sigut}, \& {Pradhan}}]{Rodriguez04}
{Rodr{\'{\i}}guez-Ardila}, A., {Pastoriza}, M.~G., {Viegas}, S., {Sigut},
  T.~A.~A., \& {Pradhan}, A.~K. 2004, \aap, 425, 457

\bibitem[{{Rodr{\'{\i}}guez-Ardila} {et~al.}(2005){Rodr{\'{\i}}guez-Ardila},
  {Riffel}, \& {Pastoriza}}]{Rodriguez05}
{Rodr{\'{\i}}guez-Ardila}, A., {Riffel}, R., \& {Pastoriza}, M.~G. 2005,
  \mnras, 364, 1041

\bibitem[{{Romero-Ca{\~n}izales} {et~al.}(2017){Romero-Ca{\~n}izales},
  {Alberdi}, {Ricci}, {Ar{\'e}valo}, {P{\'e}rez-Torres}, {Conway}, {Beswick},
  {Bondi}, {Muxlow}, {Argo}, {Bauer}, {Efstathiou}, {Herrero-Illana},
  {Mattila}, \& {Ryder}}]{Romero-Canizales17}
{Romero-Ca{\~n}izales}, C., {Alberdi}, A., {Ricci}, C., {et~al.} 2017, \mnras,
  467, 2504

\bibitem[{{Rupke} {et~al.}(2002){Rupke}, {Veilleux}, \& {Sanders}}]{Rupke02}
{Rupke}, D.~S., {Veilleux}, S., \& {Sanders}, D.~B. 2002, \apj, 570, 588

\bibitem[{{Rupke} {et~al.}(2005{\natexlab{a}}){Rupke}, {Veilleux}, \&
  {Sanders}}]{Rupke05a}
---. 2005{\natexlab{a}}, \apjs, 160, 87

\bibitem[{{Rupke} {et~al.}(2005{\natexlab{b}}){Rupke}, {Veilleux}, \&
  {Sanders}}]{Rupke05b}
---. 2005{\natexlab{b}}, \apjs, 160, 115

\bibitem[{{Rupke} \& {Veilleux}(2011)}]{Rupke11}
{Rupke}, D.~S.~N., \& {Veilleux}, S. 2011, \apjl, 729, L27

\bibitem[{{Rupke} \& {Veilleux}(2013)}]{Rupke13}
---. 2013, \apj, 768, 75

\bibitem[{{Sanders} {et~al.}(2003){Sanders}, {Mazzarella}, {Kim}, {Surace}, \&
  {Soifer}}]{Sanders03}
{Sanders}, D.~B., {Mazzarella}, J.~M., {Kim}, D.-C., {Surace}, J.~A., \&
  {Soifer}, B.~T. 2003, \aj, 126, 1607

\bibitem[{{Sanders} {et~al.}(1988){Sanders}, {Soifer}, {Elias}, {Madore},
  {Matthews}, {Neugebauer}, \& {Scoville}}]{Sanders88}
{Sanders}, D.~B., {Soifer}, B.~T., {Elias}, J.~H., {et~al.} 1988, \apj, 325, 74

\bibitem[{{Scoville} {et~al.}(1982){Scoville}, {Hall}, {Ridgway}, \&
  {Kleinmann}}]{Scoville82}
{Scoville}, N.~Z., {Hall}, D.~N.~B., {Ridgway}, S.~T., \& {Kleinmann}, S.~G.
  1982, \apj, 253, 136

\bibitem[{{Smaji{\'c}} {et~al.}(2015){Smaji{\'c}}, {Moser}, {Eckart}, {Busch},
  {Combes}, {Garc{\'{\i}}a-Burillo}, {Valencia-S.}, \& {Horrobin}}]{Smajic15}
{Smaji{\'c}}, S., {Moser}, L., {Eckart}, A., {et~al.} 2015, A\&A, 583, A104

\bibitem[{{Soto} {et~al.}(2012){Soto}, {Martin}, {Prescott}, \&
  {Armus}}]{Soto12}
{Soto}, K.~T., {Martin}, C.~L., {Prescott}, M.~K.~M., \& {Armus}, L. 2012, ApJ,
  757, 86

\bibitem[{{Spoon} {et~al.}(2007){Spoon}, {Marshall}, {Houck}, {Elitzur}, {Hao},
  {Armus}, {Brandl}, \& {Charmandaris}}]{Spoon07}
{Spoon}, H.~W.~W., {Marshall}, J.~A., {Houck}, J.~R., {et~al.} 2007, \apjl,
  654, L49

\bibitem[{{Spoon} {et~al.}(2013){Spoon}, {Farrah}, {Lebouteiller},
  {Gonz{\'a}lez-Alfonso}, {Bernard-Salas}, {Urrutia}, {Rigopoulou},
  {Westmoquette}, {Smith}, {Afonso}, {Pearson}, {Cormier}, {Efstathiou},
  {Borys}, {Verma}, {Etxaluze}, \& {Clements}}]{Spoon13}
{Spoon}, H.~W.~W., {Farrah}, D., {Lebouteiller}, V., {et~al.} 2013, \apj, 775,
  127

\bibitem[{{Springel} \& {Hernquist}(2003)}]{Springel03}
{Springel}, V., \& {Hernquist}, L. 2003, \mnras, 339, 289

\bibitem[{{Sternberg} \& {Dalgarno}(1989)}]{Sternberg89}
{Sternberg}, A., \& {Dalgarno}, A. 1989, \apj, 338, 197

\bibitem[{{Stierwalt} {et~al.}(2013){Stierwalt}, {Armus}, {Surace}, {Inami},
  {Petric}, {Diaz-Santos}, {Haan}, {Charmandaris}, {Howell}, {Kim}, {Marshall},
  {Mazzarella}, {Spoon}, {Veilleux}, {Evans}, {Sanders}, {Appleton}, {Bothun},
  {Bridge}, {Chan}, {Frayer}, {Iwasawa}, {Kewley}, {Lord}, {Madore},
  {Melbourne}, {Murphy}, {Rich}, {Schulz}, {Sturm}, {Vavilkin}, \&
  {Xu}}]{Stierwalt13}
{Stierwalt}, S., {Armus}, L., {Surace}, J.~A., {et~al.} 2013, ApJS, 206, 1

\bibitem[{{Stierwalt} {et~al.}(2014){Stierwalt}, {Armus}, {Charmandaris},
  {Diaz-Santos}, {Marshall}, {Evans}, {Haan}, {Howell}, {Iwasawa}, {Kim},
  {Murphy}, {Rich}, {Spoon}, {Inami}, {Petric}, \& {U}}]{Stierwalt14}
{Stierwalt}, S., {Armus}, L., {Charmandaris}, V., {et~al.} 2014, \apj, 790, 124

\bibitem[{{Torrey} {et~al.}(2012){Torrey}, {Cox}, {Kewley}, \&
  {Hernquist}}]{Torrey12}
{Torrey}, P., {Cox}, T.~J., {Kewley}, L., \& {Hernquist}, L. 2012, \apj, 746,
  108

\bibitem[{{U} {et~al.}(2012){U}, {Sanders}, {Mazzarella}, {Evans}, {Howell},
  {Surace}, {Armus}, {Iwasawa}, {Kim}, {Casey}, {Vavilkin}, {Dufault},
  {Larson}, {Barnes}, {Chan}, {Frayer}, {Haan}, {Inami}, {Ishida},
  {Kartaltepe}, {Melbourne}, \& {Petric}}]{U12}
{U}, V., {Sanders}, D.~B., {Mazzarella}, J.~M., {et~al.} 2012, \apjs, 203, 9

\bibitem[{{U} {et~al.}(2013){U}, {Medling}, {Sanders}, {Max}, {Armus},
  {Iwasawa}, {Evans}, {Kewley}, \& {Fazio}}]{U13}
{U}, V., {Medling}, A., {Sanders}, D., {et~al.} 2013, \apj, 775, 115

\bibitem[{{V{\"a}is{\"a}nen} {et~al.}(2017){V{\"a}is{\"a}nen}, {Reunanen},
  {Kotilainen}, {Mattila}, {Johansson}, {Ramphul}, {Romero-Ca{\~n}izales}, \&
  {Kuncarayakti}}]{Vaisanen17}
{V{\"a}is{\"a}nen}, P., {Reunanen}, J., {Kotilainen}, J., {et~al.} 2017, MNRAS,
  471, 2059

\bibitem[{{van Dam} {et~al.}(2006){van Dam}, {Bouchez}, {Le Mignant},
  {Johansson}, {Wizinowich}, {Campbell}, {Chin}, {Hartman}, {Lafon}, {Stomski},
  \& {Summers}}]{vanDam06}
{van Dam}, M.~A., {Bouchez}, A.~H., {Le Mignant}, D., {et~al.} 2006, \pasp,
  118, 310

\bibitem[{{van der Werf} {et~al.}(1993){van der Werf}, {Genzel}, {Krabbe},
  {Blietz}, {Lutz}, {Drapatz}, {Ward}, \& {Forbes}}]{vanderWerf93}
{van der Werf}, P.~P., {Genzel}, R., {Krabbe}, A., {et~al.} 1993, \apj, 405,
  522

\bibitem[{{Vardoulaki} {et~al.}(2015){Vardoulaki}, {Charmandaris}, {Murphy},
  {Diaz-Santos}, {Armus}, {Evans}, {Mazzarella}, {Privon}, {Stierwalt}, \&
  {Barcos-Mu{\~n}oz}}]{Vardoulaki15}
{Vardoulaki}, E., {Charmandaris}, V., {Murphy}, E.~J., {et~al.} 2015, \aap,
  574, A4

\bibitem[{{Veilleux} {et~al.}(2005){Veilleux}, {Cecil}, \&
  {Bland-Hawthorn}}]{Veilleux05}
{Veilleux}, S., {Cecil}, G., \& {Bland-Hawthorn}, J. 2005, \araa, 43, 769

\bibitem[{{Veilleux} {et~al.}(1995){Veilleux}, {Kim}, {Sanders}, {Mazzarella},
  \& {Soifer}}]{Veilleux95}
{Veilleux}, S., {Kim}, D.-C., {Sanders}, D.~B., {Mazzarella}, J.~M., \&
  {Soifer}, B.~T. 1995, \apjs, 98, 171

\bibitem[{{Veilleux} \& {Osterbrock}(1987)}]{Veilleux87}
{Veilleux}, S., \& {Osterbrock}, D.~E. 1987, ApJS, 63, 295

\bibitem[{{Veilleux} {et~al.}(2013){Veilleux}, {Mel{\'e}ndez}, {Sturm},
  {Gracia-Carpio}, {Fischer}, {Gonz{\'a}lez-Alfonso}, {Contursi}, {Lutz},
  {Poglitsch}, {Davies}, {Genzel}, {Tacconi}, {de Jong}, {Sternberg}, {Netzer},
  {Hailey-Dunsheath}, {Verma}, {Rupke}, {Maiolino}, {Teng}, \&
  {Polisensky}}]{Veilleux13}
{Veilleux}, S., {Mel{\'e}ndez}, M., {Sturm}, E., {et~al.} 2013, \apj, 776, 27

\bibitem[{{Wizinowich} {et~al.}(2000){Wizinowich}, {Acton}, {Shelton},
  {Stomski}, {Gathright}, {Ho}, {Lupton}, {Tsubota}, {Lai}, {Max}, {Brase},
  {An}, {Avicola}, {Olivier}, {Gavel}, {Macintosh}, {Ghez}, \&
  {Larkin}}]{Wizinowich00}
{Wizinowich}, P., {Acton}, D.~S., {Shelton}, C., {et~al.} 2000, \pasp, 112, 315

\bibitem[{{Wizinowich} {et~al.}(2006){Wizinowich}, {Le Mignant}, {Bouchez},
  {Campbell}, {Chin}, {Contos}, {van Dam}, {Hartman}, {Johansson}, {Lafon},
  {Lewis}, {Stomski}, {Summers}, {Brown}, {Danforth}, {Max}, \&
  {Pennington}}]{Wizinowich06}
{Wizinowich}, P.~L., {Le Mignant}, D., {Bouchez}, A.~H., {et~al.} 2006, \pasp,
  118, 297

\bibitem[{{Yuan} {et~al.}(2010){Yuan}, {Kewley}, \& {Sanders}}]{Yuan10}
{Yuan}, T.-T., {Kewley}, L.~J., \& {Sanders}, D.~B. 2010, \apj, 709, 884

\end{thebibliography}

\begin{longrotatetable}
   \begin{deluxetable*}{lccccccccccc}
    \centering
    \tabletypesize{\scriptsize}
    \tablewidth{0pt}
    \tablecolumns{12}
    \tablecaption{Keck OSIRIS Observations\label{tbl:obs}}
    \tablehead{ % column headings
      \colhead{Galaxy Name} &
      \colhead{Right Ascension} &
      \colhead{Declination} &
      \colhead{$\log L_{\rm IR}$} &
%      \colhead{$\log L_{\rm IR} [8-1000\mu m]$} &
      \colhead{Redshift} &
      \colhead{Physical Scale} &
      \colhead{Filter} &
      \colhead{Plate Scale} & 
      \colhead{FOV} & 
      \colhead{Obs Date} & 
      \colhead{$t_{\rm exp}$} &
      \colhead{PA} \\
      \colhead{} & 
      \colhead{(J2000)} & 
      \colhead{(J2000)} & 
      \colhead{($L_\odot$)} & 
      \colhead{} & 
      \colhead{(kpc/arcsec)} &
      \colhead{} & 
      \colhead{(mas)} & 
      \colhead{(arcsec$^2$)} &
      \colhead{(YYYYMMDD)} & 
      \colhead{(min)} &
      \colhead{(deg)} \\
      \colhead{(1)} &
      \colhead{(2)} &
      \colhead{(3)} &
      \colhead{(4)} &
      \colhead{(5)} &
      \colhead{(6)} &
      \colhead{(7)} &
      \colhead{(8)} &
      \colhead{(9)} &
      \colhead{(10)} &
      \colhead{(11)} &
      \colhead{(12)}
    }
    \startdata
    UGC 08058 / Mrk 231 & 12:56:14.231 & 56:52:25.250 & 12.53 & 0.0433 & 0.856 & Kbb & 35 & 0.56$\times$2.24 & 20110523 & 36 &  45 \\   %0.0422
    IRAS F17207$-$0014 & 17:23:22.010 & -00:17:00.200 & 12.46$^a$ & 0.0432 & 0.878 & Kcb & 100 & 1.6$\times$6.4 & 20110523-24 &  60 & 0 \\ %0.0428
    UGC 08696 / Mrk 273 & 13:44:42.140 & 55:53:13.700 & 12.18 & 0.0380 & 0.775 & Kbb & 100 & 1.6$\times$6.4 & 20110522 & 50 & 0 \\ %0.0378
    IRAS F22491$-$1808 & 22:51:49.220 & -17:52:23.400 & 12.19 & 0.0781 & 1.467 & Kcb & 100 & 1.6$\times$6.4 & 20101114 & 20 & 150 \\
    IRAS F15250+3608 & 15:26:59.443 & 35:58:37.010 & 12.07 & 0.0563 & 1.103 & Kbb & 50 & 0.8$\times$3.2 & 20110523 & 80 & 120 \\
    UGC 05101 & 09:35:51.611 & 61:21:11.600 &  12.00 & 0.0390 & 0.793 & Kcb & 100 & 1.6$\times$6.4 & 20101114 & 40  & 70 \\ %0.0394
    VV 340a & 14:57:00.701 & 24:37:02.220 & 11.79 & 0.0344 & 0.710 & Kcb & 100 & 1.6$\times$6.4 & 20130518 & 20 & 185 \\ %0.0335
    IRAS F01364$-$1042 & 01:38:52.882 & -10:27:11.480 & 11.79 & 0.0490 & 0.930 & Kcb & 100 & 1.6$\times$6.4 & 20101113-14 & 100  & 30 \\ %0.0482
                                   &  & &         &            & & Kbb & 35 & 0.56$\times$2.24 & 20121001 & 20 & 30 \\
    UGC 08387 & 13:20:35.350 & 34:08:21.750 & 11.72 & 0.0239 & 0.507 & Kcb & 100 & 1.6$\times$6.4 & 20130518 & 30 &140 \\ %0.0233
    CGCG 436$-$030 & 01:20:02.634 & 14:21:42.260 & 11.68 & 0.0315 & 0.610 & Kbb & 35 & 0.56$\times$2.24 & 20120930 & 30 & 0\\ %0.0312
%                             &           &             & Kcb & 100 & 1.6$\times$6.4 & & \\
    NGC 6670E & 18:33:37.617 & 59:53:23.280 & 11.65$^a$ &  0.0291 & 0.592 & Kbb & 35 & 0.56$\times$2.24 & 20140719 & 60 & 280 \\
    IRAS F06076$-$2139N & 06:09:45.800 & -21:40:23.640 & 11.65$^a$ & 0.0374 & 0.742 & Kbb & 35 & 0.56$\times$2.24 & 20141112 & 40  & -10 \\
    IRAS F18090+0130E & 18:11:38.380 & 01:31:39.820 & 11.65$^a$ & 0.0286 & 0.611 & Kbb & 35 & 0.56$\times$2.24 & 20150529 & 60 & 20 \\
    IRAS F18090+0130W & 18:11:33.367 & 01:31:42.370 & 11.65$^a$ & 0.0292 & 0.611 & Kbb & 35 & 0.56$\times$2.24 & 20150529 & 80 & 110 \\
    III Zw 035 & 01:44:30.537 & 17:06:08.900 & 11.62 & 0.0278 & 0.547 & Kcb & 100 & 1.6$\times$6.4 & 20110110 & 20 & 90 \\ %0.0274
                    & & &           &            & & Kbb & 35 & 0.56$\times$2.24 & 20111210 &100 & 38 \\ %t_int = 90?
    IRAS F20351+2521 & 20:37:17.743 & 25:31:37.750 & 11.61$^a$ & 0.0344 & 0.683 & Kcb & 100 & 1.6$\times$6.4 & 20110522 & 40 & 0 \\ %0.0337
    NGC 2623 & 08:38:24.087 & 25:45:16.590 & 11.58 & 0.0196 & 0.393 & Kcb & 100 & 1.6$\times$6.4 & 20110110 & 50 & -50 \\ %0.0185
    NGC 7469N & 23:03:17.985 & 08:53:37.750 & 11.58 & 0.0163 & 0.332 & Kbb & 35 & 0.56$\times$2.24 & 20141112 & 80 & 130 \\
    NGC 6090 & 16:11:40.865 & 52:27:27.640 & 11.55 & 0.0303 & 0.626 & Kcb & 100 & 1.6$\times$6.4 & 20110524 & 20 & 150 \\ %0.0293
    NGC 7674W & 23:27:56.726 & 08:46:44.660 & 11.51 & 0.0289 & 0.574 & Kbb & 35 & 0.56$\times$2.24 & 20141112 & 50 & 110 \\
    IRAS F03359+1523 & 03:38:47.070 & 15:32:53.740 & 11.51 & 0.0365 & 0.690 & Kcb & 100 & 1.6$\times$6.4 & 20101114 & 60 & 75 \\ %0.0354
                           & &        &           &            & & Kbb & 35 & 0.56$\times$2.24 & 20141112 & 60 & 75 \\                                   
    MCG +08$-$11$-$002 & 05:40:43.783 & 49:41:42.150 & 11.46$^a$ & 0.0195 & 0.391 & Kcb & 100 & 1.6$\times$6.4 & 20120102 & 50 & 70 \\ %0.0192
                                 &  & &                   &           &  & Kbb & 35 & 0.56$\times$2.24 & 20120102 & 90 & 70 
    \enddata
    \tablecomments{Column 1: Galaxy name, following the naming convention of the Revised Bright Galaxy Sample~\cite[][]{Sanders03}; Column 2 \& 3: Precise right ascension and declination locations for OSIRIS pointing; Column 4: Logarithmic infrared luminosity defined as $\log L_{\rm IR} [8-1000\mu m]$ as adopted from~\cite{U12} and $^a$\cite{Armus09}; Column 5: Redshift from fitting emission lines; Column 6: Physical scale in kpc/arcsec; Column 7: OSIRIS K broad- or narrow-band filter used; Column 8: Plate scale in milliarcsec; Column 9: Field of view in arcsec$^2$; Column 10:  Observation dates; Column 11: Total exposure time on target in minutes; Column 12: Position Angle of OSIRIS field of view}
    \end{deluxetable*}
\end{longrotatetable}
  
\begin{longrotatetable}
 \begin{deluxetable*}{lcccccccc}
    \centering
    \tabletypesize{\scriptsize}
    \tablewidth{0pt}
    \tablecolumns{8} %
    \tablecaption{Median \molhy~and \brg~Line Fluxes and Ratios\label{tbl:fluxes}}
    \tablehead{ % column headings
      \colhead{Galaxy Name$^\dagger$} &
      \colhead{H$_2$ 1$-$0 S(1) Flux} &
      \colhead{1$-$0 S(3)/1$-$0 S(1)} &
      \colhead{1$-$0 S(2)/1$-$0 S(1)} &
      \colhead{1$-$0 S(0)/1$-$0 S(1)} & 
      \colhead{2$-$1 S(1)/1$-$0 S(1)} & 
      \colhead{\brg~Flux} &
      \colhead{\brd/\brg} \\
      \colhead{} &
      \colhead{(10$^{-16}$ erg s$^{-1}$ cm$^{-2}$)} & 
      \colhead{} & 
      \colhead{} & 
      \colhead{} & 
      \colhead{} & 
      \colhead{(10$^{-16}$ erg s$^{-1}$ cm$^{-2}$)} & 
      \colhead{}
    }
    \startdata
   UGC 08058 (35mas) &  0.18$\pm$0.07 & 0.92$\pm$0.03 & 0.48$\pm$0.03 & 0.32$\pm$0.95 & 0.17$\pm$0.03 &  0.14$\pm$0.01 & 0.07$\pm$0.47 \\
         IRAS F17207$-$0014 &  0.48$\pm$0.01 & 0.92:: & 0.48:: & 0.25$\pm$0.01 & 0.18$\pm$0.01 &  0.30$\pm$0.01 & 0.09$\pm$0.09 \\
           UGC 08696 &  1.01$\pm$0.01 & 1.04$\pm$0.01 & 0.38$\pm$0.01 & 0.27$\pm$0.01 & 0.16$\pm$0.01 &  0.46$\pm$0.27 & 0.16$\pm$0.07 \\
         IRAS F22941-1808 &  0.17$\pm$0.91 & 0.75$\pm$0.08 & 0.31$\pm$0.22 & \ldots & \ldots &  0.15$\pm$0.78 & 0.04$\pm$1.49 \\
 IRAS F15250+3608 (50mas) &  0.10:: & 0.85$\pm$0.03 & 0.35$\pm$0.02 & \ldots & \ldots &  0.11$\pm$0.06 & 0.07:: \\
           UGC 05101 &  0.52$\pm$0.01 & 1.18$\pm$0.04 & 0.39$\pm$0.02 & 0.36$\pm$0.01 & 0.22$\pm$0.01 &  0.29$\pm$0.02 & 0.09$\pm$0.46 \\
             VV 340a &  0.12:: & 0.92$\pm$0.01 & 0.48$\pm$0.01 & 0.26$\pm$0.03 & 0.17$\pm$0.02 &  0.05$\pm$0.01 & 0.07$\pm$0.78 \\
         IRAS F01364$-$1042 &  0.69$\pm$0.05 & 0.79$\pm$0.02 & 0.37$\pm$0.02 & 0.27$\pm$0.05 & 0.19$\pm$0.02 &  0.38$\pm$0.34 & 0.03$\pm$0.99 \\
         IRAS F01364$-$1042 (35mas) &  0.31$\pm$0.02 & 0.88$\pm$0.03 & 0.43$\pm$0.03 & 0.32$\pm$0.10 & 0.21$\pm$0.44 &  0.31$\pm$0.01 & 0.05$\pm$0.34 \\
           UGC 08387 &  0.30:: & 0.92$\pm$0.02 & 0.44$\pm$0.01 & 0.32$\pm$0.06 & 0.23$\pm$0.04 &  0.25$\pm$0.01 & 0.21$\pm$0.39 \\
CGCG 436$-$030 (35mas) &  0.40$\pm$0.03 & 0.71$\pm$0.06 & 0.33$\pm$0.22 & 0.44$\pm$0.13 & 0.35$\pm$0.04 &  0.58$\pm$0.01 & 0.15$\pm$0.01 \\
   NGC 6670E (35mas) & \ldots & 0.83$\pm$0.02 & 0.39$\pm$0.01 & 0.35$\pm$0.01 & 0.30$\pm$0.01 & \ldots & 0.11:: \\
IRAS F06076$-$2139N (35mas) &  0.07:: & 0.98$\pm$0.02 & 0.34$\pm$0.01 & 0.29$\pm$0.01 & 0.22$\pm$0.02 &  0.07:: & 0.04$\pm$0.12 \\
       IRAS F18090+0130E & \ldots & 1.00$\pm$0.05 & 0.38$\pm$0.02 & 0.31$\pm$0.21 & 0.17$\pm$0.30 & \ldots & 0.10$\pm$0.25 \\
        IRAS F18090+0130W & \ldots & 0.96$\pm$0.05 & 0.28$\pm$0.04 & 0.24$\pm$0.06 & 0.18$\pm$0.05 & \ldots & 0.18$\pm$0.06 \\
          III Zw 035 &  1.42$\pm$0.02 & 0.90$\pm$0.02 & 0.35$\pm$0.02 & 0.24$\pm$0.01 & 0.17$\pm$0.01 &  0.49$\pm$0.03 & 0.12$\pm$0.02 \\
          III Zw 035 (35mas) &  0.14:: & 0.74$\pm$0.01 & 0.32$\pm$0.01 & 0.22$\pm$0.01 & 0.14$\pm$0.01 &  0.04$\pm$0.09 & 0.03$\pm$0.29 \\
         IRAS F20351+2521 &  0.29$\pm$0.01 & 0.76$\pm$0.03 & 0.38$\pm$0.03 & 0.26$\pm$0.02 & 0.27$\pm$0.02 &  0.52$\pm$0.01 & 0.15:: \\
            NGC 2623 &  1.22$\pm$0.01 & 0.90$\pm$0.01 & 0.33$\pm$0.01 & 0.24$\pm$0.01 & 0.15$\pm$0.01 &  1.11$\pm$0.63 & 0.18$\pm$0.19 \\
   NGC 7469N (35mas) &  0.02:: & 0.92$\pm$0.04 & 0.49$\pm$0.03 & 0.27$\pm$0.02 & 0.55$\pm$0.04 &  0.02:: & 0.02:: \\
            NGC 6090 &  0.10$\pm$0.01 & 0.90$\pm$0.07 & 0.36$\pm$0.04 & 0.51$\pm$0.05 & 0.30$\pm$0.05 &  0.12$\pm$0.01 & 0.05$\pm$0.05 \\
   NGC 7674W (35mas) &  0.17:: & 1.23$\pm$0.14 & 0.37$\pm$0.04 & 0.25$\pm$0.02 & 0.22$\pm$0.11 &  0.18$\pm$0.01 & 0.21$\pm$0.06 \\
         IRAS F03359+1523 &  0.14$\pm$0.08 & 0.69$\pm$0.34 & 0.39$\pm$0.09 & 0.43$\pm$0.04 & 0.36$\pm$0.03 &  0.25:: & 0.07$\pm$0.01 \\
         IRAS F03359+1523  (35scale)&  0.06:: & 0.56$\pm$0.03 & 0.36$\pm$0.03 & 0.40$\pm$0.02 & 0.31$\pm$0.02 &  0.09:: & 0.06:: \\
      MCG +08$-$11$-$002 &  0.26$\pm$0.30 & 0.87$\pm$0.02 & 0.44$\pm$0.01 & 0.29$\pm$0.01 & 0.21$\pm$0.01 &  0.34$\pm$0.03 & 0.10$\pm$0.08 \\
      MCG +08$-$11$-$002  (35scale)&  0.09$\pm$0.20 & 0.94$\pm$0.15 & 0.34$\pm$0.01 & 0.26$\pm$0.01 & 0.22$\pm$0.01 &  0.14$\pm$0.12 & 0.03$\pm$0.01 
    \enddata
    \tablecomments{$\dagger$All galaxies refer to the 100mas data unless specified otherwise. ::Upper limit}
    \end{deluxetable*}
\end{longrotatetable}

\begin{longrotatetable}
 \begin{deluxetable*}{lcccccccccc}
    \centering
    \tabletypesize{\footnotesize}
    \tablewidth{0pt}
    \tablecolumns{11} %
    \tablecaption{Derived Quantities\label{tbl:derivedprop}}
    \tablehead{ % column headings
      \colhead{Galaxy Name$^\dagger$} &
      \colhead{Area} &
      \colhead{F$_{\rm total}$ (\molhy)} &
      \colhead{M$_{\rm total}$ (\molhy)} &
      \colhead{A$_{2.2\micron}$} &
      \colhead{L(\brg)} &
      \colhead{L$^{\rm corr}$(\brg)} &
      \colhead{$\Sigma_{\rm SFR}$} &
      \colhead{$\Sigma^{\rm corr}_{\rm SFR}$} &
      \colhead{$T_{\rm vib}$} &
      \colhead{$T_{\rm rot}$} \\
      \colhead{} &
      \colhead{(kpc$^2$)} &
      \colhead{(10$^{-16}$ erg s$^{-1}$ cm$^{-2}$)} & 
      \colhead{(10$^3$ M$_\odot$)} & 
      \colhead{} &
      \colhead{(10$^6$ L$_\odot$)} & 
      \colhead{(10$^6$ L$_\odot$)} & 
      \colhead{(M$_\odot$ yr$^{-1}$ kpc$^{-2}$)} & 
      \colhead{(M$_\odot$ yr$^{-1}$ kpc$^{-2}$)} & 
      \colhead{(K)} &
      \colhead{(K)} 
    }
    \startdata
   UGC 08058 (35mas) & 0.32 &  49.5$\pm$9.2 &    9.1$\pm$1.7 &    0.0$\pm$2.6 &   2.7$\pm$0.1 &    2.7$\pm$0.1 &   26.8$\pm$1.8 &   26.8$\pm$5.6 &  2670$\pm$190 & 1540$\pm$640 \\
         IRAS F17207$-$0014 & 1.39 & 111.6$\pm$11.2 &  360.0$\pm$36.1 &    3.1$\pm$2.3 &    6.2$\pm$0.1 &  109.5$\pm$1.4 &   14.0$\pm$2.3 &  247.7$\pm$29.5 & 2790$\pm$65 & 2250$\pm$160 \\
           UGC 08696 & 1.44 & 123.0$\pm$20.3 &   89.6$\pm$14.8 &    1.2$\pm$2.3 &    2.2$\pm$0.1 &   10.7$\pm$0.3 &    6.7$\pm$1.3 &   33.0$\pm$6.3 & 2590$\pm$50 & 1380$\pm$60 \\
         IRAS F22941-1808 & 3.08 & 20.0$\pm$2.7 &  283.3$\pm$37.8 &    5.5$\pm$0.1 &    0.4$\pm$0.1 &    9.0$\pm$1.4 &    0.4$\pm$0.4 &    9.2$\pm$3.0 & \dots & \dots \\
 IRAS F15250+3608 (50mas) & 0.43 &  22.8$\pm$2.1 &   57.6$\pm$5.4 &    2.3$\pm$2.2 &    2.8$\pm$0.1 &   22.7$\pm$0.4 &   20.3$\pm$0.9 &  163.9$\pm$17.0 & \dots & \dots \\
           UGC 05101 & 1.04 & 123.0$\pm$20.3 &   89.6$\pm$14.8 &    1.7$\pm$3.1 &    2.2$\pm$0.1 &   10.7$\pm$0.3 &    6.7$\pm$1.3 &   33.0$\pm$6.3 &  3100$\pm$100 & 1080$\pm$70 \\
             VV 340a & 0.78 & 9.7$\pm$1.7 &   28.5$\pm$5.1 &    0.0$\pm$1.5 &    0.1$\pm$0.0 &    3.8$\pm$0.2 &    0.6$\pm$0.1 &   15.2$\pm$2.3 &  2720$\pm$190 & 2150$\pm$420 \\
         IRAS F01364$-$1042 & 1.14 & 97.2$\pm$13.5 &  745.9$\pm$103.8 &    3.8$\pm$2.2 &    3.8$\pm$0.1 &  123.4$\pm$4.1 &   10.4$\pm$1.8 &  339.1$\pm$50.3  & 2890$\pm$180 & 1350$\pm$330 \\
 IRAS F01364$-$1042 (35mas) & 0.12 & 28.8$\pm$5.3 &   41.9$\pm$7.6 &    2.0$\pm$2.7 &    3.2$\pm$0.1 &   19.8$\pm$0.4 &   86.1$\pm$1.5 &  532.4$\pm$81.0  & 2990$\pm$330 & 1310$\pm$500 \\
           UGC 08387 & 1.10 &  111.3$\pm$15.4 &   14.8$\pm$2.0 &    1.5$\pm$2.1 &    3.1$\pm$0.0 &    7.4$\pm$0.1 &    8.7$\pm$1.1 &   21.1$\pm$2.4 &  3180$\pm$320 & 1380$\pm$320 \\
CGCG 436$-$030 (35mas) & 0.07 & 8.4$\pm$2.2 &    1.3$\pm$0.3 &    0.0$\pm$3.8 &    6.2$\pm$0.1 &   10.1$\pm$0.1 &  273.5$\pm$2.7 &  445.9$\pm$61.6 & 4120$\pm$380 & 790$\pm$400 \\
   NGC 6670E (35mas) &  0.12 & \dots &  \dots & 0.8$\pm$2.4 & \dots &   \dots & \dots &  \dots & 3750$\pm$90 & 1100$\pm$50 \\
IRAS F06076$-$2139N (35mas) & 0.47 &  38.5$\pm$6.2 &   29.0$\pm$4.6 &    1.9$\pm$3.4 &    2.5$\pm$0.0 &   13.7$\pm$0.1 &   16.5$\pm$1.2 &   91.8$\pm$14.7 & 3090$\pm$140 & 1140$\pm$50 \\
        IRAS F18090+0130E & 0.22 &  \dots &  \dots & 0.0$\pm$1.7 &  \dots &  \dots & \dots &  \dots & 2690$\pm$230 & 1200$\pm$910 \\
        IRAS F18090+0130W & 0.12 &  \dots &  \dots & 0.0$\pm$1.3 & \dots &  \dots & \dots &  \dots & 2780$\pm$350 & 1140$\pm$340 \\
          III Zw 035 & 0.25 &   92.1$\pm$10.3 &   11.8$\pm$1.3 &    0.6$\pm$2.9 &    1.5$\pm$0.0 &    2.6$\pm$0.0 &   18.7$\pm$0.3 &   32.3$\pm$2.2 &  2720$\pm$110 & 1450$\pm$140 \\
  III Zw 035 (35mas) & 0.18 &   61.2$\pm$4.9 &  86.9$\pm$7.0 &    3.2$\pm$3.6 &    0.7$\pm$0.0 &   12.8$\pm$0.1 &   11.7$\pm$0.3 &  222.8$\pm$27.5 &  2460$\pm$40 & 1530$\pm$70 \\
         IRAS F20351+2521 & 0.41 &  23.1$\pm$4.4 &   23.1$\pm$4.4 &    3.0$\pm$4.2 &    3.2$\pm$0.0 &   27.8$\pm$0.3 &   24.3$\pm$1.0 &  212.2$\pm$20.6 &  3460$\pm$180 & 1480$\pm$220 \\
            NGC 2623 & 0.28 & 244.6$\pm$19.9 &   44.7$\pm$3.6 &    2.4$\pm$1.6 &    4.9$\pm$0.0 &   24.4$\pm$0.2 &   54.5$\pm$1.1 &  272.2$\pm$19.5 &  2570$\pm$50 & 1380$\pm$60 \\
   NGC 7469N (35mas) & 0.03 &   1.5$\pm$0.3 &    0.2$\pm$0.0 &    1.8$\pm$3.3 &   0.1$\pm$0.0 &    0.4$\pm$0.0 &    7.8$\pm$0.0 &   41.9$\pm$4.6 &   6260$\pm$470 & 2080$\pm$410 \\
            NGC 6090 & 0.95 &  5.3$\pm$0.8 &    1.7$\pm$0.2 &    0.0$\pm$3.2 &    0.7$\pm$0.0 &    2.4$\pm$0.0 &    2.3$\pm$0.3 &    8.1$\pm$1.1 &  3680$\pm$370 &  750$\pm$80 \\
   NGC 7674W (35mas) & 0.08 &  26.0$\pm$5.4 &    0.9$\pm$0.2 &    0.0$\pm$2.9 &    1.2$\pm$0.0 &    0.5$\pm$0.0 &   48.9$\pm$0.7 &   21.2$\pm$3.9 &  3060$\pm$860 & 1460$\pm$270 \\
         IRAS F03359+1523 &  0.95 &  21.6$\pm$4.1 &   11.9$\pm$2.3 &    1.7$\pm$3.3 &    4.0$\pm$0.0 &   16.9$\pm$0.2 &   13.1$\pm$1.5 &   55.7$\pm$6.7 &  4200$\pm$260 & 900$\pm$190 \\
 IRAS F03359+1523 (35mas) &  0.11 &   9.3$\pm$1.9 &    7.0$\pm$1.4 &    1.9$\pm$2.4 &    1.3$\pm$0.0 &    7.9$\pm$0.0 &   37.0$\pm$0.1 &  215.6$\pm$2.7 &  3790$\pm$180 & 910$\pm$60 \\
      MCG +08$-$11$-$002 &  0.44 &  62.6$\pm$9.9 &   23.3$\pm$3.7 &    2.5$\pm$1.9 &    1.9$\pm$0.0 &   19.8$\pm$0.2 &   13.8$\pm$0.7 &  142.1$\pm$15.9 &  3020$\pm$90 & 1560$\pm$110 \\
MCG +08$-$11$-$002 (35mas) & 0.08 & 25.6$\pm$4.5 &   22.8$\pm$4.0 &    3.5$\pm$3.3 &    1.2$\pm$0.0 &   30.7$\pm$0.2 &   52.0$\pm$0.4 & 1277.7$\pm$120.5 & 3050$\pm$90 & 1280$\pm$100 
    \enddata
    \tablecomments{$\dagger$All galaxies refer to the 100mas data
      unless specified otherwise.}
    \end{deluxetable*}
\end{longrotatetable}

%\clearpage

  \begin{deluxetable*}{lcccc}
    \centering
    \tabletypesize{\footnotesize}
    \tablewidth{0pt}
    \tablecolumns{5}
    \tablecaption{\replaced{Integrated}{Statistical} \molhy/\brg~Line Ratios\label{tbl:h2brgratios}}
    \tablehead{ % column headings
      \colhead{Galaxy Name$^\dagger$} &
      \colhead{Maximum (visual)} &
      \colhead{Median} &
      \colhead{Mean} &
      \colhead{Variance} 
    }
    \startdata
          UGC 08058 (35mas)  &  4.20 &  1.54 &  2.56 &  2.96 \\
         IRAS F17207$-$0014  &  4.85 &  1.30 &  1.41 &  0.62 \\
         UGC 08696  &  5.30 &  1.64 &  1.84 &  0.79 \\
         IRAS F22491$-$1808  &  1.63 &  1.63 &  1.63 &  \ldots \\
          IRAS F15250+3608 (50mas)  &  2.86 &  1.13 &  1.26 &  0.50 \\
         UGC 05101  &  3.71 &  1.31 &  1.46 &  0.63 \\
         \hline
         \textbf{ULIRGs} & 3.76 & 1.43 & 1.69 & 0.92 \\
         \hline
          VV 340a  &  1.92 &  0.99 &  1.00 &  0.55 \\
         IRAS F01364$-$1042  &  2.17 &  1.22 &  1.27 &  0.39 \\
          IRAS F01364$-$1042 (35mas)  &  2.21 &  0.88 &  0.97 &  0.41 \\
         UGC 08387  &  1.67 &  0.80 &  0.81 &  0.44 \\
          CGCG 436$-$030 (35mas)  &  1.17 &  0.54 &  0.68 &  0.31 \\
         NGC 6670E (35mas)  &  1.22 &  0.34 &  0.40 &  0.21 \\
          IRAS F06076 (35mas)  &  3.72 &  0.90 &  0.97 &  0.40 \\
         IRAS F18090+0130E (35mas)  &  1.23 &  0.59 &  0.64 &  0.30 \\
         IRAS F18090+0130W (35mas)  &  0.98 &  0.38 &  0.40 &  0.17 \\
        III Zw 035  &  4.27 &  2.25 &  2.19 &  1.09 \\
         III Zw 035 (35mas)  &  6.00 &  2.01 &  2.40 &  1.26 \\
         IRAS F20351+2521  &  1.85 &  0.58 &  0.56 &  0.18 \\
         NGC 2623  &  2.04 &  0.99 &  1.02 &  0.28 \\
         NGC 7469N (35mas)  &  1.18 &  0.74 &  0.75 &  0.35 \\
         NGC 6090  &  1.06 &  0.55 &  0.67 &  0.43 \\
         NGC 7674W (35mas)  &  2.65 &  0.83 &  0.89 &  0.45 \\
         IRAS F03359+1523  &  1.20 &  0.51 &  0.58 &  0.30 \\
          IRAS F03359+1523 (35mas)  &  3.58 &  0.58 &  0.60 &  0.21 \\
          MCG +08$-$11$-$002  &  2.20 &  0.70 &  0.74 &  0.31 \\
          MCG +08$-$11$-$002 (35mas)  &  1.46 &  0.56 &  0.59 &  0.22 \\
          \hline
          \textbf{LIRGs$^\star$} & 2.12 & 0.77 & 0.83 & 0.39
    \enddata
    \tablecomments{$\dagger$All galaxies refer to the 100mas data unless specified otherwise.}
    {$\star$For galaxies observed in both 35mas and 100mas scales, only the values at the 35mas scale have been
      incorporated into the LIRG statistics.}
    \end{deluxetable*}

  \begin{deluxetable*}{lccccc}
    \centering
    \tabletypesize{\footnotesize}
    \tablewidth{0pt}
    \tablecolumns{6}
    \tablecaption{Summary Table\label{tbl:summary}}
    \tablehead{ % column headings
      \colhead{Galaxy Name$^\dagger$} &
      \colhead{Merger Class} &
      \colhead{Nuclear Separation (kpc)} &
      \colhead{AGN} &
      \twocolhead{Feedback Signature} \\
      \colhead{} &
      \colhead{} &
      \colhead{} &
      \colhead{} &
      \colhead{from OSIRIS} &
      \colhead{from literature} \\
      \colhead{(1)} &
      \colhead{(2)} &
      \colhead{(3)} &
      \colhead{(4)} &
      \colhead{(5)} &
      \colhead{(6)} 
    }
    \startdata
          UGC 08058 (35mas)  & 5 & single$^a$ &  Y$^1$ & Y & Y$^i$ \\
         IRAS F17207$-$0014  & 5 & 0.20$^b$ &  N$^b$ & Y$^{b}$ & N$^{i,iii}$,Y$^{ii}$ \\
         UGC 08696  & 5 & 0.75$^c$ &  Y$^1$ & Y$^{c}$ & Y$^i$ \\
         IRAS F22491$-$1808  & 5 & 2.20$^{a,d}$ & N$^{2,3}$ & Y & N$^i$,Y$^{iii}$ \\
          IRAS F15250+3608 (50mas)  & 5 & single$^a$ & Y(buried)$^{2,3}$ & (Y) & Y$^i$ \\
         UGC 05101  & 5 & single$^a$ & Y$^1$ & Y & N$^i$ \\
          VV 340a  & 1 & single$^a$ & Y$^1$ & N & -- \\
         IRAS F01364$-$1042  & 5 & single$^a$ & unclear$^4$ & (Y) & -- \\
          IRAS F01364$-$1042 (35mas)  & 5 & single$^a$ & unclear$^4$ & (Y) & -- \\
         UGC 08387  & 5 & 0.45$^e$ & Y$^8$ & N & -- \\
          CGCG 436$-$030 (35mas)  & 2 & 36.23$^e$ & unclear$^4$ & N & -- \\
         NGC 6670E (35mas)  & 2 & 1.04$^e$ & N$^{5,6}$ & (Y) & -- \\
          IRAS F06076$-$2139N (35mas)  & 3 & 6.70$^f$ & N$^6$ & N & -- \\
         IRAS F18090+0130E (35mas)  & 2 & 49.35$^e$ & N$^6$ & N & -- \\
         IRAS F18090+0130W (35mas)  & 2 & 49.35$^e$ & N$^6$ & N & -- \\
        III Zw 035  & 3 & 4.99$^e$ & N$^6$,CT$^7$ & Y & -- \\
         III Zw 035 (35mas)  & 3 & 4.99$^e$ & N$^6$,CT$^7$ & Y & -- \\
         IRAS F20351+2521  & 0 & single$^{a,e}$ &  N$^6$ & N & -- \\
         NGC 2623  & 5 & single$^{a,e}$ & Y$^8$ & (Y) & -- \\
         NGC 7469N (35mas)  & 2 &  26.76$^e$ & Y$^8$(S) & N & -- \\
         NGC 6090  & 4 &  4.33$^e$ &  N$^9$ & N & -- \\
         NGC 7674W (35mas)  & 1 & 19.99$^e$ & Y$^8$ & (maybe) & -- \\
         IRAS F03359+1523  & 3 &  7.94$^e$ & N$^6$,unclear$^4$ & (maybe) & -- \\
          IRAS F03359+1523 (35mas)  & 3 &  7.94$^e$ & N$^6$,unclear$^4$ & (maybe) & -- \\
          MCG +08$-$11$-$002  & 6 & 0.13$^g$ & N$^{8,g}$ & N & -- \\
          MCG +08$-$11$-$002 (35mas)  & 6 & 0.13$^g$ & N$^{8,g}$ & N & --
    \enddata
    \tablecomments{Column 1: Galaxy name ($\dagger$All galaxies refer to the 100mas data unless specified otherwise); 
Column 2: Merger classification adopted from~\cite{Haan11}, see text
for details;
Column 3: Nuclear separation between double nuclei, where detected
(Galaxies listed as ``single'' has only one kinematic nucleus known to
us from this data set and literature. References:
$^a$\cite{Medling14}; $^b$\cite{Medling15_ir17207}; $^c$\cite{U13};
$^d$this OSIRIS work; $^e$\cite{Haan11}; $^f$\emph{HST}-ACS images;
$^g$\cite{Davies16}); 
Column 4: Detection of any AGN signature -- Y: AGN dominated; N:
Starburst dominated; Y(buried): obscured AGN; unclear: likely a
composite (References:
$^1$\cite{Iwasawa11}; $^2$\cite{Yuan10}; $^3$\cite{Imanishi16};
$^4$\cite{Vardoulaki15}; $^5$\cite{Mudd14}; $^6$\cite{Koss13};
$^7$\cite{Gonzalez15}; $^8$\cite{Petric11}; $^9$\cite{Cortijo17}); for
NGC 7469, the southern nucleus is one identified as a Seyfert;
Column 5: Detection of feedback signature in this OSIRIS work:
Y = shock evidence from elevated \molhy/\brg~ratio ($>$ 2); (Y/maybe) =
kinematic signature of turbulent gas to be presented in forthcoming
paper; 
Column 6: Detection of feedback signature from literature: $^i$OH molecular outflows from~\emph{Herschel Space
Observatory}~\cite[][]{Veilleux13}; $^{ii}$Na {\sc i} D interstellar
absorption from ground-based telescopes~\cite[][]{Rupke05a,Rupke05b}; 
and $^{iii}$hot molecular gas observed with
VLT/SINFONI~\cite[][]{Emonts17}.
Most studies cover a higher infrared luminosity range than is targeted
here by our GOALS/KOALA survey so the sample overlap only extends to ULIRGs.
}
    \end{deluxetable*}

\clearpage

\appendix

\section{Notes on Individual Objects}

\replaced{Here we present the 6-panel figure similar to Figure~\ref{fig:6panel}
for the entire sample, along with notes on individual galaxies.}{Here
we include detailed notes on the individual galaxies as relevant for
interpreting our OSIRIS maps.}

\explain{Thanks to the tip from the formatting editor, all the images
  previously placed in the Appendix have now been moved to Figure 2 as
  part of the online figure set.} 

\subsection{UGC 08058}

UGC 08058, or Mrk 231, is the most luminous ULIRG in GOALS, and hence,
in this KOALA sample. As a bona fide QSO, it has been found to host
molecular outflows as an indicator of AGN feedback in previous
infrared and submillimeter work~\cite[][]{Gonzalez17,Alatalo15}. With
velocities well exceeding 500 km s$^{-1}$ $\sim$ 1kpc away from the
center, one might expect to see high velocity gas close to the
ionizing source in the central kpc region. However, the
\replaced{SNR}{S/N} in our
emission lines are weak as the near-infrared spectrum is dominated by
the AGN continuum.  Thus, we were not able to directly detect outflow
signatures in the \molhy~or \brg~line kinematics close to the center, though a
coherent region $\sim$ 250 pc SE of the nucleus is seen to feature
elevated \molhy/\brg~ratios.

% \begin{figure*}[htb]
%   \centering
%   \includegraphics[width=.65\textwidth,trim={0 8.5cm 0 0},clip]{fig_UGC8058_35scale.eps}
%   \caption{Six-panel figure showing the \emph{(left to right, top to
%       bottom)} $K$-band continuum, \molhy~1$-$0 S(1) flux, \brg~flux,
%     \molhy/\brg, differential velocity, and \molhy~velocity dispersion
%     maps, similar to Figure~\ref{fig:6panel}, but for UGC
%     08058. Scale bar %and compass rose 
%     is illustrated in the first panel, and the continuum contours are
%     displayed in all the panels to indicate the position of the infrared nucleus. A
%     signal-to-noise cutoff has been applied for easier visualization purposes.}
%   \label{fig:6panel_app1}
% \end{figure*}

\subsection{IRAS F17207$-$0014}

IRAS F17207$-$0014 is a late-stage merging ULIRG with two kinematically
distinct nuclei, the collision of whose ISM has likely induced shocks
tracking the base of a collimated outflow~\cite[see details in our
previous work in][]{Medling15_ir17207}. \added{High-resolution Plateau de
Bure interferometer data had also found a CO molecular outflow
plausibly associated with a hidden AGN in the western nucleus, as suggested
by~\cite{GarciaBurillo15}}. From the SFR map, we find that
the sites for star formation as traced by \brg~are to the west of the
outflow base. 

% \begin{figure*}[htb]
%   \centering
%   \includegraphics[width=.65\textwidth,trim={0 8.5cm 0 0},clip]{fig_IR17207_100scale.eps}
%   \caption{Six-panel figure similar to Figure \ref{fig:6panel_app1}, but
%   for IRAS F17207$-$0014.}
%   \label{fig:6panel_app2}
% \end{figure*}

\subsection{UGC 08696}

UGC 08696, Mrk 273, is another ULIRG in the late-stage merging phases
analyzed in detail in our previous work~\cite[][]{U13}. We have found
evidence for a molecular outflow originating from a plausible obscured
AGN nucleus in the north (instead of from the X-ray bright AGN nucleus in
the southwest). The multiphase and multi-scale nature of its outflow
has been detailed in other studies of the warm and cold ionized and molecular
gases~\cite[e.g.][]{Rupke13,Veilleux13,Cicone14,Aladro18}. Here, we can
see that the outflow is traced by shocked 
\molhy~gas in the northern nucleus. 

% \begin{figure*}[htb]
%   \centering
%   \includegraphics[width=.65\textwidth,trim={0 8.5cm 0 0},clip]{fig_UGC8696_100scale.eps}
%   \caption{Six-panel figure similar to Figure \ref{fig:6panel_app1}, but
%     for UGC 08696.}
%   \label{fig:6panel_app3}
% \end{figure*}

\subsection{IRAS F22491$-$1808}

IRAS F22491$-$1808 is a ULIRG system with two kinematically distinct nuclei in
the $K$-band continuum and in \paa~emission approximately 2.2 kpc
apart (see Paper I for detailed flux and kinematics map of \molhy~and
\paa).  Since the \molhy~and \paa~do not share the same kinematics,
the presence of a strongly streaming or outflowing shocked gas has
been conjectured. Indeed, \molhy~outflows have been detected
by~\cite{Emonts17} from the eastern galaxy, which is the \replaced{dimmer}{fainter}
component in the system \added{and co-spatial with our maximum \molhy/\brg~spaxels}. 

% \begin{figure*}[htb]
%   \centering
%   \includegraphics[width=.65\textwidth,trim={0 8.5cm 0 0},clip]{fig_IR22491_100scale.eps}
%   \caption{Six-panel figure similar to Figure \ref{fig:6panel_app1}, but
%     for IRAS F22491$-$1808.}
%   \label{fig:6panel_app4}
% \end{figure*}

\subsection{IRAS F15250+3608}
IRAS F15250+3608 is the only galaxy in Paper I for which we did not
see rotation in either stars or gas. \deleted{In this case, the differential
velocity map should not be interpreted in a similar manner as the
others but has been included for completeness.} There is potentially
shocked \molhy~gas near the nucleus reaching velocity dispersion
upward of 150 km s$^{-1}$, which could be worth follow-up observations
to confirm. 

% \begin{figure*}[htb]
%   \centering
%   \includegraphics[width=.65\textwidth,trim={0 8.5cm 0 0},clip]{fig_IR15250_50scale.eps}
%   \caption{Six-panel figure similar to Figure \ref{fig:6panel_app1}, but
%     for IRAS F15250+3608.}
%   \label{fig:6panel_app5}
% \end{figure*}

\subsection{UGC 05101}
UGC 5101 features a Compton-Thick AGN with strong
mid-infrared~\cite[][]{Armus04,Armus07} and X-ray
signature~\cite[][]{Imanishi03,Ptak03,Gonzalez09} but lacks optical
signatures~\cite[][]{Yuan10}. In our OSIRIS maps, the AGN continuum
dominates though \molhy~and \brg~emission \deleted{can} both
\deleted{be seen to} peak at 
the same location. The \molhy~gas is particularly bright near the
nucleus such that enhanced \molhy/\brg~spaxels can be seen coming out
of the center, and may be interpreted as a shock candidate. 

% \begin{figure*}[htb]
%   \centering
%   \includegraphics[width=.65\textwidth,trim={0 8.5cm 0 0},clip]{fig_UGC5101_100scale.eps}
%   \caption{Six-panel figure similar to Figure \ref{fig:6panel_app1}, but
%     for UGC 05101.}
%   \label{fig:6panel_app6}
% \end{figure*}

\subsection{VV 340a}
VV 340a is at an early stage of merging as categorized
in~\cite{Haan11}. It hosts a Compton-Thick AGN that was detected
by~\emph{Chandra} but not
by~\emph{Swift}-BAT~\cite[][]{Iwasawa11,Koss13}. Its large scale disk
features a solid dust lane while its nuclear disk is seen in the
continuum and \molhy~emission (see nuclear disk properties in Paper
I). We see no coherent structure in the \molhy/\brg~map that could be
associated with outflows, which, if warranted by the depth of the
data, could place an upper limit on how early outflows are triggered
in the nuclei of merging progenitors.

% \begin{figure*}[htb]
%   \centering
%   \includegraphics[width=.65\textwidth,trim={0 8.5cm 0 0},clip]{fig_VV340a_100scale.eps}
%   \caption{Six-panel figure similar to Figure \ref{fig:6panel_app1}, but
%     for VV 340a.}
%   \label{fig:6panel_app7}
% \end{figure*}

\subsection{IRAS F01364$-$1042}
IRAS F01364$-$1042 is a LIRG with a clear rotating gas disk detected in
\molhy~and \paa~as detailed in Paper I. In the 100mas scale data,
extended \molhy~to the south between the major and minor axis has been
seen, lending potential support to an outflowing structure traced by
\molhy. In the 35mas scale data, % (which is XXX deg rotated with respect to the 100mas scale data), 
the continuum is resolved into east-west extended emission
with the enhanced \molhy/\brg~spaxels originating from the center. The spatial correlation of this excited molecular gas
with prominent non-Keplerian motion and elevated velocity dispersion
as mentioned in the text lends support that there could be
molecular outflows present, to be confirmed by follow-up high angular
resolution observations.

% \begin{figure*}[htb]
%   \centering
%   \includegraphics[width=.65\textwidth,trim={0 8.5cm 0 0},clip]{fig_IR01364_100scale.eps}
%   \caption{Six-panel figure similar to Figure \ref{fig:6panel_app1}, but
%     for IRAS F01364$-$1042 (100mas).}
%   \label{fig:6panel_app8}
% \end{figure*}

% \begin{figure*}[htb]
%   \centering
%   \includegraphics[width=.65\textwidth,trim={0 8.5cm 0 0},clip]{fig_IR01364_35scale.eps}
%   \caption{Six-panel figure similar to Figure \ref{fig:6panel_app1},
%     but for IRAS F01364$-$1042 (35mas).}
%   \label{fig:6panel_app9}
% \end{figure*}

\subsection{UGC 08387}
UGC 08387, or IC 883, has been identified as an AGN host by the detection of the \nev~line
from IRS spectra~\cite[][]{Petric11}. A parsec-scale radio jet
detected using VLBI further suggests that the nucleus plays host to
AGN activity~\cite[][]{Romero-Canizales17}. Though the mosaicking of the nucleus of
this galaxy is unfortunately incomplete \deleted{as mentioned in Paper I} due to
observing conditions, we included the data analysis here for
completeness. Even though the central AGN is not pictured within the
frame, we see a hint of what may be low-velocity, excited
\molhy~on the outskirts of the molecular gas disk. Due possibly to
heating by AGN photoionization, the \molhy/\brg~map may indicate a layer
of \molhy~gas heated by the Seyfert nucleus worthy of future follow-up
observations perpendicular to the gas disk. 

% \begin{figure*}[htb]
%   \centering
%   \includegraphics[width=.65\textwidth,trim={0 8.5cm 0 0},clip]{fig_UGC8387_100scale.eps}
%   \caption{Six-panel figure similar to Figure \ref{fig:6panel_app1}, but
%     for UGC 08387.}
%   \label{fig:6panel_app10}
% \end{figure*}

\subsection{CGCG 436$-$030}
CGCG 436$-$030 is a LIRG with a star-forming clump to the northwest of
the nucleus, as demonstrated in its \brg~flux map well modeled in
Paper I. In this galaxy, the \molhy~gas is relatively weak and does
not appear to be shocked heated by the nearby star-forming clump. Both
\molhy~and \brg~display rotational kinematic signatures that follow
each other fairly consistently. 

% \begin{figure*}[htb]
%   \centering
%   \includegraphics[width=.65\textwidth,trim={0 8.5cm 0 0},clip]{fig_CGCG436_35scale.eps}
%   \caption{Six-panel figure similar to Figure \ref{fig:6panel_app1}, but
%     for CGCG 436$-$030.}
%   \label{fig:6panel_app11}
% \end{figure*}

\subsection{NGC 6670E}
NGC 6670E is the eastern component of a merging galaxy pair with
a companion $\sim$ 17 kpc away~\cite[see resolved X-ray
emission in][]{Mudd14}, but the measured separation between the two nuclei
as detected in $H$-band \emph{HST}-NICMOS images within the eastern
galaxy is $\sim$1 kpc~\cite[][]{Haan11}. There are two caveats regarding the 
analysis of this galaxy: 1. the OSIRIS data analyzed in this work was
taken after Paper I was published, and thus we do not yet have the
quantitative properties of the nuclear disks; 2. this was one of three
nuclei for which we did not have proper calibration frames of our
standard stars and thus we were unable to flux calibrate this galaxy
in the consistent manner we did the others. For this reason, the units
on the continuum, \molhy, and \brg~maps are not properly displayed,
and the intensity of the line and continuum should be treated in
relative units. We were also unable to produce a SFR map since we did
not have a properly calibrated map of the \brg~luminosity. 

Nonetheless, it is interesting that the gas peaks for both \molhy~and
\brg~are displaced from the main continuum peak, residing instead at
an off-nucleus star-forming clump $\sim$ 200 pc west of the $K$-band
continuum peak. The \molhy~does not appear to be excited, though it
displays high velocity dispersion that may be traced to the gas peak
and could be a candidate for star-formation driven outflows in future
studies. 

% \begin{figure*}[htb]
%   \centering
%   \includegraphics[width=.65\textwidth,trim={0 8.5cm 0 0},clip]{fig_NGC6670E_35scale.eps}
%   \caption{Six-panel figure similar to Figure \ref{fig:6panel_app1}, but
%     for NGC 6670E.}
%   \label{fig:6panel_app12}
% \end{figure*}

\subsection{IRAS F06076$-$2139N}
IRAS F06076 is a system of two clearly distinct galaxies with
projected nuclear separation of $\sim$ 6.7 kpc
\replaced{from}{according to} HST ACS
imaging. Large-scale, optical IFS data from VLT/VIMOS show that at
low-intermediate masses and with projected velocities of $\sim$ 550 km
s$^{-1}$, the two galaxies are unlikely to ever
merge~\cite[][]{Arribas08}. The northern nucleus, which is the one our
OSIRIS FOV covers, is dominated by star formation as indicated by
optical line ratio and velocity dispersion~\cite[][]{Rich15}. Our
35mas data resolves the $K$-band continuum emission into two peaks,
and illustrates that the \brg~gas is extended over this region. The
\molhy~gas is more concentrated at the continuum peaks, with velocity
dispersion $\sim$ 110 km s$^{-1}$ in this inner 200 pc region. 

% \begin{figure*}[htb]
%   \centering
%   \includegraphics[width=.65\textwidth,trim={0 8.5cm 0 0},clip]{fig_IR06076_35scale.eps}
%   \caption{Six-panel figure similar to Figure \ref{fig:6panel_app1}, but
%     for IRAS F06076$-$2139N.}
%   \label{fig:6panel_app13}
% \end{figure*}

\subsection{IRAS F18090+0130 E / W}
IRAS F18090+0130 hosts two progenitor galaxies 49 kpc apart~\cite[][]{Haan11}. At the
early stage of merging, this system was observed but not detected
in~\emph{Swift} BAT~\cite[][]{Koss13}. The OSIRIS observations here
provide a baseline of what early-stage galaxy mergers may feature
before they become morphologically disturbed: the extended continuum is
resolved into what appear like spiral arms well traced by the gas in
both nuclei. Both the molecular and ionized gas appears extended but
not heated, given the lack of significant nuclear activity in the
respective regions. Like NGC 6670E, we lacked proper calibration
frames for these galaxies and thus were unable to flux calibrate the
OSIRIS data. 

% \begin{figure*}[htb]
%   \centering
%   \includegraphics[width=.65\textwidth,trim={0 8.5cm 0 0},clip]{fig_IR18090A_35scale.eps}
%   \caption{Six-panel figure similar to Figure \ref{fig:6panel_app1}, but
%     for IRAS F18090+0130E.}
%   \label{fig:6panel_app14}
% \end{figure*}

% \begin{figure*}[htb]
%   \centering
%   \includegraphics[width=.65\textwidth,trim={0 8.5cm 0 0},clip]{fig_IR18090B_35scale.eps}
%   \caption{Six-panel figure similar to Figure \ref{fig:6panel_app1}, but
%     for IRAS F18090+0130W.}
%   \label{fig:6panel_app15}
% \end{figure*}

%\clearpage 
\subsection{III Zw 035}
III Zw 035 is a LIRG that shows strong extension in \molhy~emission
along the minor axis of the galaxy coupled with non-Keplerian
kinematics and enhanced velocity dispersion in the same region. The
100mas scale data show a hint of this \added{extension} in the \molhy~maps, but the
molecular outflow is best resolved in the 35mas scale maps in Figure
\ref{fig:6panel}. The \molhy~peak is not only displaced from that of
the $K$-band continuum, but is shown to be shock-excited in a coherent
region in the \molhy/\brg~map. 

% \begin{figure*}[htb]
%   \centering
%   \includegraphics[width=.65\textwidth,trim={0 8.5cm 0 0},clip]{fig_IIIZw035_100scale.eps}
%   \caption{Six-panel figure similar to Figure \ref{fig:6panel_app1}, but
%     for III Zw 035.}
%   \label{fig:6panel_app16}
% \end{figure*}

\subsection{IRAS F20351+2521}
IRAS F20351+2521 is a LIRG with galactic-scale spiral arms that feed
into the clumpy nature as sampled within the OSIRIS FOV. As was found
in Paper I, its \molhy~and \brg~emission show similar kinematics, as
this early-stage LIRG shows no sign of outflowing or shock-heated
gas. 

% \begin{figure*}[htb]
%   \centering
%   \includegraphics[width=.65\textwidth,trim={0 8.5cm 0 0},clip]{fig_IR20351_100scale.eps}
%   \caption{Six-panel figure similar to Figure \ref{fig:6panel_app1}, but
%     for IRAS F20351+2521.}
%   \label{fig:6panel_app17}
% \end{figure*}

\subsection{NGC 2623}
NGC 2623, a well-studied LIRG \replaced{identified to host}{hosting} both off-nuclear star
clusters and an AGN, represents a prototypical advanced merger with
twin tidal tails extending from a single
nucleus~\cite[][]{Sanders03,Evans08}. Its $K$-band IFS data feature
smooth flux profiles and kinematically rotating kinematics in all
tracers (Paper I). Its \molhy~gas is extended relative to the
\brg~emission, and may have a hint of $>$ 150 km s$^{-1}$ dispersion
of \molhy~gas along the minor axis of the rotating disk. \replaced{Higher
resolution data will be needed to identify potential outflows in this
nucleus.}{It is a potential outflow candidates to be resolved by
higher resolution data in the future.}

% \begin{figure*}[htb]
%   \centering
%   \includegraphics[width=.65\textwidth,trim={0 8.5cm 0 0},clip]{fig_NGC2623_100scale.eps}
%   \caption{Six-panel figure similar to Figure \ref{fig:6panel_app1}, but
%     for NGC 2623.}
%   \label{fig:6panel_app18}
% \end{figure*}

\subsection{NGC 7469N}

NGC 7469 is a Seyfert 1 \added{system} with a disturbed companion $\sim$ 26 kpc
away. It features a $\sim$10$^7$ M$_\odot$ black hole measured
virially from reverberation mapping~\cite[][]{Peterson14} surrounded
by a star-forming ring~\cite[][]{Diaz-Santos07}. The southern nucleus
\deleted{has been observed to} features coronal line biconical
outflows~\cite[see details in][]{Muller-Sanchez11}. Here we have imaged
the northern component of this merger system with OSIRIS. At
0$\farcs$035/spaxel resolution, the nucleus appears compact with
extended diffuse emission in the $K$-band continuum. Interestingly, the peak of
the \brg~flux is offset ($\sim0\farcs2$, or 65pc) from the continuum
peak; its rotational kinematics likely signifies an off-nucleus star
cluster rather than outflowing ionized gas.

% \begin{figure*}[htb]
%   \centering
%   \includegraphics[width=.65\textwidth,trim={0 8.5cm 0 0},clip]{fig_NGC7469N_35scale.eps}
%   \caption{Six-panel figure similar to Figure \ref{fig:6panel_app1}, but
%     for NGC 7469N.}
%   \label{fig:6panel_app19}
% \end{figure*}

\subsection{NGC 6090}
NGC 6090 has a companion nucleus 4 kpc away~\cite[][]{Haan11}. The
main nuclus that we targeted with OSIRIS features large-scale spiral arms traced primarily by
\brg~flux as star formation sites.~\cite{Chisholm16} have detected a
galactic outflow with mass outflow rate of 2.3 M$_\odot$ yr$^{-1}$
using \added{the} \emph{HST} Cosmic Origins Spectrograph. As noted in Paper I, extracting
kinematic information for the gas has proven to be challenging given
the relative lack of gas in the near-infrared range. \added{The} \molhy/\brg~line
ratio map also provides no evidence for the presence of shocked gas in this nucleus. 

% \begin{figure*}[htb]
%   \centering
%   \includegraphics[width=.65\textwidth,trim={0 8.5cm 0 0},clip]{fig_NGC6090_100scale.eps}
%   \caption{Six-panel figure similar to Figure \ref{fig:6panel_app1}, but
%     for NGC 6090.}
%   \label{fig:6panel_app20}
% \end{figure*}

\subsection{NGC 7674W}
NGC 7674W is a famous Seyfert galaxy recently found to be a sub-parsec
binary black hole candidate with projected nuclear separation of
0.35 pc between radio cores~\cite[][]{Kharb17}.  It also has a
companion 20 kpc away~\cite[][]{Haan11}. Its $K$-band
continuum emission is compact while the \brg~gas emission is
extended. There is a relative lack of  \molhy~at the peak of the
$K$-band continuum, \replaced{because that location is
  dominated}{possibly masked} by strong \sivi~emission\replaced{(to be
  presented in a future paper)}{at that location}. While the 
\molhy/\brg~ratio does not reach the canonical value of 2 for shocks,
it does appear enhanced in a coherent region that also features highly
dispersed ($\gtrsim$ 200 km s$^{-1}$) gas. This gas is likely heated
by photoionization by the AGN in the nucleus; a detailed study of the
kinematics will follow. 

% \begin{figure*}[htb]
%   \centering
%   \includegraphics[width=.65\textwidth,trim={0 8.5cm 0 0},clip]{fig_NGC7674W_35scale.eps}
%   \caption{Six-panel figure similar to Figure \ref{fig:6panel_app1}, but
%     for NGC 7674W.}
%   \label{fig:6panel_app21}
% \end{figure*}

\subsection{IRAS F03359+1523}
IRAS F03359+1523 is an early-stage merger with a large, nearly edge-on
disk that has yet to be disrupted (as evidenced from the analysis of
the 100mas data in Paper I). Here, we zoom into the nucleus of the
galaxy with 35mas data. The distribution of the continuum and gas are
consistent with the larger scale data. However, the higher-resolution
data shows that enhanced \molhy/\brg~may correlate with a region of
high \molhy~velocity dispersion ($\sim$ 200 cm s$^{-1}$) even though
the gas has not yet reached \replaced{the}{our conservative} shock threshold.

% \begin{figure*}[htb]
%   \centering
%   \includegraphics[width=.65\textwidth,trim={0 8.5cm 0 0},clip]{fig_IR03359_100scale.eps}
%   \caption{Six-panel figure similar to Figure \ref{fig:6panel_app1}, but
%     for IRAS F03359+1523.}
%   \label{fig:6panel_app22}
% \end{figure*}

% \begin{figure*}[htb]
%   \centering
%   \includegraphics[width=.65\textwidth,trim={0 8.5cm 0 0},clip]{fig_IR03359_35scale.eps}
%   \caption{Six-panel figure similar to Figure \ref{fig:6panel_app1}, but for IRAS F03359+1523.}
%   \label{fig:6panel_app23}
% \end{figure*}

\subsection{MCG +08$-$11$-$002}
MCG +08$-$11$-$002 is a late-stage merging LIRG with two distinct
\replaced{nuclei separated by 133 pc}{clumps}, with the ages of its
stellar clusters thoroughly analyzed in our previous
work~\cite[][]{Davies16}. The extended 
\brg~emission relative to that of the $K$-band continuum illustrates
the strong and clumpy nuclear star formation (Paper I). In the finer
resolution 35mas data taken since Paper I, the two continuum peaks are
better resolved to a separation of 110 pc apart. The \brg~emission
traces the bright but more extended continuum \replaced{peak}{nucleus}
in the northeastern 
direction, leaving the compact continuum peak, \added{likely a star
  cluster}, void of ionized 
gas. The \molhy~gas seems to concentrate on the compact
\replaced{nucleus}{star cluster} but
we see no evidence for shocked or outflowing molecular gas from the
OSIRIS data.

% \begin{figure*}[htb]
%   \centering
%   \includegraphics[width=.65\textwidth,trim={0 8.5cm 0 0},clip]{fig_MCG+08-11-002_100scale.eps}
%   \caption{Six-panel figure similar to Figure \ref{fig:6panel_app1}, but
%     for MCG +08$-$11$-$002.}
%   \label{fig:6panel_app24}
% \end{figure*}

% \begin{figure*}[htb]
%   \centering
%   \includegraphics[width=.65\textwidth,trim={0 8.5cm 0 0},clip]{fig_MCG+08-11-002_35scale.eps}
%   \caption{Six-panel figure similar to Figure \ref{fig:6panel_app1}, but for MCG +08$-$11$-$002.}
%   \label{fig:6panel_app25}
% \end{figure*}

\listofchanges

\end{document}